\pdfoutput=1
\documentclass[aps,pra,twocolumn,superscriptaddress,amsmath,amssymb,floatfix]{revtex4-2}

\usepackage{graphicx}
\usepackage{dcolumn}
\usepackage{bm}
\usepackage{physics}
\usepackage{booktabs}
\usepackage{xcolor}
\usepackage{hyperref}
\usepackage[utf8]{inputenc}
\usepackage[T1]{fontenc}
\hypersetup{
    colorlinks=true,
    linkcolor=blue,
    filecolor=magenta,
    urlcolor=blue,
    citecolor=blue
}

\begin{document}

\title{Multiphoton interference and internal phase references in classical shadow tomography of noisy linear-optical networks}

\author{Attila Baumann}
\affiliation{Institute for Physics and Astronomy, Eötvös Loránd University, Pázmány Péter sétány 1/A, Budapest 1117, Hungary}

\author{Zsolt Kis}
\affiliation{Institute for Solid State Physics and Optics, HUN-REN Wigner Research Centre for Physics, Konkoly-Thege Miklós út 29-33, Budapest 1121, Hungary}

\author{János Koltai}
\affiliation{Institute for Physics and Astronomy, Eötvös Loránd University, Pázmány Péter sétány 1/A, Budapest 1117, Hungary}

\author{Gábor Vattay}
\email[Contact author: ]{gabor.vattay@ttk.elte.hu}
\affiliation{Institute for Physics and Astronomy, Eötvös Loránd University, Pázmány Péter sétány 1/A, Budapest 1117, Hungary}

\date{\today}

\begin{abstract}
Photon-number-resolving detectors record mode occupations while erasing optical phase, yet random linear-optical interferometry enables full reconstruction of multiphoton quantum states. Extending classical shadow tomography to passive $M$-mode photonic networks, however, raises two physical questions: how bosonic interference without photon-photon interactions reshapes information scrambling, and why static interferometer crosstalk halts tomographic convergence at a shot-independent error floor. Solving the two-photon classical shadow channel analytically, we show that its spectrum collapses to three optical visibility eigenvalues, $\{1, \frac{1}{M+1}, \frac{2}{M(M+3)}\}$, which govern the transmission of conserved photon flux, single-photon first-order optical coherence ($G^{(1)}$), and irreducible two-photon Hong-Ou-Mandel correlations ($G^{(2)}$), respectively. Local optical phase invariance conserves the spatial photon-number imbalance $\vec{\Delta} \in \mathbb{Z}^M$ between ket and bra, partitioning the state space into decoupled population, spectator-assisted, and mode-disjoint coherence sectors. Physically, a shared photon in a spectator mode acts as a co-propagating internal phase reference through the interferometer, whereas mode-disjoint superpositions, including two-photon NOON states and dual-rail photonic qubits, lack any internal phase reference. Under static thermo-optic crosstalk and directional-coupler errors modeled by Gaussian Unitary Ensemble perturbations, ensemble averaging generates a number-conserving bosonic mode-hopping Lindblad dissipator confined within each $\vec{\Delta}$ sector. This microscopic dynamics reveals a counterintuitive quantum-optical reversal: Hong-Ou-Mandel bunched coherences ($\ket{i,i}\bra{i,j}$) exhibit both lower statistical shot noise and a smaller coherent-crosstalk error floor than phase-insensitive unbunched coincidence populations ($\ket{i,j}\bra{i,j}$) for every $M \ge 2$. Finally, inverting the noise-calibrated superselection blocks cancels multi-step intra-sector crosstalk and restores $1/N$ shot-noise scaling across $N$ experimental shots.
\end{abstract}

\maketitle

% ==============================================================================
% SECTION I: INTRODUCTION
% ==============================================================================
\section{Introduction}
\label{sec:introduction}

Programmable multiport Mach-Zehnder interferometer meshes~\cite{Reck1994, Clements2016} on integrated photonic chips form the optical backbone of Boson Sampling experiments~\cite{Aaronson2013}, linear-optical quantum computing with dual-rail qubits~\cite{Knill2001, Flamini2018}, and multiphoton quantum interferometry~\cite{Boto2000}. Certifying the multiphoton density matrix $\rho$ prepared inside an $M$-mode network with $n$ indistinguishable photons rapidly outgrows conventional quantum state tomography as the symmetric Fock space $\mathcal{H}_{\text{sym}}^{(n)}$ expands to dimension $D_n = \binom{M+n-1}{n}$~\cite{Preskill2018}. Even in the two-photon sector ($n=2$), $\mathcal{H}_{\text{sym}}^{(2)}$ has dimension $D = \binom{M+1}{2} = M(M+1)/2$, requiring $D^2 - 1 = \mathcal{O}(M^4)$ independent real observables to specify a general state.

Classical shadow tomography~\cite{Huang2020, Struchalin2021} bypasses full measurement tomographies by applying random unitary transformations prior to single-shot projective detection. In multi-qubit processors, random Clifford circuits scramble all traceless operators at a single depolarizing rate because the Clifford group forms a projective unitary 2-design. Passive linear-optical networks obey markedly different physics~\cite{Oszmaniec2017, Hearth2024, Gandhari2024, Conrad2026}. Beam splitters and phase shifters act solely on single-photon creation operators, $a_i^\dagger \to \sum_{p=1}^M U_{pi} a_p^\dagger$, without direct photon-photon interactions. On $\mathcal{H}_{\text{sym}}^{(n)}$, this single-particle transformation $U \in \mathcal{U}(M)$ induces the symmetric tensor representation $\Gamma(U) = \text{Sym}^n(U)$, whose matrix elements are permanents of $n \times n$ submatrices of $U$. Indistinguishable photons correlate across ports purely through single-particle wave interference and Hong-Ou-Mandel bosonic bunching~\cite{Hong1987}, which couples collision states (multiple photons coalescing into one waveguide) to coincidence states with $\sqrt{2}$ bosonic stimulation weights. Sampling $U$ from the Haar measure $d\mu(U)$ on $\mathcal{U}(M)$ therefore fails to form a unitary 2-design on $\mathcal{H}_{\text{sym}}^{(n)}$, producing a non-diagonal shadow channel $\mathbf{S}^{(0)}$ whose brute-force numerical matrix inversion scales as $\mathcal{O}(D^6) = \mathcal{O}(M^{12})$ for $n=2$.

A second physical question arises at the optical readout stage. Photon-number-resolving (PNR) detectors count output photons in each spatial mode without an external homodyne local oscillator, erasing the optical phase of every Fock basis projection. How phase-blind photon counting reconstructs off-diagonal multiphoton phases, and why certain quantum superpositions degrade much faster than others, becomes especially pressing in imperfect hardware. Reconfigurable silicon nitride and silica-on-silicon meshes experience thermo-optic phase crosstalk, directional-coupler splitting tolerances, and voltage-calibration drift~\cite{HardwareHorizon2026}. Unlike incoherent photon loss or depolarizing qubit channels~\cite{Koh2022, Chen2021, Liao2022, Turkeshi2024}, these static hardware imperfections distort each programmed interferometer coherently as $\tilde{U} = U \exp(i\sigma H)$, where $\sigma$ sets the error scale and $H$ is a Hermitian crosstalk generator. In a recent study~\cite{HardwareHorizon2026}, we observed that coherent circuit perturbations arrest the statistical $1/N$ shot-noise decay of photonic state reconstruction at a shot-independent error floor.

In this article, we solve the classical shadow channel of noisy linear-optical networks analytically for two-photon states ($n=2$) and identify the quantum-optical mechanisms that control interferometric visibility, phase sensing, and coherent-noise resilience. Evaluating the Haar-averaged interferometer channel via unitary Weingarten calculus~\cite{Collins2006} and the Bose-Mesner algebra of the Johnson scheme $J(M,2)$, we show that the $D^2 \times D^2$ shadow superoperator $\mathbf{S}^{(0)}$ possesses only three distinct eigenvalues: $\lambda_0 = 1$, $\lambda_1 = \frac{1}{M+1}$, and $\lambda_2 = \frac{2}{M(M+3)}$. Physically, this three-level spectrum mirrors Glauber's hierarchy of quantum-optical coherence orders~\cite{Glauber1963}. The unit eigenvalue $\lambda_0 = 1$ expresses photon-number conservation ($G^{(0)}$). The intermediate eigenvalue $\lambda_1 = \frac{1}{M+1}$ spans the $(M^2-1)$-dimensional subspace of single-photon first-order optical coherences $G^{(1)}_{ij} = a_i^\dagger a_j - \frac{\delta_{ij}}{M}\hat{n}$, whose interferometric visibility decays inversely with the port count $M$. The smallest eigenvalue $\lambda_2 = \frac{2}{M(M+3)}$ spans the remaining $(D^2 - M^2)$-dimensional space of irreducible two-photon correlations ($G^{(2)}$), whose visibility decays quadratically as $\sim 2/M^2$. Projecting onto these three coherence orders yields closed-form rational expressions in $M$ for every entry of $(\mathbf{S}^{(0)})^{-1}$, eliminating numerical matrix inversion.

To explain how phase-blind PNR detectors sense multiphoton phases, we show that invariance of the random circuit ensemble under local phase shifts $U(1)^{\otimes M} \subset \mathcal{U}(M)$ imposes a dynamical superselection rule for arbitrary photon number $n$: the shadow channel conserves the spatial photon-number imbalance vector $\vec{\Delta} \in \mathbb{Z}^M$, which records the mode-by-mode photon difference between ket and bra. For $n=2$, $\vec{\Delta}$ decouples the state space into one $D \times D$ Population block ($\vec{\Delta} = \vec{0}$), $M(M-1)$ spectator-assisted Social Coherence blocks of dimension $M$ (where ket and bra share one photon), and $M(M-1)(M^2-M+2)/4$ isolated $1 \times 1$ Lonely Coherence blocks (where ket and bra share no modes). Physically, a shared photon in a spectator mode acts as a co-propagating internal phase reference traversing the same interferometer as the transitioning photon. Mode-disjoint superpositions, such as two-photon NOON states $(\ket{i,i} + e^{i\phi}\ket{j,j})/\sqrt{2}$~\cite{Boto2000} and dual-rail two-qubit coherences $\ket{i,j}\bra{k,l}$~\cite{Knill2001}, share zero photons between ket and bra, carry zero $G^{(1)}$ weight, and have no internal phase reference.

Averaging the coherently perturbed circuit $\tilde{U} = U\exp(i\sigma H)$ over the Gaussian Unitary Ensemble ($\text{GUE}(M)$) maps static phase and coupler crosstalk onto a number-conserving Gorini-Kossakowski-Sudarshan-Lindblad (GKSL) dissipator $\mathcal{D}(\rho) = 2(M+1)\rho - \sum_{p,q=1}^M L_{pq}\rho L_{pq}^\dagger$ driven by bosonic mode-hopping operators $L_{pq} = a_p^\dagger a_q$. Because $\mathcal{D}$ commutes with local phase rotations, coherent crosstalk diffuses probability and phase strictly within each $\vec{\Delta}$ block. Evaluating both statistical shot noise across the seven spatial collision topologies ($\mathcal{T}_1$--$\mathcal{T}_7$) and coherent hopping bias across the three block sizes $d(\vec{\Delta}) \in \{D, M, 1\}$ groups all transitions into five Quantum Memory Families and uncovers a counterintuitive ordering reversal: a phase-sensitive Hong-Ou-Mandel bunched coherence ($\ket{i,i}\bra{i,j}$) achieves both lower statistical variance and a smaller coherent-crosstalk floor ($2M(2M+1)\sigma^4$) than a phase-insensitive unbunched coincidence population ($\ket{i,j}\bra{i,j}$, floor $2(2M^2+M+2)\sigma^4$) for all $M \ge 2$. Bosonic bunching concentrates three of the four ket-bra photon amplitudes into a single waveguide, halving the number of parasitic hopping channels while retaining a shared reference photon. When $\sigma$ is known from hardware characterization, inverting the noisy $d(\vec{\Delta}) \times d(\vec{\Delta})$ blocks resums intra-sector hopping to all orders, suppressing the $\mathcal{O}(\sigma^4)$ error floor to $\mathcal{O}(\sigma^8)$ and restoring $1/N$ convergence.

Section~\ref{sec:shadow_protocol} formulates the two-photon shadow protocol and evaluates the forward channel via Weingarten calculus. Section~\ref{sec:superselection} derives the local phase superselection rule and explains the physical role of spectator photons versus NOON and dual-rail states. Section~\ref{sec:exact_inversion} establishes the Glauber optical coherence hierarchy ($G^{(0)}, G^{(1)}, G^{(2)}$) and derives the rational inverse catalog. Section~\ref{sec:coherent_noise} derives the bosonic mode-hopping Lindblad dissipator for GUE circuit crosstalk and compares active calibration methods. Section~\ref{sec:duality_catalog} analyzes the shot-noise versus crosstalk duality, the five Quantum Memory Families, and the Hong-Ou-Mandel protection mechanism. Section~\ref{sec:numerical_verification} presents Monte Carlo benchmarks for $M=4$ across bunched, spectator-assisted, and NOON-like superpositions, and Section~\ref{sec:conclusion} discusses experimental implications and extensions to $n$-photon Boson Sampling. Detailed derivations appear in Appendices~\ref{app:weingarten}--\ref{app:dissipator}.

% ==============================================================================
% SECTION II: BOSONIC CLASSICAL SHADOW PROTOCOL AND TRANSITION SUPEROPERATOR
% ==============================================================================
\section{Bosonic Classical Shadow Protocol and the Transition Superoperator}
\label{sec:shadow_protocol}

\subsection{Symmetric Fock Space and Induced Unitary Representation}
\label{subsec:fock_space}

Consider an $M$-mode linear optical interferometer injected with $n$ indistinguishable photons. Let $a_m^\dagger$ and $a_m$ ($m \in \{1, \dots, M\}$) denote the bosonic creation and annihilation operators for spatial mode $m$, satisfying the canonical commutation relations $[a_l, a_m^\dagger] = \delta_{lm}$ and $[a_l, a_m] = [a_l^\dagger, a_m^\dagger] = 0$. The $n$-photon quantum state resides in the symmetric Fock subspace $\mathcal{H}_{\text{sym}}^{(n)}$ spanned by occupation-number states $\ket{n_1, n_2, \dots, n_M} = \prod_{m=1}^M \frac{(a_m^\dagger)^{n_m}}{\sqrt{n_m!}}\ket{\text{vac}}$ with total photon number $\sum_{m=1}^M n_m = n$. The dimension of $\mathcal{H}_{\text{sym}}^{(n)}$ is $\binom{M+n-1}{n}$.

While our symmetry analysis applies to general photon number $n$ (preserving $n$ exclusively for photon number and reserving $N$ for the number of experimental measurement shots), we work out the complete analytical inversion for the two-photon sector ($n=2$), where the Fock space dimension is
\begin{equation}
    D = \binom{M+1}{2} = \frac{M(M+1)}{2} \text{.}
    \label{eq:hilbert_dim}
\end{equation}
We use Greek indices $\alpha, \beta, \mu, \nu, \kappa \in \{1, \dots, D\}$ to label basis states in $\mathcal{H}_{\text{sym}}^{(2)}$. Each basis index $\alpha$ corresponds bijectively to an unordered two-element mode multiset $\{i, j\}$ with $1 \le i \le j \le M$, and has a mode-occupation vector $\vec{n}(\alpha) \in \{0, 1, 2\}^M$ with components $n_m(\alpha) = \delta_{m,i} + \delta_{m,j}$. Depending on whether the two photons occupy the same spatial mode or distinct modes, the basis of $\mathcal{H}_{\text{sym}}^{(2)}$ splits into two geometric subspaces:
\begin{enumerate}
    \item \textbf{Collision Subspace ($\mathcal{C}$):} Both photons occupy the same mode ($i = j$). There are $|\mathcal{C}| = M$ such states,
    \begin{equation}
        \ket{i, i} = \frac{1}{\sqrt{2}} (a_i^\dagger)^2 \ket{\text{vac}} \text{,}
        \label{eq:collision_state_def}
    \end{equation}
    for which the occupation vector satisfies $\langle \vec{n}(\alpha), \vec{n}(\alpha) \rangle = \sum_{m=1}^M n_m(\alpha)^2 = 4$.
    \item \textbf{Distinct Subspace ($\mathcal{D}$):} The two photons occupy different modes ($1 \le i < j \le M$). There are $|\mathcal{D}| = \binom{M}{2} = M(M-1)/2$ such states,
    \begin{equation}
        \ket{i, j} = a_i^\dagger a_j^\dagger \ket{\text{vac}} \text{,}
        \label{eq:distinct_state_def}
    \end{equation}
    for which $\langle \vec{n}(\alpha), \vec{n}(\alpha) \rangle = 2$.
\end{enumerate}
Combining Eqs.~\eqref{eq:collision_state_def} and~\eqref{eq:distinct_state_def}, any two-photon Fock state $\ket{\nu}$ with $\nu = \{i, j\}$ is written in unified form as $\ket{\nu} = \frac{1}{\sqrt{1+\delta_{ij}}} a_i^\dagger a_j^\dagger \ket{\text{vac}}$.

A passive linear optical circuit implements a single-particle unitary matrix $U \in \mathcal{U}(M)$ that transforms the creation operators according to $U a_i^\dagger U^\dagger = \sum_{k=1}^M U_{ki} a_k^\dagger$. Acting on a two-photon input state $\ket{\nu} = \ket{i, j}$, the induced unitary operator $V(U) \equiv \Gamma(U)$ on $\mathcal{H}_{\text{sym}}^{(2)}$ yields
\begin{equation}
    V(U)\ket{i, j} = \frac{1}{\sqrt{1+\delta_{ij}}} \sum_{k,l=1}^M U_{ki} U_{lj} a_k^\dagger a_l^\dagger \ket{\text{vac}} \text{.}
\end{equation}
Projecting onto an output basis state $\bra{\mu} = \bra{k, l}$ with $k \le l$ and using $\bra{k, l} a_p^\dagger a_q^\dagger \ket{\text{vac}} = \frac{\delta_{kp}\delta_{lq} + \delta_{lp}\delta_{kq}}{\sqrt{1+\delta_{kl}}}$ gives the symmetrized permanent representation of $V(U) \in \mathcal{U}(D)$:
\begin{equation}
    V_{\mu\nu}(U) = \bra{\mu} V(U) \ket{\nu} = \frac{U_{ki}U_{lj} + U_{li}U_{kj}}{\sqrt{(1+\delta_{kl})(1+\delta_{ij})}} \text{.}
    \label{eq:V_munu}
\end{equation}
The map $\Gamma: \mathcal{U}(M) \to \mathcal{U}(D)$ is a group homomorphism, satisfying $V(U_1 U_2) = V(U_1)V(U_2)$ and $V(U^\dagger) = V^\dagger(U)$. Because $\Gamma(\mathcal{U}(M))$ is a proper subgroup of $\mathcal{U}(D)$ for all $M \ge 2$, averaging over $U \sim \text{Haar}(M)$ does not reproduce the Haar measure on $\mathcal{U}(D)$, and the collision factors $\sqrt{1+\delta_{ij}}$ introduce mode-overlap weights into the shadow channel.

\subsection{Projective Measurement, Raw Snapshots, and the Expected Shadow Signal}
\label{subsec:shadow_snapshots}

In each experimental shot $s \in \{1, \dots, N\}$ of the classical shadow protocol, a random single-particle unitary $U_s \sim \text{Haar}(M)$ is dialed into the optical interferometer, rotating the input two-photon density matrix $\rho$ to $V(U_s) \rho V^\dagger(U_s)$. Photon-number-resolving detectors at the $M$ output ports then perform a projective measurement in the Fock basis $\{\ket{\kappa}\}_{\kappa=1}^D$. By Born's rule, the conditional probability of observing the two-photon outcome $\kappa = \{m_1, m_2\}$ given the circuit unitary $U$ is
\begin{equation}
\begin{split}
    p(\kappa \mid U) &= \mathbb{E}_{\text{shots}}\left[\rho^{\text{exp}}_{\kappa\kappa} \mid U\right] = \bra{\kappa} V(U) \rho V^\dagger(U) \ket{\kappa} \\
    &= \sum_{\alpha,\beta=1}^D V_{\kappa\alpha}(U) \rho_{\alpha\beta} V^*_{\kappa\beta}(U) \text{,}
\end{split}
\label{eq:born_rule}
\end{equation}
where $\rho^{\text{exp}} = \ket{\kappa}\bra{\kappa}$ denotes the single-shot post-measurement projector.

Following the classical shadow procedure, the observed outcome $\ket{\kappa}\bra{\kappa}$ is rotated backward in software by applying the inverse unitary $V^\dagger(U)$ to form the single-shot pre-inversion snapshot matrix $I^{(\text{snap})}$:
\begin{equation}
    I^{(\text{snap})} = V^\dagger(U) \ket{\kappa}\bra{\kappa} V(U) \text{,} \quad I^{(\text{snap})}_{\mu\nu} = V^*_{\kappa\mu}(U) V_{\kappa\nu}(U) \text{.}
    \label{eq:snapshot_def}
\end{equation}
Over $N$ independent measurement shots $\{(U_s, \kappa_s)\}_{s=1}^N$, the empirical shadow signal is the sample average $\hat{I}_{\mu\nu} = \frac{1}{N}\sum_{s=1}^N I^{(\text{snap}, s)}_{\mu\nu}$. In the limit $N \to \infty$, $\hat{I}_{\mu\nu}$ converges to the expected shadow signal $\overline{I}_{\mu\nu} = \mathbb{E}_{U, \kappa}[I^{(\text{snap})}_{\mu\nu}]$:
\begin{equation}
\begin{split}
    \overline{I}_{\mu\nu} &= \mathbb{E}_{U \sim \text{Haar}(M)} \left[ \sum_{\kappa=1}^D V^*_{\kappa\mu}(U) \, p(\kappa \mid U) \, V_{\kappa\nu}(U) \right] \\
    &= \sum_{\alpha,\beta=1}^D \rho_{\alpha\beta} \sum_{\kappa=1}^D \mathbb{E}_{U} \left[ V_{\kappa\alpha}(U) V_{\kappa\nu}(U) V^*_{\kappa\mu}(U) V^*_{\kappa\beta}(U) \right] \text{.}
\end{split}
\label{eq:I_munu_exp}
\end{equation}
Equation~\eqref{eq:I_munu_exp} defines a linear map $\overline{I}_{\mu\nu} = \sum_{\alpha,\beta=1}^D S^{(0)}_{\mu\nu, \alpha\beta} \rho_{\alpha\beta}$ between the target density matrix $\rho$ and the expected shadow signal $\overline{I}$, governed by the rank-4 forward transition superoperator
\begin{equation}
    S^{(0)}_{\mu\nu, \alpha\beta} = \sum_{\kappa=1}^D \int_{\mathcal{U}(M)} V_{\kappa\alpha}(U) V_{\kappa\nu}(U) V^*_{\kappa\mu}(U) V^*_{\kappa\beta}(U) \, d\mu(U) \text{.}
    \label{eq:S0_tensor_def}
\end{equation}
Reconstructing the quantum state $\rho$ from the measured shadow signal $\overline{I}$ requires inverting this linear superoperator:
\begin{equation}
    \rho_{\alpha\beta} = \sum_{\mu,\nu=1}^D \left( \mathbf{S}^{(0)} \right)^{-1}_{\alpha\beta, \mu\nu} \overline{I}_{\mu\nu} \text{.}
    \label{eq:global_reconstruction}
\end{equation}

\subsection{Evaluation via Degree-8 Unitary Weingarten Calculus}
\label{subsec:weingarten_evaluation}

To express $S^{(0)}_{\mu\nu,\alpha\beta}$ in terms of single-particle matrix elements $U_{rc}$, we assign spatial mode pairs to the five Fock indices: measurement outcome $\kappa = \{m_1, m_2\}$, input bra $\alpha = \{i_1, i_2\}$, input ket $\beta = \{j_1, j_2\}$, output bra $\mu = \{k_1, k_2\}$, and output ket $\nu = \{l_1, l_2\}$. Substituting Eq.~\eqref{eq:V_munu} for each of the four factors in Eq.~\eqref{eq:S0_tensor_def} collects the square-root symmetrization factors into a single normalization denominator:
\begin{equation}
\begin{split}
    \mathcal{N}_{\kappa, \mu, \nu, \alpha, \beta} &= (1+\delta_{m_1 m_2})^2 \\
    &\quad \times \big[ (1+\delta_{i_1 i_2})(1+\delta_{l_1 l_2}) \\
    &\quad\quad \times (1+\delta_{k_1 k_2})(1+\delta_{j_1 j_2}) \big]^{1/2} \text{,}
\end{split}
\label{eq:norm_factor}
\end{equation}
where the factor $(1+\delta_{m_1 m_2})$ appears squared because the measurement outcome $\kappa$ enters all four matrix elements $V_{\kappa\alpha}$, $V_{\kappa\nu}$, $V^*_{\kappa\mu}$, and $V^*_{\kappa\beta}$.

Each two-photon matrix element in the numerator is a sum over the symmetric group $S_2$ permuting the two output detector modes $(m_1, m_2)$. Defining the non-conjugated degree-4 polynomial $\Sigma_U(\kappa, \alpha, \nu)$ and the conjugated degree-4 polynomial $\Sigma_{U^*}(\kappa, \mu, \beta)$ via four independent permutations $w = (\pi_1, \pi_2, \pi_3, \pi_4) \in \Omega \equiv S_2 \times S_2 \times S_2 \times S_2$,
\begin{align}
    \Sigma_U(\kappa, \alpha, \nu) &= \sum_{\pi_1, \pi_2 \in S_2} U_{m_{\pi_1(1)} i_1} U_{m_{\pi_1(2)} i_2} \nonumber \\
    &\quad \times U_{m_{\pi_2(1)} l_1} U_{m_{\pi_2(2)} l_2} \text{,} \label{eq:Sigma_U} \\
    \Sigma_{U^*}(\kappa, \mu, \beta) &= \sum_{\pi_3, \pi_4 \in S_2} U^*_{m_{\pi_3(1)} k_1} U^*_{m_{\pi_3(2)} k_2} \nonumber \\
    &\quad \times U^*_{m_{\pi_4(1)} j_1} U^*_{m_{\pi_4(2)} j_2} \text{,} \label{eq:Sigma_Ustar}
\end{align}
their product expands into $|\Omega| = 2^4 = 16$ degree-8 Haar integral monomials (written out in full in Eq.~\eqref{eq:supp_16_integrals} of the Supplemental Material):
\begin{equation}
\begin{split}
    S^{(0)}_{\mu\nu, \alpha\beta} &= \sum_{\kappa=1}^D \frac{1}{\mathcal{N}_{\kappa, \mu, \nu, \alpha, \beta}} \sum_{w \in \Omega} \int_{\mathcal{U}(M)} \left( \prod_{x=1}^4 U_{r^{(w)}_x c_x} \right) \\
    &\quad \times \left( \prod_{y=1}^4 U^*_{{r'}^{(w)}_y c'_y} \right) d\mu(U) \text{.}
\end{split}
\label{eq:S0_16_sum}
\end{equation}
Here, for each permutation tuple $w = (\pi_1, \pi_2, \pi_3, \pi_4) \in \Omega$, the row multi-indices $\mathbf{r}^{(w)}, {\mathbf{r}'}^{(w)} \in \{1, \dots, M\}^4$ contain the permuted detector modes, while the column multi-indices $\mathbf{c}, \mathbf{c}' \in \{1, \dots, M\}^4$ contain the external state modes and are independent of $w$:
\begin{align}
    \mathbf{r}^{(w)} &= \big( m_{\pi_1(1)}, \, m_{\pi_1(2)}, \, m_{\pi_2(1)}, \, m_{\pi_2(2)} \big) \text{,} \label{eq:row_vec_U} \\
    {\mathbf{r}'}^{(w)} &= \big( m_{\pi_3(1)}, \, m_{\pi_3(2)}, \, m_{\pi_4(1)}, \, m_{\pi_4(2)} \big) \text{,} \label{eq:row_vec_Ustar} \\
    \mathbf{c} &= \big( i_1, \, i_2, \, l_1, \, l_2 \big) \equiv \alpha \uplus \nu \text{,} \label{eq:col_vec_U} \\
    \mathbf{c}' &= \big( k_1, \, k_2, \, j_1, \, j_2 \big) \equiv \mu \uplus \beta \text{.} \label{eq:col_vec_Ustar}
\end{align}

By unitary Weingarten calculus of degree $q=4$~\cite{Collins2006}, each degree-8 Haar integral over $\mathcal{U}(M)$ contracts into a double sum over permutations $\sigma, \tau \in S_4$:
\begin{equation}
\begin{split}
    &\int_{\mathcal{U}(M)} \left( \prod_{x=1}^4 U_{r_x c_x} \right) \left( \prod_{y=1}^4 U^*_{r'_y c'_y} \right) d\mu(U) \\
    &\quad = \sum_{\sigma, \tau \in S_4} \delta_\sigma(\mathbf{r}, \mathbf{r}') \, \delta_\tau(\mathbf{c}, \mathbf{c}') \, \text{Wg}(\sigma^{-1}\tau, M) \text{,}
\end{split}
\label{eq:weingarten_general}
\end{equation}
where $\delta_\sigma(\mathbf{r}, \mathbf{r}') = \prod_{x=1}^4 \delta_{r_x, r'_{\sigma(x)}}$ and $\delta_\tau(\mathbf{c}, \mathbf{c}') = \prod_{x=1}^4 \delta_{c_x, c'_{\tau(x)}}$ enforce index-by-index mode matching, and $\text{Wg}(\pi, M)$ is the unitary Weingarten function of $S_4$ (given explicitly in Appendix~\ref{app:weingarten}). Because $\mathbf{c}$ and $\mathbf{c}'$ do not depend on $w \in \Omega$, $\delta_\tau(\mathbf{c}, \mathbf{c}')$ factors outside the sum over $\Omega$, yielding the exact closed-form Weingarten representation of the transition superoperator:
\begin{equation}
\begin{split}
    S^{(0)}_{\mu\nu, \alpha\beta} &= \sum_{\kappa=1}^D \frac{1}{\mathcal{N}_{\kappa, \mu, \nu, \alpha, \beta}} \sum_{\tau \in S_4} \delta_\tau(\mathbf{c}, \mathbf{c}') \\
    &\quad \times \left( \sum_{\sigma \in S_4} \text{Wg}(\sigma^{-1}\tau, M) \sum_{w \in \Omega} \delta_\sigma\big(\mathbf{r}^{(w)}, {\mathbf{r}'}^{(w)}\big) \right) \text{.}
\end{split}
\label{eq:S0_weingarten_master}
\end{equation}

% ==============================================================================
% SECTION III: LOCAL PHASE SUPERSELECTION AND INTERNAL OPTICAL PHASE REFERENCES
% ==============================================================================
\section{Local Phase Superselection and Internal Optical Phase References}
\label{sec:superselection}

\subsection{Maximal Torus Phase Rotations and the Photon Imbalance Vector}
\label{subsec:maximal_torus}

Evaluating Eq.~\eqref{eq:S0_weingarten_master} across all $D^4$ index quadruples $(\mu, \nu, \alpha, \beta)$ and inverting the resulting $D^2 \times D^2$ matrix numerically would require $\mathcal{O}(D^6) = \mathcal{O}(M^{12})$ operations. However, the presence of the column Kronecker delta $\delta_\tau(\mathbf{c}, \mathbf{c}')$ in Eq.~\eqref{eq:S0_weingarten_master} reflects a continuous gauge symmetry of the Haar measure that renders almost all entries of $\mathbf{S}^{(0)}$ identically zero.

Consider the maximal torus $U(1)^{\otimes M} \subset \mathcal{U}(M)$ consisting of diagonal single-particle phase-shift matrices
\begin{equation}
    D(\vec{\theta}) = \text{diag}\big(e^{i\theta_1}, e^{i\theta_2}, \dots, e^{i\theta_M}\big) \text{,}
    \label{eq:diagonal_torus}
\end{equation}
parameterized by $M$ independent phase angles $\vec{\theta} = (\theta_1, \dots, \theta_M) \in [0, 2\pi)^M$. On the $n$-photon symmetric Fock space $\mathcal{H}_{\text{sym}}^{(n)}$, $D(\vec{\theta})$ induces the diagonal unitary operator
\begin{equation}
    U_{\vec{\theta}} \equiv \Gamma\big(D(\vec{\theta})\big) = \exp\left( i \sum_{m=1}^M \theta_m \hat{n}_m \right) \text{,}
    \label{eq:fock_torus_operator}
\end{equation}
where $\hat{n}_m = a_m^\dagger a_m$ is the photon number operator for mode $m$.

Under the adjoint action $\text{Ad}_{U_{\vec{\theta}}}(\rho) = U_{\vec{\theta}} \rho U_{\vec{\theta}}^\dagger$, any $n$-photon transition operator $\ket{\alpha}\bra{\beta}$ between occupation states $\ket{\alpha}$ and $\ket{\beta}$ is an eigenoperator of $\text{Ad}_{U_{\vec{\theta}}}$:
\begin{equation}
\begin{split}
    \text{Ad}_{U_{\vec{\theta}}}\big(\ket{\alpha}\bra{\beta}\big) &= U_{\vec{\theta}} \ket{\alpha}\bra{\beta} U_{\vec{\theta}}^\dagger \\
    &= e^{i \sum_{m=1}^M \theta_m [n_m(\alpha) - n_m(\beta)]} \ket{\alpha}\bra{\beta} \text{.}
\end{split}
\label{eq:adjoint_eigenvalue_general}
\end{equation}
We define the \textit{photon imbalance vector} $\vec{\Delta}_{\alpha\beta} \in \mathbb{Z}^M$ associated with the density-matrix element $\rho_{\alpha\beta} = \bra{\alpha}\rho\ket{\beta}$ as the difference between the mode-occupation vectors of $\alpha$ and $\beta$:
\begin{equation}
    \Delta_m(\alpha, \beta) = n_m(\alpha) - n_m(\beta) \text{.}
    \label{eq:delta_general_n}
\end{equation}
For the two-photon case ($n=2$) with $\alpha = \{i_1, i_2\}$ and $\beta = \{j_1, j_2\}$, Eq.~\eqref{eq:delta_general_n} takes the explicit form
\begin{equation}
    \Delta_m(\alpha, \beta) = (\delta_{m,i_1} + \delta_{m,i_2}) - (\delta_{m,j_1} + \delta_{m,j_2}) \text{.}
    \label{eq:delta_vector_definition}
\end{equation}
Because both $\ket{\alpha}$ and $\ket{\beta}$ contain exactly $n$ photons ($\sum_{m=1}^M n_m(\alpha) = \sum_{m=1}^M n_m(\beta) = n$), every photon imbalance vector satisfies two conservation laws for arbitrary photon number $n$:
\begin{equation}
    \sum_{m=1}^M \Delta_m(\alpha, \beta) = 0 \text{,} \qquad \|\vec{\Delta}_{\alpha\beta}\|_1 = \sum_{m=1}^M |\Delta_m(\alpha, \beta)| \le 2n \text{.}
    \label{eq:delta_conservation_laws}
\end{equation}
With this definition, Eq.~\eqref{eq:adjoint_eigenvalue_general} simplifies to the compact eigenvalue relation
\begin{equation}
    \text{Ad}_{U_{\vec{\theta}}}\big(\ket{\alpha}\bra{\beta}\big) = e^{i \vec{\theta} \cdot \vec{\Delta}_{\alpha\beta}} \ket{\alpha}\bra{\beta} \text{.}
    \label{eq:adjoint_eigenvalue}
\end{equation}

\subsection{Proof of the Dynamical Superselection Rule}
\label{subsec:superselection_proof}

The normalized Haar measure $d\mu(U)$ on $\mathcal{U}(M)$ is invariant under right multiplication by any fixed unitary matrix $D(\vec{\theta}) \in U(1)^{\otimes M}$: $d\mu(U D(\vec{\theta})) = d\mu(U)$. Under the substitution $U \to U D(\vec{\theta})$, the induced Fock unitary transforms as $V(U D(\vec{\theta})) = V(U) U_{\vec{\theta}}$, and its matrix elements acquire the phases $V_{\kappa\alpha}(U D(\vec{\theta})) = V_{\kappa\alpha}(U) e^{i\vec{\theta}\cdot\vec{n}(\alpha)}$. Substituting $U \to U D(\vec{\theta})$ into the integral definition~\eqref{eq:S0_tensor_def} of $S^{(0)}_{\mu\nu, \alpha\beta}$ yields
\begin{equation}
    S^{(0)}_{\mu\nu, \alpha\beta} = e^{i \vec{\theta} \cdot (\vec{\Delta}_{\alpha\beta} - \vec{\Delta}_{\mu\nu})} S^{(0)}_{\mu\nu, \alpha\beta} \text{.}
    \label{eq:phase_invariance_identity}
\end{equation}
Equivalently, at the superoperator level, right-invariance of the Haar measure implies that $\mathbf{S}^{(0)}$ commutes with the adjoint action of the maximal torus, $[\mathbf{S}^{(0)}, \text{Ad}_{U_{\vec{\theta}}}] = 0$ for all $\vec{\theta} \in [0, 2\pi)^M$. Integrating Eq.~\eqref{eq:phase_invariance_identity} over the $M$-dimensional torus $\frac{1}{(2\pi)^M}\int_{[0,2\pi)^M} d^M\theta$ projects onto the Fourier mode $\vec{\Delta}_{\alpha\beta} - \vec{\Delta}_{\mu\nu} = \vec{0}$ and establishes the \textit{Dynamical Superselection Rule}:
\begin{equation}
    S^{(0)}_{\mu\nu, \alpha\beta} = 0 \quad \text{whenever} \quad \vec{\Delta}_{\alpha\beta} \neq \vec{\Delta}_{\mu\nu} \text{.}
    \label{eq:superselection_main}
\end{equation}

In terms of the Weingarten expansion~\eqref{eq:S0_weingarten_master}, the condition $\vec{\Delta}_{\alpha\beta} = \vec{\Delta}_{\mu\nu}$ is equivalent to $\vec{n}(\alpha) + \vec{n}(\nu) = \vec{n}(\mu) + \vec{n}(\beta)$, which states that the column multi-indices $\mathbf{c} = \alpha \uplus \nu$ and $\mathbf{c}' = \mu \uplus \beta$ are identical as multisets, so that at least one permutation $\tau \in S_4$ satisfies $\delta_\tau(\mathbf{c}, \mathbf{c}') = 1$.

Because of Eq.~\eqref{eq:superselection_main}, the global transition superoperator decomposes into a direct sum of uncoupled block matrices indexed by the conserved imbalance vector $\vec{\Delta}$:
\begin{equation}
    \mathbf{S}^{(0)} = \bigoplus_{\vec{\Delta}} \mathbf{S}^{(0)}_{\vec{\Delta}} \text{,} \qquad \left(\mathbf{S}^{(0)}\right)^{-1} = \bigoplus_{\vec{\Delta}} \left(\mathbf{S}^{(0)}_{\vec{\Delta}}\right)^{-1} \text{.}
    \label{eq:direct_sum_decomposition}
\end{equation}
To reconstruct a specific density-matrix element $\rho_{\alpha\beta}$ with target imbalance vector $\vec{\Delta}^* \equiv \vec{\Delta}_{\alpha\beta}$, one only needs to invert the sub-block $\mathbf{S}^{(0)}_{\vec{\Delta}^*}$ acting on the preimage sector
\begin{equation}
    \mathcal{V}_{\vec{\Delta}^*} = \left\{ (\mu, \nu) \in \{1, \dots, D\}^2 \;\Big|\; \vec{\Delta}_{\mu\nu} = \vec{\Delta}^* \right\} \text{,}
    \label{eq:preimage_sector}
\end{equation}
reducing Eq.~\eqref{eq:global_reconstruction} to the localized block estimator
\begin{equation}
    \rho_{\alpha\beta} = \sum_{(\mu,\nu) \in \mathcal{V}_{\vec{\Delta}^*}} \left( \mathbf{S}^{(0)}_{\vec{\Delta}^*} \right)^{-1}_{\alpha\beta, \mu\nu} \overline{I}_{\mu\nu} \text{.}
    \label{eq:decoupled_estimator}
\end{equation}

\subsection{Classification of Superselection Sectors and Block Degeneracies}
\label{subsec:sector_classification}

\begin{figure}[htbp]
    \centering
    \includegraphics[width=\linewidth]{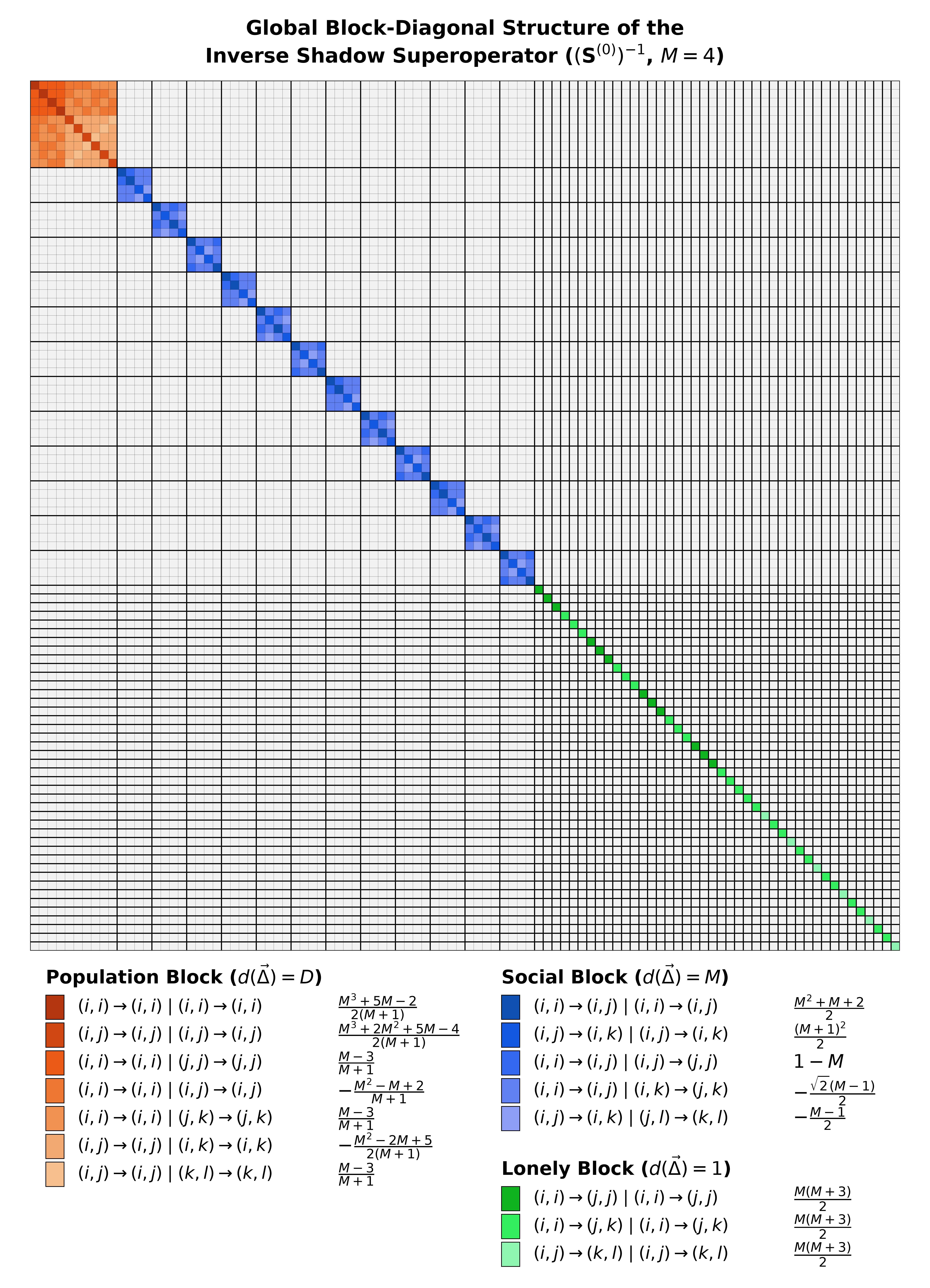}
    \caption{\textbf{Block-diagonal decomposition of the inverse shadow superoperator $(\mathbf{S}^{(0)})^{-1}$ for $M=4$ ($D=10$).} Continuous gauge invariance under $U(1)^{\otimes M}$ partitions the $100 \times 100$ transition space into one $10 \times 10$ Population Block ($\vec{\Delta} = \vec{0}$), twelve $4 \times 4$ Social Coherence Blocks ($d(\vec{\Delta}) = 4$), and 42 isolated $1 \times 1$ Lonely Coherence Blocks ($d(\vec{\Delta}) = 1$).}
    \label{fig:global_inverse_matrix}
\end{figure}

The dimension of each superselection block is given by the degeneracy $d(\vec{\Delta}) = |\mathcal{V}_{\vec{\Delta}}|$, which counts the number of basis transitions $(\mu, \nu)$ sharing the same imbalance vector $\vec{\Delta}$. For two photons ($n=2$), the $\ell_1$-norm constraint $\|\vec{\Delta}\|_1 \le 4$ and zero-sum constraint $\sum_{m=1}^M \Delta_m = 0$ restrict $d(\vec{\Delta})$ to only three possible values: $d(\vec{\Delta}) \in \{D, M, 1\}$, illustrated for $M=4$ in Fig.~\ref{fig:global_inverse_matrix} and summarized in Table~\ref{tab:delta_classification}.

\begin{table*}[htbp]
\centering
\caption{Classification of the $D^2$ two-photon density-matrix elements by their superselection sector $\vec{\Delta}$, block dimension $d(\vec{\Delta})$, and sector multiplicity across $M$ spatial modes.}
\label{tab:delta_classification}
\begin{ruledtabular}
\begin{tabular}{lccc}
Architecture Class & $\|\vec{\Delta}\|_1$ & Block Size $d(\vec{\Delta})$ & Number of Sectors \\
\hline
\textbf{I. Populations ($\mathcal{P}$)} & $0$ & $D = \frac{M(M+1)}{2}$ & $1$ \\
\textbf{II. Social Coherences ($\mathcal{C}_S$)} & $2$ & $M$ & $M(M-1)$ \\
\textbf{III. Lonely Coherences ($\mathcal{C}_L$)} & $4$ & $1$ & $\frac{M(M-1)(M^2-M+2)}{4}$ \\
\end{tabular}
\end{ruledtabular}
\end{table*}

We examine the three dynamical architectures in turn:
\begin{enumerate}
    \item \textbf{Populations ($\mathcal{P}$, $\|\vec{\Delta}\|_1 = 0$, $d = D$):} The imbalance vector vanishes identically, $\vec{\Delta} = (0, 0, \dots, 0)$, which occurs if and only if $\mu = \nu$. This single sector contains all $D = M(M+1)/2$ diagonal population elements $\ket{i,j}\bra{i,j}$ ($1 \le i \le j \le M$), which mix with one another under Haar-random interferometry through multi-photon interference.
    \item \textbf{Social Coherences ($\mathcal{C}_S$, $\|\vec{\Delta}\|_1 = 2$, $d = M$):} The bra and ket share exactly one photon in a common spectator mode while the second photon transitions between two distinct modes $i \neq j$. The imbalance vector $\vec{\Delta}$ has $+1$ at position $i$, $-1$ at position $j$, and $0$ in the remaining $M-2$ modes. For each ordered pair $(i, j)$ with $i \neq j$ (giving $M(M-1)$ distinct sectors), the sector $\mathcal{V}_{\vec{\Delta}}$ has dimension $d(\vec{\Delta}) = M$, spanned by two collision-assisted transitions $\ket{i,i}\bra{i,j}$ and $\ket{i,j}\bra{j,j}$ together with $M-2$ spectator-mode transitions $\ket{i,k}\bra{j,k}$ ($k \notin \{i, j\}$).
    \item \textbf{Lonely Coherences ($\mathcal{C}_L$, $\|\vec{\Delta}\|_1 = 4$, $d = 1$):} The bra $\alpha$ and ket $\beta$ share no common spatial modes ($\alpha \cap \beta = \emptyset$). Because all four creation and annihilation indices are fixed by the non-zero entries of $\vec{\Delta}$, each such transition forms an isolated $1 \times 1$ scalar block ($d(\vec{\Delta}) = 1$). Depending on whether $\alpha$ and $\beta$ are collision or distinct states, Lonely Coherences divide into three sub-orbits:
    \begin{itemize}
        \item $\vec{\Delta} \in \text{Perm}(+2, -2, 0, \dots, 0)$: transitions $\ket{i,i}\bra{j,j}$ between two collision states, giving $M(M-1)$ blocks;
        \item $\vec{\Delta} \in \text{Perm}(+2, -1, -1, 0, \dots, 0)$ or $\text{Perm}(+1, +1, -2, 0, \dots, 0)$: transitions $\ket{i,i}\bra{j,k}$ or $\ket{j,k}\bra{i,i}$ between a collision state and a disjoint distinct state, giving $2 \times M\binom{M-1}{2} = M(M-1)(M-2)$ blocks;
        \item $\vec{\Delta} \in \text{Perm}(+1, +1, -1, -1, 0, \dots, 0)$: transitions $\ket{i,j}\bra{k,l}$ between four mutually distinct modes, giving $\binom{M}{2}\binom{M-2}{2} = M(M-1)(M-2)(M-3)/4$ blocks.
    \end{itemize}
    Summing the three sub-orbits yields $M(M-1)\big[1 + (M-2) + (M-2)(M-3)/4\big] = M(M-1)(M^2-M+2)/4$ isolated $1 \times 1$ sectors.
\end{enumerate}

Summing the dimensions of all superselection sectors verifies completeness across the $D^2$-dimensional Liouville space:
\begin{equation}
\begin{split}
    &D + M(M-1)\cdot M + \frac{M(M-1)(M^2-M+2)}{4}\cdot 1 \\
    &\quad = \frac{M^2(M+1)^2}{4} = D^2 \text{.}
\end{split}
\end{equation}
Similarly, the total number of non-zero entries in the $D^2 \times D^2$ matrix $\mathbf{S}^{(0)}$ (and in its inverse $(\mathbf{S}^{(0)})^{-1}$) is reduced from $D^4 = \mathcal{O}(M^8)$ to
\begin{equation}
\begin{split}
    N_{\text{nz}}(M) &= D^2 + M^3(M-1) \\
    &\quad + \frac{M(M-1)(M^2-M+2)}{4} \\
    &= \frac{M(3M^3 - M^2 + 3M - 1)}{2} \text{,}
\end{split}
\end{equation}
which equals $N_{\text{nz}}(4) = 374$ out of $10{,}000$ entries for $M=4$.

\subsection{Phase-Blind Detection, Spectator References, and NOON versus Dual-Rail Coherences}
\label{subsec:phase_references}

The block decomposition in Table~\ref{tab:delta_classification} clarifies how phase-blind photon-number-resolving (PNR) detectors measure multiphoton quantum phases without an external homodyne local oscillator. At the output ports of the interferometer, PNR detection projects onto occupation states $\ket{\kappa}\bra{\kappa}$ that commute with all local phase rotations $U_{\vec{\theta}}$, erasing optical phase at the detector plane. Phase sensitivity enters through the preceding interferometer $U \in \mathcal{U}(M)$, which interferes distinct input modes before detection so that input relative phases modulate the output coincidence probabilities $p(\kappa \mid U)$, while the classical back-projection $V^\dagger(U)\ket{\kappa}\bra{\kappa}V(U)$ restores the spatial phase signature $\vec{\Delta}_{\alpha\beta}$.

The superselection rule~\eqref{eq:superselection_main} separates off-diagonal multiphoton coherences into two physically distinct interferometric regimes depending on whether ket and bra share a photon:
\begin{itemize}
    \item \textbf{Spectator-Assisted Internal Phase Reference ($\mathcal{C}_S$, $\|\vec{\Delta}\|_1 = 2$, $d = M$):} In transitions $\ket{i,k}\bra{j,k}$ (or collision-assisted transitions $\ket{i,i}\bra{i,j}$), one photon is in a superposition between waveguides $i$ and $j$, while the second photon occupies a shared waveguide ($k$ or $i$) common to both ket and bra. Physically, this shared photon acts as a co-propagating internal phase reference (an internal local oscillator) traversing the same random interferometer $U$ as the signal photon. Because all $M$ choices of the reference waveguide carry the identical phase charge $\Delta_i = +1, \Delta_j = -1$, random beam splitting mixes them coherently inside an $M \times M$ block.
    \item \textbf{Reference-Free Two-Photon Interference ($\mathcal{C}_L$, $\|\vec{\Delta}\|_1 = 4$, $d = 1$):} The three Lonely Coherence sub-orbits correspond to prominent nonclassical states in quantum optics and photonic quantum computing: (i) $\ket{i,i}\bra{j,j}$ represents the off-diagonal phase coherence of a two-photon NOON state $(\ket{2}_i\ket{0}_j + e^{i\phi}\ket{0}_i\ket{2}_j)/\sqrt{2}$~\cite{Boto2000}; (ii) $\ket{i,i}\bra{j,k}$ represents Hong-Ou-Mandel coherence between a bunched photon pair and a disjoint coincidence pair~\cite{Hong1987}; and (iii) $\ket{i,j}\bra{k,l}$ represents two-qubit entangling coherences of dual-rail photonic qubits~\cite{Knill2001, Flamini2018}, where logical states are encoded by single photons in disjoint waveguide pairs. Because $\alpha \cap \beta = \emptyset$ in all three sub-orbits, neither photon remains in a common waveguide to serve as a phase reference. Extracting their phase requires simultaneous two-photon interference of both particles, isolating every such coherence in a $1 \times 1$ scalar sector.
\end{itemize}

% ==============================================================================
% SECTION IV: GLAUBER OPTICAL COHERENCE HIERARCHY AND ANALYTICAL INVERSION
% ==============================================================================
\section{Glauber Optical Coherence Hierarchy and Analytical Channel Inversion}
\label{sec:exact_inversion}

\subsection{Eight-Index Multisets and the 26 Non-Zero Equivalence Classes}
\label{subsec:26_classes}

Even after decomposing $\mathbf{S}^{(0)}$ into blocks of size $D$, $M$, and $1$, numerical inversion of the $D \times D$ Population Block would still scale as $\mathcal{O}(D^3) = \mathcal{O}(M^6)$. We bypass numerical inversion altogether by exploiting the permutation symmetry of the Haar measure under the symmetric group $S_M$ acting on the $M$ mode labels.

Each matrix element $S^{(0)}_{\mu\nu, \alpha\beta}$ depends on eight spatial mode indices: $\alpha = \{i_1, i_2\}$, $\beta = \{j_1, j_2\}$, $\mu = \{k_1, k_2\}$, and $\nu = \{l_1, l_2\}$. Because $d\mu(P_\pi U P_\pi^\dagger) = d\mu(U)$ for any mode-permutation matrix $P_\pi$ ($\pi \in S_M$), the integral~\eqref{eq:S0_weingarten_master} depends only on the partition of the eight-index multiset $S_8 = \alpha \uplus \beta \uplus \mu \uplus \nu$ into subsets of equal mode indices, together with the pairing of indices inside $\alpha, \beta, \mu, \nu$. As cataloged in Table~\ref{tab:supp_22_partitions} of the Supplemental Material, there are 22 integer partitions of 8. The superselection rule $\alpha \uplus \nu = \mu \uplus \beta$ requires every mode index in $S_8$ to appear an even number of times, eliminating 17 partitions and leaving only the five all-even partitions $[8]$, $[6,2]$, $[4,4]$, $[4,2,2]$, and $[2,2,2,2]$.

Enumerating the distinct index pairings within these five partitions yields 26 non-zero transition equivalence classes. Contracting Eq.~\eqref{eq:S0_weingarten_master} over the detector outcomes $\kappa \in \{1, \dots, D\}$ (carried out explicitly in Appendix~\ref{app:weingarten}) groups these 26 classes into ten Weingarten Integral Groups $\mathcal{O}_1$--$\mathcal{O}_{10}$, summarized in Table~\ref{tab:forward_integral_groups}.

\begin{table*}[htbp]
\centering
\caption{The 26 non-zero equivalence classes of the forward classical shadow superoperator $\mathbf{S}^{(0)}$, categorized by their 8-index multiset partition and grouped into the ten Weingarten Integral Groups $\mathcal{O}_1$--$\mathcal{O}_{10}$ derived in Appendix~\ref{app:weingarten}.}
\label{tab:forward_integral_groups}
\renewcommand{\arraystretch}{1.35}
\begin{ruledtabular}
\begin{tabular}{cccl}
Partition Class & Integral Group & Transition Geometries ($\alpha \to \beta \mid \mu \to \nu$) & Exact Analytical Value $S^{(0)}_{\mu\nu, \alpha\beta}$ \\
\hline
$[8]$ & $\mathcal{O}_1$ & $(i,i)\to(i,i) \mid (i,i)\to(i,i)$ & $\displaystyle \frac{8}{(M+1)(M+3)}$ \\
\hline
$[6, 2]$ & $\mathcal{O}_2$ & $(i,i)\to(i,i) \mid (i,j)\to(i,j)$, \;\; $(i,j)\to(i,j) \mid (i,i)\to(i,i)$, & $\displaystyle \frac{4}{(M+1)(M+3)}$ \\
 & & $(i,i)\to(i,j) \mid (i,i)\to(i,j)$, \;\; $(i,j)\to(i,i) \mid (i,j)\to(i,i)$ & \\
\hline
$[4, 4]$ & $\mathcal{O}_3$ & $(i,i)\to(i,i) \mid (j,j)\to(j,j)$, \;\; $(i,i)\to(j,j) \mid (i,i)\to(j,j)$ & $\displaystyle \frac{2}{M(M+3)}$ \\
 & $\mathcal{O}_4$ & $(i,i)\to(i,j) \mid (i,j)\to(j,j)$, \;\; $(i,j)\to(i,i) \mid (j,j)\to(i,j)$, & $\displaystyle \frac{2(M-1)}{M(M+1)(M+3)}$ \\
 & & $(i,j)\to(j,j) \mid (i,i)\to(i,j)$ & \\
 & $\mathcal{O}_5$ & $(i,j)\to(i,j) \mid (i,j)\to(i,j)$ & $\displaystyle \frac{2(3M+1)}{M(M+1)(M+3)}$ \\
\hline
$[4, 2, 2]$ & $\mathcal{O}_6$ & $(i,i)\to(i,i) \mid (j,k)\to(j,k)$, \;\; $(i,j)\to(i,j) \mid (k,k)\to(k,k)$, & $\displaystyle \frac{2}{M(M+3)}$ \\
 & & $(i,i)\to(j,k) \mid (i,i)\to(j,k)$, \;\; $(i,j)\to(k,k) \mid (i,j)\to(k,k)$ & \\
 & $\mathcal{O}_7$ & $(i,i)\to(i,j) \mid (i,k)\to(j,k)$, \;\; $(i,j)\to(i,i) \mid (j,k)\to(i,k)$, & $\displaystyle \frac{\sqrt{2}(M-1)}{M(M+1)(M+3)}$ \\
 & & $(i,j)\to(i,k) \mid (j,k)\to(k,k)$, \;\; $(i,j)\to(j,k) \mid (i,i)\to(i,k)$, & \\
 & & $(i,j)\to(j,k) \mid (i,k)\to(k,k)$ & \\
 & $\mathcal{O}_8$ & $(i,j)\to(i,j) \mid (i,k)\to(i,k)$, \;\; $(i,j)\to(i,k) \mid (i,j)\to(i,k)$ & $\displaystyle \frac{3M+1}{M(M+1)(M+3)}$ \\
\hline
$[2, 2, 2, 2]$ & $\mathcal{O}_9$ & $(i,j)\to(i,j) \mid (k,l)\to(k,l)$, \;\; $(i,j)\to(k,l) \mid (i,j)\to(k,l)$ & $\displaystyle \frac{2}{M(M+3)}$ \\
 & $\mathcal{O}_{10}$ & $(i,j)\to(i,k) \mid (j,l)\to(k,l)$, \;\; $(i,j)\to(j,k) \mid (i,l)\to(k,l)$ & $\displaystyle \frac{M-1}{M(M+1)(M+3)}$ \\
\end{tabular}
\end{ruledtabular}
\end{table*}

\subsection{Commutative Bose-Mesner Algebra and Spectral Decomposition of the Population Block}
\label{subsec:population_inversion}

For general $n$-photon states, the $\binom{M}{n}$ collision-free populations form the vertex set of the Johnson graph $J(M,n)$, where two vertices are at distance $r$ if their mode sets share $n-r$ common modes. Because $S_M$ acts distance-transitively on $J(M,n)$, any $S_M$-invariant operator on the population sector lies in the commutative Bose-Mesner algebra spanned by the adjacency matrices of $J(M,n)$ coupled with the collision sectors.

For $n=2$, inspecting the forward Population Block $\mathbf{S}^{(0)}_{\vec{0}}$ in Table~\ref{tab:forward_integral_groups} (comprising classes $\mathcal{O}_1, \mathcal{O}_2, \mathcal{O}_3, \mathcal{O}_5, \mathcal{O}_6, \mathcal{O}_8, \mathcal{O}_9$) simplifies the block structure directly in terms of mode occupations. Let $\mathbf{N} \in \mathbb{R}^{D \times M}$ denote the mode-occupation incidence matrix whose $\alpha$-th row is the occupation vector $\vec{n}(\alpha) = (n_1(\alpha), \dots, n_M(\alpha))$ with $n_m(\alpha) = \delta_{m,i} + \delta_{m,j}$ for $\alpha = \{i, j\}$. The inner product $[\mathbf{N}\mathbf{N}^\top]_{\mu\alpha} = \langle \vec{n}(\mu), \vec{n}(\alpha) \rangle$ counts the weighted mode overlap between states $\mu$ and $\alpha$:
\begin{equation}
    [\mathbf{N}\mathbf{N}^\top]_{\mu\alpha} = \begin{cases}
        4 & \text{if } \mu = \alpha = (i,i) \;\; (\mathcal{O}_1)\text{,} \\
        2 & \text{if } \mu = \alpha = (i,j) \;\; (\mathcal{O}_5) \\
          & \quad \text{or } \{\mu,\alpha\} = \{(i,i),(i,j)\} \;\; (\mathcal{O}_2)\text{,} \\
        1 & \text{if } \mu = (i,j), \, \alpha = (i,k), \, j \neq k \;\; (\mathcal{O}_8)\text{,} \\
        0 & \text{if } \mu \cap \alpha = \emptyset \;\; (\mathcal{O}_3, \mathcal{O}_6, \mathcal{O}_9)\text{.}
    \end{cases}
    \label{eq:NNT_overlap}
\end{equation}
Comparing Eq.~\eqref{eq:NNT_overlap} with the rational expressions for $\mathcal{O}_1, \mathcal{O}_2, \mathcal{O}_3, \mathcal{O}_5, \mathcal{O}_6, \mathcal{O}_8, \mathcal{O}_9$ in Table~\ref{tab:forward_integral_groups} shows that the entire $D \times D$ forward Population Block is given by the exact matrix identity
\begin{equation}
\begin{split}
    \mathbf{S}^{(0)}_{\vec{0}} &= \frac{1}{M(M+1)(M+3)} \\
    &\quad \times \Big[ 2(M+1)(I_D + \mathbf{J}_D) + (M-1)\, \mathbf{N}\mathbf{N}^\top \Big] \text{,}
\end{split}
\label{eq:S0_pop_gram}
\end{equation}
where $I_D$ is the $D \times D$ identity matrix and $\mathbf{J}_D = \mathbf{1}_D \mathbf{1}_D^\top$ is the $D \times D$ all-ones matrix.

As proved in Appendix~\ref{app:inversion}, the three symmetric matrices $I_D$, $\mathbf{J}_D$, and $\mathbf{N}\mathbf{N}^\top$ commute pairwise because $\mathbf{N}\mathbf{1}_M = 2\mathbf{1}_D$, $\mathbf{N}^\top \mathbf{1}_D = (M+1)\mathbf{1}_M$, and $\mathbf{N}^\top \mathbf{N} = (M+2)I_M + \mathbf{J}_M$. Consequently, $\mathbb{R}^D$ decomposes into three mutually orthogonal eigenspaces with orthogonal idempotents $E_0$ (rank $1$), $E_1$ (rank $M-1$), and $E_2$ (rank $M(M-1)/2$):
\begin{align}
    E_0 &= \frac{1}{D} \mathbf{J}_D \text{,} \label{eq:idempotent_E0} \\
    E_1 &= \frac{1}{M+2}\left(\mathbf{N}\mathbf{N}^\top - \frac{4}{M}\mathbf{J}_D\right) \text{,} \label{eq:idempotent_E1} \\
    E_2 &= I_D - E_0 - E_1 \text{.} \label{eq:idempotent_E2}
\end{align}
Evaluating Eq.~\eqref{eq:S0_pop_gram} on each idempotent yields the three eigenvalues of $\mathbf{S}^{(0)}_{\vec{0}}$:
\begin{equation}
    \lambda_0 = 1 \text{,} \qquad \lambda_1 = \frac{1}{M+1} \text{,} \qquad \lambda_2 = \frac{2}{M(M+3)} \text{.}
    \label{eq:three_eigenvalues}
\end{equation}
Notice that $\lambda_0 = 1$ reflects trace preservation ($\Tr(\overline{I}) = \Tr(\rho) = 1$). Because $E_0, E_1, E_2$ are orthogonal projectors, the inverse of the Population Block is obtained immediately by inverting the three eigenvalues:
\begin{equation}
\begin{split}
    \left(\mathbf{S}^{(0)}_{\vec{0}}\right)^{-1} &= \lambda_0^{-1} E_0 + \lambda_1^{-1} E_1 + \lambda_2^{-1} E_2 \\
    &= \frac{M(M+3)}{2}\, I_D - \frac{M-1}{2}\, \mathbf{N}\mathbf{N}^\top + \frac{M-3}{M+1}\, \mathbf{J}_D \text{.}
\end{split}
\label{eq:S0_pop_inverse_closed}
\end{equation}
Substituting the four overlap values of $[\mathbf{N}\mathbf{N}^\top]_{\mu\alpha} \in \{4, 2, 1, 0\}$ from Eq.~\eqref{eq:NNT_overlap} into Eq.~\eqref{eq:S0_pop_inverse_closed} immediately generates all seven closed-form rational expressions for the Inverse Population Block in Part I of Table~\ref{tab:exact_inverse_formulas_final}.

\subsection{Rank-One Structure of the Social Coherence Blocks and Complete Inverse Catalog}
\label{subsec:social_lonely_inversion}

A similar algebraic reduction holds for the $M \times M$ Social Coherence Blocks ($\|\vec{\Delta}\|_1 = 2$) and the $1 \times 1$ Lonely Coherence Blocks ($\|\vec{\Delta}\|_1 = 4$).

In a Social Coherence Block with $\Delta_i = +1$ and $\Delta_j = -1$ ($i \neq j$), we order the $M$ basis transitions by placing the two collision-assisted transitions $\ket{i,i}\bra{i,j}$ and $\ket{i,j}\bra{j,j}$ first, followed by the $M-2$ spectator-mode transitions $\ket{i,k}\bra{j,k}$ ($k \notin \{i, j\}$). Defining the vector $\mathbf{v} = (\sqrt{2}, \sqrt{2}, 1, 1, \dots, 1)^\top \in \mathbb{R}^M$ with squared norm $\mathbf{v}^\top \mathbf{v} = 2 + 2 + (M-2) = M+2$, the outer product $[\mathbf{v}\mathbf{v}^\top]_{ab}$ takes the values $2$ between collision-assisted states ($\mathcal{O}_2, \mathcal{O}_4$), $\sqrt{2}$ between a collision-assisted state and a spectator state ($\mathcal{O}_7$), and $1$ between two spectator states ($\mathcal{O}_8, \mathcal{O}_{10}$). Inspecting Table~\ref{tab:forward_integral_groups} shows that every $M \times M$ Social Coherence Block is a rank-one perturbation of the identity matrix $I_M$:
\begin{equation}
    \mathbf{S}^{(0)}_{\text{soc}} = \frac{2}{M(M+3)}\, I_M + \frac{M-1}{M(M+1)(M+3)}\, \mathbf{v}\mathbf{v}^\top \text{.}
    \label{eq:S0_soc_rank1}
\end{equation}
Equation~\eqref{eq:S0_soc_rank1} has eigenvalue $\lambda_1 = \frac{1}{M+1}$ (with multiplicity $1$, along $\mathbf{v}$) and eigenvalue $\lambda_2 = \frac{2}{M(M+3)}$ (with multiplicity $M-1$, on the subspace orthogonal to $\mathbf{v}$). Applying the Sherman-Morrison formula (Appendix~\ref{app:inversion}) yields the exact inverse of every Social Coherence Block:
\begin{equation}
    \left(\mathbf{S}^{(0)}_{\text{soc}}\right)^{-1} = \frac{M(M+3)}{2}\, I_M - \frac{M-1}{2}\, \mathbf{v}\mathbf{v}^\top \text{.}
    \label{eq:S0_soc_inverse_closed}
\end{equation}
Evaluating Eq.~\eqref{eq:S0_soc_inverse_closed} for $v_a v_b \in \{2, \sqrt{2}, 1\}$ produces all five rational expressions in Part II of Table~\ref{tab:exact_inverse_formulas_final}.

Finally, for each $1 \times 1$ Lonely Coherence Block ($\mathcal{O}_3, \mathcal{O}_6, \mathcal{O}_9$), the forward channel is the scalar $\lambda_2 = \frac{2}{M(M+3)}$, so its inverse is simply $\lambda_2^{-1} = \frac{M(M+3)}{2}$ (Part III of Table~\ref{tab:exact_inverse_formulas_final}). Thus, across the entire $D^2 \times D^2$ transition space, $\mathbf{S}^{(0)}$ has only the three distinct eigenvalues $\{\lambda_0, \lambda_1, \lambda_2\}$ of Eq.~\eqref{eq:three_eigenvalues}, and every inverse element is an explicit rational function of $M$.

\subsection{Physical Origin of the Three-Eigenvalue Spectrum: Glauber Coherence Orders \texorpdfstring{$G^{(0)}, G^{(1)}, G^{(2)}$}{G(0), G(1), G(2)}}
\label{subsec:glauber_hierarchy}

Why does the $D^2 \times D^2$ transition superoperator collapse to just three eigenvalues across arbitrary port counts $M$? Unlike multi-qubit Clifford circuits that mix single- and multi-body operators indiscriminately, passive linear optics implements non-interacting single-particle transformations $U \in \mathcal{U}(M)$. Under the adjoint action of $\mathcal{U}(M)$, the two-photon operator space $\mathcal{L}(\mathcal{H}_{\text{sym}}^{(2)})$ splits into three irreducible sectors matching Glauber's hierarchy of quantum-optical coherence orders~\cite{Glauber1963}, with dimensions summing to $D^2$:
\begin{equation}
    D^2 = 1 + (M^2 - 1) + \frac{M^2(M-1)(M+3)}{4} \text{.}
    \label{eq:glauber_dimension_split}
\end{equation}

Each eigenvalue in Eq.~\eqref{eq:three_eigenvalues} is the interferometric shadow visibility of a distinct optical correlation order:
\begin{enumerate}
    \item \textbf{Zeroth-Order Photon Flux ($G^{(0)}$, dimension $1$, $\lambda_0 = 1$):} Spanned by the identity $I^{(2)} = \frac{1}{2}\hat{n}$ (projector $E_0$ in the Population Block). Lossless beam splitters conserve total photon number on every shot, transmitting total flux without attenuation ($\lambda_0 = 1$).
    \item \textbf{Single-Photon First-Order Coherence ($G^{(1)}$, dimension $M^2 - 1$, $\lambda_1 = \frac{1}{M+1}$):} Spanned by the traceless one-body bosonic bilinears $\hat{G}^{(1)}_{ij} = a_i^\dagger a_j - \frac{\delta_{ij}}{M}\hat{n}$ generating $\mathfrak{su}(M)$ (which for a two-port beam splitter $M=2$ reduce to the three Stokes polarization operators). Its $M-1$ diagonal mode-intensity imbalances span $E_1$ inside the Population Block. For $i \neq j$, projecting the Mach-Zehnder phase-coherence operator $a_i^\dagger a_j$ onto the two-photon Fock basis gives
    \begin{equation}
    \begin{split}
        a_i^\dagger a_j \Big|_{\mathcal{H}_{\text{sym}}^{(2)}} &= \sqrt{2}\ket{i,i}\bra{i,j} + \sqrt{2}\ket{i,j}\bra{j,j} \\
        &\quad + \sum_{k \notin \{i,j\}} \ket{i,k}\bra{j,k} \text{,}
    \end{split}
    \label{eq:G1_vector_v}
    \end{equation}
    whose coefficient vector in the $\Delta_i = +1, \Delta_j = -1$ sector is precisely $\mathbf{v} = (\sqrt{2}, \sqrt{2}, 1, \dots, 1)^\top$ in Eq.~\eqref{eq:S0_soc_rank1}, with the $\sqrt{2}$ weights originating from Hong-Ou-Mandel bosonic stimulation~\cite{Hong1987}. Combining the $M-1$ diagonal generators in $E_1$ with one vector $\mathbf{v}$ from each of the $M(M-1)$ Social Coherence blocks accounts for all $(M-1) + M(M-1) = M^2 - 1$ components of $G^{(1)}$. Because $G^{(1)}$ observables depend only on single-photon wave interference, their shadow visibility $\lambda_1 = \frac{1}{M+1}$ decays linearly as $1/M$, matching that of a single photon in $M$ modes.
    \item \textbf{Irreducible Two-Photon Correlations ($G^{(2)}$, dimension $\frac{M^2(M-1)(M+3)}{4}$, $\lambda_2 = \frac{2}{M(M+3)}$):} Spanned by normal-ordered two-body correlations orthogonal to all single-photon observables. This sector comprises $\frac{M(M-1)}{2}$ intensity-intensity coincidence correlations ($E_2$ in the Population Block), $M(M-1)^2$ spectator-conditioned phase correlations ($\mathbf{v}^\perp$ in the Social Coherence Blocks, which measure how the interferometric phase between $i$ and $j$ depends on which waveguide carries the second photon), and all $\frac{M(M-1)(M^2-M+2)}{4}$ Lonely Coherence blocks ($d=1$). Because NOON states ($\ket{i,i}\bra{j,j}$) and dual-rail photonic qubits ($\ket{i,j}\bra{k,l}$) satisfy $\bra{\alpha}a_p^\dagger a_q\ket{\beta} = 0$, they carry zero first-order optical coherence ($G^{(1)} = 0$) and reside entirely in the $G^{(2)}$ sector, causing their shadow visibility to decay quadratically with mode count as $\lambda_2 = \frac{2}{M(M+3)} \sim \frac{2}{M^2}$.
\end{enumerate}

\begin{table*}[htbp]
\centering
\caption{Complete catalog of exact analytical inverse elements $[(\mathbf{S}^{(0)})^{-1}]_{\alpha\beta,\mu\nu}$ derived from Eqs.~\eqref{eq:S0_pop_inverse_closed} and~\eqref{eq:S0_soc_inverse_closed}, organized by superselection block degeneracy $d(\vec{\Delta}) \in \{D, M, 1\}$.}
\label{tab:exact_inverse_formulas_final}
\renewcommand{\arraystretch}{2.0}
\begin{ruledtabular}
\begin{tabular}{cccl}
$d(\vec{\Delta})$ & Transition Geometry: $\alpha \to \beta \mid \mu \to \nu$ & Matrix Entry & Exact Analytical Inverse $\left[ (\mathbf{S}^{(0)})^{-1} \right]$ \\
\hline
\multicolumn{4}{c}{\textbf{I. Population Block ($\vec{\Delta} = \vec{0}$)}} \\
\hline
$D$ & $(i,i)\to(i,i) \mid (i,i)\to(i,i)$ & Diagonal & $\displaystyle \frac{M^{3} + 5 M - 2}{2 (M + 1)}$ \\
$D$ & $(i,j)\to(i,j) \mid (i,j)\to(i,j)$ & Diagonal & $\displaystyle \frac{M^{3} + 2 M^{2} + 5 M - 4}{2 (M + 1)}$ \\
$D$ & $(i,i)\to(i,i) \mid (j,j)\to(j,j)$ & Off-Diagonal & $\displaystyle \frac{M - 3}{M + 1}$ \\
$D$ & $(i,i)\to(i,i) \mid (i,j)\to(i,j)$ & Off-Diagonal & $\displaystyle - \frac{M^{2} - M + 2}{M + 1}$ \\
$D$ & $(i,i)\to(i,i) \mid (j,k)\to(j,k)$ & Off-Diagonal & $\displaystyle \frac{M - 3}{M + 1}$ \\
$D$ & $(i,j)\to(i,j) \mid (i,k)\to(i,k)$ & Off-Diagonal & $\displaystyle - \frac{M^{2} - 2 M + 5}{2 (M + 1)}$ \\
$D$ & $(i,j)\to(i,j) \mid (k,l)\to(k,l)$ & Off-Diagonal & $\displaystyle \frac{M - 3}{M + 1}$ \\
\hline
\multicolumn{4}{c}{\textbf{II. Social Coherence Blocks ($\|\vec{\Delta}\|_1 = 2$)}} \\
\hline
$M$ & $(i,i)\to(i,j) \mid (i,i)\to(i,j)$ & Diagonal & $\displaystyle \frac{M^{2} + M + 2}{2}$ \\
$M$ & $(i,j)\to(i,k) \mid (i,j)\to(i,k)$ & Diagonal & $\displaystyle \frac{(M + 1)^{2}}{2}$ \\
$M$ & $(i,i)\to(i,j) \mid (i,j)\to(j,j)$ & Off-Diagonal & $\displaystyle 1 - M$ \\
$M$ & $(i,i)\to(i,j) \mid (i,k)\to(j,k)$ & Off-Diagonal & $\displaystyle - \frac{\sqrt{2} (M - 1)}{2}$ \\
$M$ & $(i,j)\to(i,k) \mid (j,l)\to(k,l)$ & Off-Diagonal & $\displaystyle - \frac{M - 1}{2}$ \\
\hline
\multicolumn{4}{c}{\textbf{III. Lonely Coherence Blocks ($\|\vec{\Delta}\|_1 = 4$)}} \\
\hline
$1$ & $(i,i)\to(j,j) \mid (i,i)\to(j,j)$ & $1\times 1$ Scalar & $\displaystyle \frac{M (M + 3)}{2}$ \\
$1$ & $(i,i)\to(j,k) \mid (i,i)\to(j,k)$ & $1\times 1$ Scalar & $\displaystyle \frac{M (M + 3)}{2}$ \\
$1$ & $(i,j)\to(k,l) \mid (i,j)\to(k,l)$ & $1\times 1$ Scalar & $\displaystyle \frac{M (M + 3)}{2}$ \\
\end{tabular}
\end{ruledtabular}
\end{table*}

% ==============================================================================
% SECTION V: COHERENT INTERFEROMETER CROSSTALK AS BOSONIC MODE-HOPPING DIFFUSION
% ==============================================================================
\section{Coherent Interferometer Crosstalk as Bosonic Mode-Hopping Diffusion}
\label{sec:coherent_noise}

\subsection{Gaussian Unitary Ensemble Perturbation Model on the Two-Photon Subspace}
\label{subsec:gue_model}

In programmable Mach-Zehnder interferometer meshes~\cite{Reck1994, Clements2016}, imperfect phase-shifter calibrations, thermo-optic crosstalk between waveguides, and directional-coupler splitting tolerances cause the physically implemented single-particle unitary $\tilde{U}$ to deviate from the target unitary $U$ assumed in classical post-processing~\cite{HardwareHorizon2026}. We model these coherent hardware imperfections as a multiplicative unitary perturbation
\begin{equation}
    \tilde{U} = U \exp(i\sigma H) \approx U \left( I_M + i\sigma H - \frac{\sigma^2}{2} H^2 \right) \text{,}
    \label{eq:perturbed_single_U}
\end{equation}
where $\sigma \ll 1$ is a dimensionless error amplitude and $H = H^\dagger \sim \text{GUE}(M)$ is a random $M \times M$ Hermitian matrix drawn from the Gaussian Unitary Ensemble, with zero mean and unit second moments:
\begin{equation}
\begin{split}
    \mathbb{E}_H[H_{pq}] &= 0 \text{,} \qquad \mathbb{E}_H[H_{pq} H_{rs}^*] = \delta_{pr}\delta_{qs} \text{,} \\
    \mathbb{E}_H[H_{ab} H_{cd}] &= \delta_{ad}\delta_{bc} \text{.}
\end{split}
\label{eq:gue_moments}
\end{equation}
Contracting Eq.~\eqref{eq:gue_moments} over a shared index yields the quadratic matrix expectation $\mathbb{E}_H[(H^2)_{ab}] = \sum_{k=1}^M \mathbb{E}_H[H_{ak}H_{kb}] = M \delta_{ab}$.

Because the Fock representation map $\Gamma: \mathcal{U}(M) \to \mathcal{U}(D_n)$ is a Lie group homomorphism, the perturbed single-particle unitary $\tilde{U} = U e^{i\sigma H}$ induces the perturbed multi-photon unitary
\begin{equation}
    \tilde{V} = \Gamma(U e^{i\sigma H}) = V(U) \exp(i\sigma \mathcal{H}) \text{,}
    \label{eq:tilde_V_exact}
\end{equation}
generated by the second-quantized Hermitian operator
\begin{equation}
    \mathcal{H} = \sum_{p,q=1}^M H_{pq} \, a_p^\dagger a_q \text{.}
    \label{eq:H_generator}
\end{equation}
Expanding $\tilde{V}$ to second order in the perturbation strength $\sigma$ gives
\begin{equation}
    \tilde{V} = V(U) \left( I^{(2)} + i\sigma \mathcal{H} - \frac{\sigma^2}{2} \mathcal{H}^2 + \mathcal{O}(\sigma^3) \right) \text{,}
    \label{eq:noisy_V_expansion}
\end{equation}
where $I^{(2)}$ denotes the identity operator on $\mathcal{H}_{\text{sym}}^{(2)}$.

\subsection{Bosonic Mode-Hopping Lindblad Dissipator and Intra-Sector Confinement}
\label{subsec:casimir_dissipator}

Before evaluating the perturbed rank-4 tensor element by element, we derive the exact operator form of the coherent noise distortion by tracking the state through a single experimental cycle. Prior to projective Fock detection, the input state $\rho$ evolves under the physical unitary $\tilde{V}$ to
\begin{equation}
    \rho_{\text{phys}}(U, H) = \tilde{V} \rho \tilde{V}^\dagger = V(U) \left( e^{i\sigma\mathcal{H}} \rho e^{-i\sigma\mathcal{H}} \right) V^\dagger(U) \text{.}
    \label{eq:rho_phys}
\end{equation}
The experimental apparatus measures $\rho_{\text{phys}}(U, H)$ in the Fock basis $\{\ket{\kappa}\}$, while the classical computer rotates $\ket{\kappa}\bra{\kappa}$ backward using the ideal assumed unitary $V^\dagger(U)$. Averaging the resulting snapshot $I^{(\text{snap})} = V^\dagger(U)\ket{\kappa}\bra{\kappa}V(U)$ over outcomes $\kappa$ and the Haar measure $d\mu(U)$ for a fixed error realization $H$ simply applies the ideal shadow channel $\mathbf{S}^{(0)}$ to the pre-rotated state $e^{i\sigma\mathcal{H}} \rho e^{-i\sigma\mathcal{H}}$:
\begin{equation}
    \mathbb{E}_{U, \kappa \mid H}\big[I^{(\text{snap})}\big] = \mathbf{S}^{(0)}\left( e^{i\sigma\mathcal{H}} \rho e^{-i\sigma\mathcal{H}} \right) \text{.}
    \label{eq:conditional_snapshot_exp}
\end{equation}
Consequently, when the uncalibrated inverse $(\mathbf{S}^{(0)})^{-1}$ is applied to the noisy shadow signal, the expected raw estimator $\mathbb{E}[\hat{\rho}_{\text{raw}}] = (\mathbf{S}^{(0)})^{-1}\overline{I}_{\text{noisy}}$ is governed by the GUE-averaged forward distortion channel $\mathcal{E}_\sigma$:
\begin{equation}
    \mathbb{E}\big[\hat{\rho}_{\text{raw}}\big] = \mathcal{E}_\sigma(\rho) \equiv \mathbb{E}_H\left[ e^{i\sigma\mathcal{H}} \rho e^{-i\sigma\mathcal{H}} \right] \text{.}
    \label{eq:forward_noise_channel}
\end{equation}

Expanding Eq.~\eqref{eq:forward_noise_channel} via the Baker-Campbell-Hausdorff series to second order in $\sigma$,
\begin{equation}
\begin{split}
    \mathcal{E}_\sigma(\rho) &= \rho + i\sigma \mathbb{E}_H\big[[\mathcal{H}, \rho]\big] \\
    &\quad - \frac{\sigma^2}{2} \mathbb{E}_H\big[ \mathcal{H}^2 \rho + \rho \mathcal{H}^2 - 2\mathcal{H}\rho\mathcal{H} \big] + \mathcal{O}(\sigma^4) \text{,}
\end{split}
\label{eq:BCH_expansion}
\end{equation}
all odd-order terms in $\sigma$ vanish identically because the GUE distribution is symmetric ($\mathbb{E}_H[H] = 0$). To evaluate the quadratic terms for an $n$-photon state, we substitute Eq.~\eqref{eq:H_generator} and contract indices using $\mathbb{E}_H[H_{ij}H_{kl}] = \delta_{il}\delta_{jk}$. The sandwich term $\mathbb{E}_H[\mathcal{H}\rho\mathcal{H}]$ contracts to the mode-exchange scattering map $C(\rho)$:
\begin{equation}
\begin{split}
    C(\rho) &\equiv \mathbb{E}_H[\mathcal{H}\rho\mathcal{H}] = \sum_{i,j,k,l=1}^M \mathbb{E}_H[H_{ij}H_{kl}] \, a_i^\dagger a_j \rho a_k^\dagger a_l \\
    &= \sum_{i,j=1}^M a_i^\dagger a_j \rho a_j^\dagger a_i \text{.}
\end{split}
\label{eq:C_scattering_map}
\end{equation}
Similarly, the quadratic operator $\mathbb{E}_H[\mathcal{H}^2]$ is a quadratic Casimir invariant of $\mathfrak{u}(M)$ on $\mathcal{H}_{\text{sym}}^{(n)}$:
\begin{equation}
\begin{split}
    \mathbb{E}_H[\mathcal{H}^2] &= \sum_{i,j=1}^M a_i^\dagger a_j a_j^\dagger a_i = \sum_{i=1}^M a_i^\dagger \left( \sum_{j=1}^M (\hat{n}_j + \delta_{ij}) \right) a_i \\
    &= \sum_{i=1}^M a_i^\dagger (\hat{n} + M) a_i = \hat{n}(\hat{n} + M - 1) \text{,}
\end{split}
\label{eq:casimir_general_n}
\end{equation}
where $\hat{n} = \sum_{j=1}^M \hat{n}_j$ is the total photon number operator and we used $a_j a_j^\dagger = \hat{n}_j + 1$ and $\hat{n} a_i = a_i(\hat{n} - 1)$. On the $n$-photon subspace $\mathcal{H}_{\text{sym}}^{(n)}$ where $\hat{n} = n I^{(n)}$, Eq.~\eqref{eq:casimir_general_n} acts as the scalar multiple $n(M + n - 1) I^{(n)}$. Specializing to two photons ($n=2$), we obtain $\mathbb{E}_H[\mathcal{H}^2] = 2(M+1)I^{(2)}$.

Substituting Eqs.~\eqref{eq:C_scattering_map} and~\eqref{eq:casimir_general_n} into Eq.~\eqref{eq:BCH_expansion} yields the effective noise superoperator $\mathcal{D}$. Introducing the $M^2$ number-conserving bosonic mode-hopping jump operators $L_{ij} = a_i^\dagger a_j$ ($L_{ij}^\dagger = L_{ji}$), $\mathcal{D}$ takes the standard Gorini-Kossakowski-Sudarshan-Lindblad (GKSL) master-equation form:
\begin{equation}
\begin{split}
    \mathcal{D}(\rho) &= \frac{1}{2} \sum_{i,j=1}^M \Big( L_{ij}^\dagger L_{ij} \rho + \rho L_{ij}^\dagger L_{ij} - 2 L_{ij} \rho L_{ij}^\dagger \Big) \\
    &= 2(M+1)\rho - \sum_{i,j=1}^M a_i^\dagger a_j \rho a_j^\dagger a_i \text{,}
\end{split}
\label{eq:dissipator_def}
\end{equation}
in terms of which the expected uncalibrated estimator satisfies
\begin{equation}
    \mathbb{E}\big[\hat{\rho}_{\text{raw}}\big] = \big(\mathcal{I} - \sigma^2 \mathcal{D}\big)\rho + \mathcal{O}(\sigma^4) \text{.}
    \label{eq:raw_estimator_distortion}
\end{equation}
Equation~\eqref{eq:dissipator_def} provides a direct physical picture of static circuit crosstalk under randomized shadow measurements. Although each circuit realization $\tilde{U} = U e^{i\sigma H}$ is unitary, ensemble averaging converts coherent phase and beam-splitter errors into open-system bosonic diffusion across the $M$ waveguides. The Casimir anti-commutator $2(M+1)\rho$ represents uniform out-scattering from the target state, while the quantum-jump recycling term $-\sum_{i,j} a_i^\dagger a_j \rho a_j^\dagger a_i$ splits into two physical contributions: diagonal phase-correlator terms ($i=j$), $-\sum_i \hat{n}_i \rho \hat{n}_i$, which cancel common-mode phase jitter whenever ket and bra share an occupied waveguide, and off-diagonal hopping terms ($i \neq j$), which transfer a shared photon simultaneously from waveguide $j$ to waveguide $i$ in both ket and bra.

In every jump term $a_i^\dagger a_j \rho a_j^\dagger a_i$ of Eq.~\eqref{eq:dissipator_def}, the creation operator $a_i^\dagger$ on the left is paired with $a_i$ on the right, and $a_j$ on the left is paired with $a_j^\dagger$ on the right. Consequently, $a_i^\dagger a_j \ket{\alpha}\bra{\beta} a_j^\dagger a_i$ carries the same photon imbalance vector $\vec{\Delta}_{\alpha\beta}$ as $\ket{\alpha}\bra{\beta}$:
\begin{equation}
    \big[\mathcal{D}, \text{Ad}_{U_{\vec{\theta}}}\big] = 0 \implies \mathcal{D} = \bigoplus_{\vec{\Delta}} \mathcal{D}_{\vec{\Delta}} \text{.}
    \label{eq:dissipator_block_diagonal}
\end{equation}
Coherent GUE crosstalk never mixes different $\vec{\Delta}$ sectors; all noise-induced bosonic hopping is confined within each superselection block.

\subsection{Quadratic Correction Tensor and Intra-Block Probability Diffusion}
\label{subsec:quadratic_tensor}

In the tensor representation, expanding the noisy shadow superoperator $\tilde{\mathbf{S}} = \mathbb{E}_H[\mathbf{S}^{(0)}(\mathcal{E}_\sigma)]$ to second order in $\sigma$ yields
\begin{equation}
    \tilde{\mathbf{S}} = \mathbf{S}^{(0)} + \sigma^2 \mathbf{S}^{(2)} + \mathcal{O}(\sigma^4) \text{,}
    \label{eq:S_tilde_expansion}
\end{equation}
where the first-order tensor vanishes ($\mathbb{E}_H[\mathbf{S}^{(1)}] = 0$) and the quadratic correction tensor $\mathbf{S}^{(2)}$ is related to the Casimir dissipator $\mathcal{D}$ by
\begin{equation}
    \mathbf{S}^{(2)} = -\mathbf{S}^{(0)} \mathcal{D} \text{,} \qquad \mathbf{S}^{(2)}_{\vec{\Delta}} = -\mathbf{S}^{(0)}_{\vec{\Delta}} \mathcal{D}_{\vec{\Delta}} \text{.}
    \label{eq:S2_from_D}
\end{equation}
Direct symbolic integration of the 576 degree-8 Haar-GUE monomials in \texttt{SymPy} confirms Eq.~\eqref{eq:S2_from_D} for all 26 non-zero transition geometries (listed explicitly in Table~\ref{tab:supp_26_complete} of the Supplemental Material). Notice that while the forward tensor $\mathbf{S}^{(0)}$ takes identical values within each of the ten Weingarten Integral Groups $\mathcal{O}_1$--$\mathcal{O}_{10}$, the action of the Casimir dissipator $\mathcal{D}$ splits groups $\mathcal{O}_2, \mathcal{O}_3, \mathcal{O}_6, \mathcal{O}_8, \mathcal{O}_9$ according to the number of modes shared between $\alpha$ and $\beta$.

\subsection{Active Mitigation: Operator Correction versus Linear Geometric Correction}
\label{subsec:active_calibration}

When the noise amplitude $\sigma$ of an optical processor has been characterized (for instance, via randomized benchmarking or single-photon interferometric visibility), Eq.~\eqref{eq:raw_estimator_distortion} and Eq.~\eqref{eq:S_tilde_expansion} motivate two distinct active error-mitigation protocols:

\begin{enumerate}
    \item \textbf{Operator Correction ($\hat{\rho}^{(\text{op})}$):} Because $\mathbf{S}^{(2)}_{\vec{\Delta}}$ preserves the superselection block structure, we construct the noise-calibrated block $\mathbf{S}^{(\text{corr})}_{\vec{\Delta}} = \mathbf{S}^{(0)}_{\vec{\Delta}} + \sigma^2 \mathbf{S}^{(2)}_{\vec{\Delta}}$ and invert it directly on the empirical shadow signal $\hat{I}$:
    \begin{equation}
    \begin{split}
        \hat{\rho}^{(\text{op})}_{\alpha\beta} &= \sum_{(\mu,\nu) \in \mathcal{V}_{\vec{\Delta}^*}} \left( \mathbf{S}^{(0)}_{\vec{\Delta}^*} + \sigma^2 \mathbf{S}^{(2)}_{\vec{\Delta}^*} \right)^{-1}_{\alpha\beta, \mu\nu} \\
        &\quad \times \left[ \frac{1}{N} \sum_{s=1}^N I^{(\text{snap}, s)}_{\mu\nu} \right] \text{.}
    \end{split}
    \label{eq:operator_correction_def}
    \end{equation}
    Using $\mathbf{S}^{(0)}_{\vec{\Delta}} + \sigma^2 \mathbf{S}^{(2)}_{\vec{\Delta}} = \mathbf{S}^{(0)}_{\vec{\Delta}}(\mathcal{I} - \sigma^2 \mathcal{D}_{\vec{\Delta}})$, the calibrated inverse factors as
    \begin{equation}
    \begin{split}
        \left( \mathbf{S}^{(0)}_{\vec{\Delta}} + \sigma^2 \mathbf{S}^{(2)}_{\vec{\Delta}} \right)^{-1} &= \big( \mathcal{I} - \sigma^2 \mathcal{D}_{\vec{\Delta}} \big)^{-1} \left( \mathbf{S}^{(0)}_{\vec{\Delta}} \right)^{-1} \\
        &= \left( \sum_{k=0}^\infty \sigma^{2k} \mathcal{D}_{\vec{\Delta}}^k \right) \left( \mathbf{S}^{(0)}_{\vec{\Delta}} \right)^{-1} \text{.}
    \end{split}
    \label{eq:neumann_exact_factorization}
    \end{equation}
    \item \textbf{Linear Geometric Correction ($\hat{\rho}^{(\text{geom})}$):} Truncating the Neumann series in Eq.~\eqref{eq:neumann_exact_factorization} at first order in $\sigma^2$ yields a linear post-processing filter applied directly to the uncalibrated estimator $\hat{\rho}_{\text{raw}} = (\mathbf{S}^{(0)})^{-1}\hat{I}$:
    \begin{equation}
    \begin{split}
        \hat{\rho}^{(\text{geom})} &= \big( \mathcal{I} + \sigma^2 \mathcal{D} \big) \hat{\rho}_{\text{raw}} \\
        &= \hat{\rho}_{\text{raw}} + \sigma^2 \Big( 2(M+1)\hat{\rho}_{\text{raw}} \\
        &\quad - \sum_{i,j=1}^M a_i^\dagger a_j \hat{\rho}_{\text{raw}} a_j^\dagger a_i \Big) \text{.}
    \end{split}
    \label{eq:geometric_correction_def}
    \end{equation}
\end{enumerate}

Comparing the expected values of the two calibrated estimators against the true state $\rho$ highlights why full block inversion is superior:
\begin{equation}
    \mathbb{E}\big[\hat{\rho}^{(\text{op})} - \hat{\rho}^{(\text{geom})}\big] = \sigma^4 \mathcal{D}^2(\rho) + \mathcal{O}(\sigma^6) \text{.}
    \label{eq:op_vs_geom_divergence}
\end{equation}
While the linear Geometric Correction cancels the leading $\mathcal{O}(\sigma^2)$ bias amplitude (reducing the MSE floor from $\mathcal{O}(\sigma^4)$ to $\mathcal{O}(\sigma^8)$ only up to two-step scattering), multiple intra-block scattering events $\sigma^4 \mathcal{D}^2(\rho)$ leave a residual saturation floor. The Operator Correction~\eqref{eq:operator_correction_def} inverts the finite-dimensional matrix $\mathbf{S}^{(0)}_{\vec{\Delta}} + \sigma^2 \mathbf{S}^{(2)}_{\vec{\Delta}}$ exactly, summing the intra-block Neumann series~\eqref{eq:neumann_exact_factorization} to infinite order and eliminating the quadratic channel bias entirely.

% ==============================================================================
% SECTION VI: SHOT-NOISE VERSUS CROSSTALK DUALITY AND HONG-OU-MANDEL PROTECTION
% ==============================================================================
\section{Shot-Noise versus Crosstalk Duality and Hong-Ou-Mandel Protection}
\label{sec:duality_catalog}

\subsection{Decomposition of Mean Squared Error and the Seven Kinematic Topologies}
\label{subsec:mse_duality}

To quantify the accuracy of reconstructing an individual density-matrix element $\rho_{\alpha\beta}$ over $N$ measurement shots in the presence of coherent circuit crosstalk $\sigma$, we evaluate the Mean Squared Error (MSE) of the uncalibrated estimator $\hat{\rho}_{\text{raw}, \alpha\beta}$. By standard bias-variance decomposition, the MSE splits into the statistical sample variance $\text{Var}(\hat{\rho}_{\text{raw}, \alpha\beta})$ and the squared systematic bias:
\begin{equation}
\begin{split}
    \text{MSE}\big(\hat{\rho}_{\text{raw}, \alpha\beta}\big) &= \mathbb{E}\left[ \big| \hat{\rho}_{\text{raw}, \alpha\beta} - \rho_{\alpha\beta} \big|^2 \right] \\
    &= \text{Var}\big(\hat{\rho}_{\text{raw}, \alpha\beta}\big) + \big| \mathbb{E}[\hat{\rho}_{\text{raw}, \alpha\beta}] - \rho_{\alpha\beta} \big|^2 \\
    &= \frac{f_{\mathcal{T}_k}(M)}{N} + \sigma^4 \big| \mathcal{D}_{\vec{\Delta}}(\rho)_{\alpha\beta} \big|^2 + \mathcal{O}(\sigma^6) \text{.}
\end{split}
\label{eq:mse_duality}
\end{equation}

Equation~\eqref{eq:mse_duality} separates reconstruction error into two distinct physical mechanisms:
\begin{itemize}
    \item \textbf{Kinematic Statistical Variance ($f_{\mathcal{T}_k}(M)/N$):} In the high-dimensional baseline regime, the leading $1/N$ statistical variance of $\hat{\rho}_{\text{raw}, \alpha\beta}$ is governed by the diagonal element of the inverse ideal superoperator in its superselection sector $\vec{\Delta}^* = \vec{\Delta}_{\alpha\beta}$:
    \begin{equation}
        f_{\mathcal{T}_k}(M) = \frac{1}{D} \left[ \left( \mathbf{S}^{(0)}_{\vec{\Delta}^*} \right)^{-1} \right]_{\alpha\beta, \alpha\beta} \text{.}
        \label{eq:variance_extraction}
    \end{equation}
    As shown in Table~\ref{tab:kinematic_topologies}, the mode-coincidence geometry of $\ket{\alpha}\bra{\beta}$ partitions the $D^2$ elements of $\rho$ into seven \textit{Kinematic Topologies} $\mathcal{T}_1$--$\mathcal{T}_7$, depending on whether $\alpha$ and $\beta$ are collision or distinct states and on the cardinality of their intersection $|\alpha \cap \beta|$.
    \item \textbf{Dynamic Coherent Bias ($\sigma^4 |\mathcal{D}_{\vec{\Delta}}(\rho)_{\alpha\beta}|^2$):} The asymptotic $N \to \infty$ saturation floor is determined by the action of the Casimir scattering dissipator $\mathcal{D}_{\vec{\Delta}}$ within the superselection block of dimension $d(\vec{\Delta}) \in \{D, M, 1\}$.
\end{itemize}

\begin{table*}[htbp]
\centering
\caption{Master Characterization Atlas of the seven Kinematic Topologies $\mathcal{T}_1$--$\mathcal{T}_7$ spanning the two-photon density matrix, showing their spatial transition geometry $\ket{\alpha}\bra{\beta}$, dynamical regime, superselection block dimension $d(\vec{\Delta})$, and element multiplicity in an $M$-mode network.}
\label{tab:kinematic_topologies}
\renewcommand{\arraystretch}{1.3}
\begin{ruledtabular}
\begin{tabular}{ccccc}
Topology & Geometry $\ket{\alpha}\bra{\beta}$ & Dynamical Regime & Block $d(\vec{\Delta})$ & Multiplicity \\
\hline
$\mathcal{T}_1$ & $\ket{i,i}\bra{j,j}$ & Lonely ($\mathcal{C}_L$) & $1$ & $M(M-1)$ \\
$\mathcal{T}_2$ & $\ket{i,i}\bra{j,k}$ & Lonely ($\mathcal{C}_L$) & $1$ & $M(M-1)(M-2)$ \\
$\mathcal{T}_3$ & $\ket{i,j}\bra{k,l}$ & Lonely ($\mathcal{C}_L$) & $1$ & $\frac{1}{4}M(M-1)(M-2)(M-3)$ \\
\hline
$\mathcal{T}_4$ & $\ket{i,i}\bra{i,j}$ & Social ($\mathcal{C}_S$) & $M$ & $2M(M-1)$ \\
$\mathcal{T}_5$ & $\ket{i,k}\bra{j,k}$ & Social ($\mathcal{C}_S$) & $M$ & $M(M-1)(M-2)$ \\
\hline
$\mathcal{T}_6$ & $\ket{i,i}\bra{i,i}$ & Population ($\mathcal{P}$) & $D$ & $M$ \\
$\mathcal{T}_7$ & $\ket{i,j}\bra{i,j}$ & Population ($\mathcal{P}$) & $D$ & $\frac{1}{2}M(M-1)$ \\
\end{tabular}
\end{ruledtabular}
\end{table*}

\subsection{Analytical Variance Prefactors and Bias Transfer Matrix Bounds}
\label{subsec:covariance_bounds}

To establish state-independent performance metrics for each transition geometry $\rho_i$ ($i \equiv (\alpha, \beta) \in \mathcal{V}_{\vec{\Delta}^*}$), we define the dimensionless bias transfer matrix $\mathbf{W}_{\vec{\Delta}^*}$ of sector $\vec{\Delta}^*$:
\begin{equation}
    \mathbf{W}_{\vec{\Delta}^*} = \left( \mathbf{S}^{(0)}_{\vec{\Delta}^*} \right)^{-1} \mathbf{S}^{(2)}_{\vec{\Delta}^*} = -\mathcal{D}_{\vec{\Delta}^*} \text{.}
    \label{eq:W_transfer_matrix}
\end{equation}
In terms of $\mathbf{W}_{\vec{\Delta}^*}$, the leading bias of element $i$ is $\mathbb{E}[\hat{\rho}_{\text{raw}, i}] - \rho_i = \sigma^2 \sum_{j=1}^{d(\vec{\Delta}^*)} (\mathbf{W}_{\vec{\Delta}^*})_{ij} \rho_j + \mathcal{O}(\sigma^4)$. By the Cauchy-Schwarz inequality, for any unit-norm state vector in sector $\vec{\Delta}^*$, the squared bias is bounded by the squared Euclidean norm of the $i$-th row of $\mathbf{W}_{\vec{\Delta}^*}$ (which also equals the exact total MSE plateau $\sum_j |\mathcal{D}_{\vec{\Delta}^*}(\ket{\alpha_i}\bra{\beta_i})_j|^2$ generated by a unit excitation of transition $i$, since $\mathcal{D}_{\vec{\Delta}^*}$ is Hermitian):
\begin{equation}
    \Pi_i(\sigma, M) = \sigma^4 \sum_{j=1}^{d(\vec{\Delta}^*)} \left| \big(\mathbf{W}_{\vec{\Delta}^*}\big)_{ij} \right|^2 = \sigma^4 \sum_{j=1}^{d(\vec{\Delta}^*)} \left| \big(\mathcal{D}_{\vec{\Delta}^*}\big)_{ij} \right|^2 \text{.}
    \label{eq:plateau_extraction}
\end{equation}

\subsection{The Five Quantum Memory Families and the Pareto Encoding Frontier}
\label{subsec:five_families}

In Appendix~\ref{app:dissipator}, we evaluate both the variance prefactor $f_{\mathcal{T}_k}(M)$ [Eq.~\eqref{eq:variance_extraction}] and the bias plateau polynomial $\Pi_i(\sigma, M)$ [Eq.~\eqref{eq:plateau_extraction}] in closed form for all seven Kinematic Topologies. Because the three Lonely Coherence topologies $\mathcal{T}_1, \mathcal{T}_2, \mathcal{T}_3$ share the same $1 \times 1$ scalar inverse $\frac{M(M+3)}{2}$ and have vanishing mode-exchange scattering $C(\ket{\alpha}\bra{\beta}) = 0$, they possess identical variance prefactors and identical bias plateaus, merging into a single family. The $D^2$ density-matrix elements thus partition into five universal \textit{Quantum Memory Families}, cataloged for general mode number $M$ (and evaluated numerically at $M=4$) in Table~\ref{tab:quantum_memory_families}.

\begin{table*}[htbp]
\centering
\caption{Classification of the five universal Quantum Memory Families across the $D^2$-dimensional two-photon state space. Closed-form analytical expressions for the statistical variance prefactor $f_{\mathcal{T}_k}(M)/N$ and the asymptotic $\mathcal{O}(\sigma^4)$ bias plateau bound $\Pi_i(\sigma, M)$ are given for arbitrary mode number $M$, alongside their exact numerical values for $M=4$ ($D=10$).}
\label{tab:quantum_memory_families}
\small
\setlength{\tabcolsep}{4pt}
\renewcommand{\arraystretch}{1.75}
\begin{ruledtabular}
\begin{tabular}{llccccc}
Family & Transition Geometry & Topology & Block $d$ & Count & Variance $f_{\mathcal{T}_k}/N$ ($M=4$) & Plateau $\Pi_i(\sigma, M)$ ($M=4$) \\
\hline
\textbf{Fam 1} & $\ket{i,i}\bra{i,i}$ & $\mathcal{T}_6$ & $D$ & $4$ & $\displaystyle \frac{M^3 + 5M - 2}{N M (M + 1)^2} \;\; \Big(\frac{0.82}{N}\Big)$ & $4M(M - 1)\,\sigma^4 \;\; \left(48\,\sigma^4\right)$ \\
\textbf{Fam 2} & $\ket{i,i}\bra{i,j}$ & $\mathcal{T}_4$ & $M$ & $24$ & $\displaystyle \frac{M^2 + M + 2}{N M (M + 1)} \;\; \Big(\frac{1.10}{N}\Big)$ & $2M(2M + 1)\,\sigma^4 \;\; \left(72\,\sigma^4\right)$ \\
\textbf{Fam 3} & $\ket{i,j}\bra{i,j}$ & $\mathcal{T}_7$ & $D$ & $6$ & $\displaystyle \frac{M^3 + 2M^2 + 5M - 4}{N M (M + 1)^2} \;\; \Big(\frac{1.12}{N}\Big)$ & $2(2M^2 + M + 2)\,\sigma^4 \;\; \left(76\,\sigma^4\right)$ \\
\textbf{Fam 4} & $\ket{i,k}\bra{j,k}$ & $\mathcal{T}_5$ & $M$ & $24$ & $\displaystyle \frac{M + 1}{N M} \;\; \Big(\frac{1.25}{N}\Big)$ & $(4M^2 + 5M + 2)\,\sigma^4 \;\; \left(86\,\sigma^4\right)$ \\
\textbf{Fam 5} & $\ket{i,i}\bra{j,j}, \, \ket{i,i}\bra{j,k}$, & $\mathcal{T}_{1,2,3}$ & $1$ & $42$ & $\displaystyle \frac{M + 3}{N (M + 1)} \;\; \Big(\frac{1.40}{N}\Big)$ & $4(M + 1)^2\,\sigma^4 \;\; \left(100\,\sigma^4\right)$ \\
 & $\ket{i,j}\bra{k,l}$ & & & & & \\
\end{tabular}
\end{ruledtabular}
\end{table*}

\begin{figure}[htbp]
    \centering
    \includegraphics[width=\linewidth]{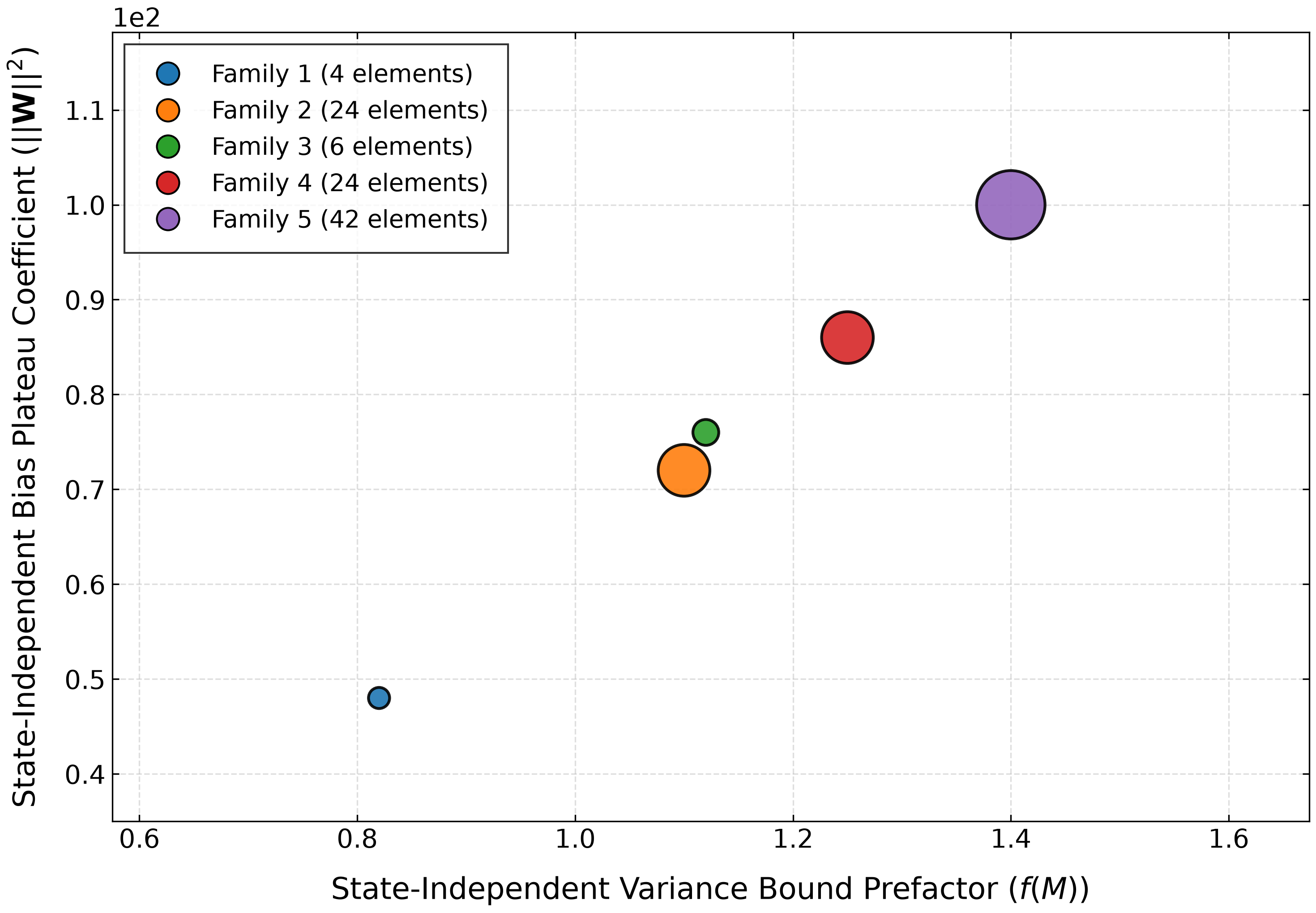}
    \caption{\textbf{Quantum Memory Frontier for $M=4$.} The five Quantum Memory Families plotted in the plane of statistical sample variance prefactor $f_{\mathcal{T}_k}(4)$ ($x$-axis) and asymptotic $\mathcal{O}(\sigma^4)$ bias plateau coefficient $\Pi_i(\sigma, 4)/\sigma^4$ ($y$-axis). Bubble areas represent the multiplicity of each family among the $D^2 = 100$ density-matrix elements.}
    \label{fig:quantum_memory_frontier}
\end{figure}

Plotting the five families in the variance-versus-bias plane (Fig.~\ref{fig:quantum_memory_frontier}) reveals a monotonic Pareto frontier governed by three quantum-optical mechanisms:
\begin{enumerate}
    \item \textbf{Maximum Crosstalk Sensitivity of NOON States and Dual-Rail Qubits (Family 5, $d = 1$):} Two-photon NOON coherences $\ket{i,i}\bra{j,j}$~\cite{Boto2000} ($\mathcal{T}_1$), bunched-to-coincidence coherences $\ket{i,i}\bra{j,k}$ ($\mathcal{T}_2$), and dual-rail two-qubit coherences $\ket{i,j}\bra{k,l}$~\cite{Knill2001} ($\mathcal{T}_3$) all belong to Family 5 ($\alpha \cap \beta = \emptyset$). Because no jump operator $L_{pq} = a_p^\dagger a_q$ can simultaneously remove a photon from $\ket{\alpha}$ and $\ket{\beta}$, the quantum-jump recycling map $C(\ket{\alpha}\bra{\beta}) = \sum_{p,q} a_p^\dagger a_q \ket{\alpha}\bra{\beta} a_q^\dagger a_p$ vanishes identically. Every waveguide phase fluctuation acts without common-mode cancellation, yielding the maximum dephasing rate $\mathcal{D}(\ket{\alpha}\bra{\beta}) = 2(M+1)\ket{\alpha}\bra{\beta}$ and the largest crosstalk plateau across all $M$, $\Pi_5(\sigma, M) = 4(M+1)^2\sigma^4$ ($100\sigma^4$ at $M=4$). Simultaneously, because Family 5 carries zero first-order coherence ($G^{(1)} = 0$), reconstruction relies exclusively on the $\lambda_2 = \frac{2}{M(M+3)}$ two-photon visibility channel, incurring the largest statistical variance $\frac{M+3}{N(M+1)}$.
    \item \textbf{Common-Mode Phase Rejection via Spectator Photons (Families 1--4, $d \in \{M, D\}$):} Whenever $\ket{\alpha}$ and $\ket{\beta}$ share an occupied waveguide, the diagonal number-correlator terms $-\sum_p \hat{n}_p \rho \hat{n}_p$ in $C(\rho)$ survive and cancel common-mode phase jitter on the shared photon. This directly lowers the diagonal decay rate of $\mathcal{D}$ from $2(M+1)$ (in Family 5) to $2M+1$ in Family 4, $2M$ in Families 2 and 3, and $2(M-1)$ in Family 1. At the same time, Families 2 and 4 overlap with the single-photon $G^{(1)}$ sector ($\lambda_1 = \frac{1}{M+1}$), reducing their statistical shot noise relative to reference-free NOON and dual-rail coherences.
    \item \textbf{Hong-Ou-Mandel Protection of Quantum Coherence over Coincidence Populations (Family 2 vs.\ Family 3):} Standard quantum-optical intuition suggests that diagonal Fock populations, which are insensitive to optical phase, should tolerate unitary interferometer errors better than off-diagonal quantum coherences. Table~\ref{tab:quantum_memory_families} shows the opposite: for every port count $M \ge 2$, the phase-carrying Hong-Ou-Mandel bunched coherence $\ket{i,i}\bra{i,j}$ (Family 2, $d=M$) achieves both lower statistical shot noise and a strictly smaller coherent-crosstalk plateau than the phase-insensitive unbunched coincidence population $\ket{i,j}\bra{i,j}$ (Family 3, $d=D$):
    \begin{equation}
    \begin{split}
        \Pi_3(\sigma, M) - \Pi_2(\sigma, M) &= 2(2M^2 + M + 2)\sigma^4 \\
        &\quad - 2M(2M + 1)\sigma^4 \\
        &= 4\sigma^4 > 0 \text{.}
    \end{split}
    \end{equation}
    At $M=4$, Family 2 saturates at $72\sigma^4$ whereas Family 3 saturates at $76\sigma^4$, alongside lower shot-noise variance ($1.10/N$ vs.\ $1.12/N$). Physically, in an unbunched coincidence $\ket{i,j}\bra{i,j}$, photons occupy two distinct waveguides $i$ and $j$, exposing both ports independently to parasitic hopping into $2(M-1)$ neighboring channels [contributing $(2M+4)\sigma^4$ to the off-diagonal leakage norm in Appendix~\ref{app:dissipator}]. In the bunched coherence $\ket{i,i}\bra{i,j}$, Hong-Ou-Mandel coalescence~\cite{Hong1987} concentrates three of the four ket-bra photon amplitudes into waveguide $i$, restricting mode hopping solely to transitions out of waveguide $i$ ($M-1$ channels, contributing $2M\sigma^4$) while preserving one shared photon in waveguide $i$ as a common-mode phase reference.
\end{enumerate}

% ==============================================================================
% SECTION VII: NUMERICAL VERIFICATION AND MULTI-STATE RECONSTRUCTION
% ==============================================================================
\section{Numerical Verification and Multi-State Reconstruction}
\label{sec:numerical_verification}

\subsection{Hybrid Symbolic-Stochastic Simulation Architecture}
\label{subsec:simulation_architecture}

To test our analytical formulas, we implemented a hybrid symbolic-stochastic simulation pipeline for an $M=4$ interferometer ($D = 10$, $D^2 = 100$). The $M=4$ architecture is the minimal network that simultaneously hosts all three superselection block dimensions ($d = 10, 4, 1$), all seven Kinematic Topologies ($\mathcal{T}_1$--$\mathcal{T}_7$), and all 26 non-zero transition equivalence classes (since $M=4$ allows four mutually distinct modes $\{i,j,k,l\}$ required by $\mathcal{T}_3$ and Integral Groups $\mathcal{O}_9, \mathcal{O}_{10}$).

In the symbolic stage, exact rational expressions for $\mathbf{S}^{(0)}_{\vec{\Delta}}$, $(\mathbf{S}^{(0)}_{\vec{\Delta}})^{-1}$, and $\mathbf{S}^{(2)}_{\vec{\Delta}}$ are evaluated at $M=4$ using \texttt{SymPy}. In the stochastic Monte Carlo stage, for each shot $s \in \{1, \dots, N\}$, a Haar-random unitary $U_s \sim \text{Haar}(4)$ is generated via QR decomposition of a Ginibre-random complex matrix, and a random Hermitian perturbation $H_s \sim \text{GUE}(4)$ satisfying Eq.~\eqref{eq:gue_moments} is sampled independently. The physical two-photon unitary $\tilde{V}_s = V(U_s \exp(i\sigma H_s))$ with $\sigma = 0.1$ is applied to the target state $\rho = \ket{\psi}\bra{\psi}$, an output Fock configuration $\kappa_s \in \{1, \dots, 10\}$ is drawn from the Born distribution $p(\kappa \mid \tilde{V}_s) = |\bra{\kappa}\tilde{V}_s\ket{\psi}|^2$, and the single-shot snapshot $I^{(\text{snap}, s)} = V^\dagger(U_s)\ket{\kappa_s}\bra{\kappa_s}V(U_s)$ is accumulated. To suppress Monte Carlo fluctuations of the MSE curves, each trajectory is averaged over 30 independent ensembles.

At each shot count $N$, we evaluate the global density-matrix MSE averaged across all $D^2 = 100$ entries:
\begin{equation}
    \text{MSE}_{\text{global}}(\hat{\rho}) = \frac{1}{D^2} \sum_{\alpha,\beta=1}^D \mathbb{E}\left[ \big| \hat{\rho}_{\alpha\beta} - \rho_{\alpha\beta} \big|^2 \right] \text{.}
    \label{eq:global_mse_def}
\end{equation}
From Table~\ref{tab:quantum_memory_families}, the state-averaged statistical variance bound across the 100 entries for $M=4$ is
\begin{equation}
\begin{split}
    \overline{f}(4) &= \frac{1}{100}\big( 4 \times 0.82 + 24 \times 1.10 + 6 \times 1.12 \\
    &\quad + 24 \times 1.25 + 42 \times 1.40 \big) = 1.252 \text{,}
\end{split}
\label{eq:state_averaged_variance}
\end{equation}
giving the universal noise-free convergence line $\text{MSE}_{\text{global}} = 1.252 / N$.

\subsection{Pure Collision State Reconstruction and Intra-Block Diffusion}
\label{subsec:collision_benchmark}

We first initialize the $M=4$ network in the pure collision state $\ket{\psi_1} = \ket{1,1}$, whose density matrix $\rho_1 = \ket{1,1}\bra{1,1}$ resides in Family 1 ($\mathcal{T}_6$) inside the $10 \times 10$ Population Block ($\vec{\Delta} = \vec{0}$). As shown in Appendix~\ref{app:dissipator} (and visualized in the full $10 \times 10$ heatmaps in Figs.~\ref{fig:supp_heatmap_target}--\ref{fig:supp_heatmap_op} of the Supplemental Material), applying the Casimir dissipator $\mathcal{D}$ to $\rho_1$ gives
\begin{equation}
    \mathcal{D}\big(\ket{1,1}\bra{1,1}\big) = 2(M-1)\ket{1,1}\bra{1,1} - 2 \sum_{p=2}^M \ket{1,p}\bra{1,p} \text{.}
    \label{eq:D_on_collision}
\end{equation}
Consequently, under coherent noise ($\sigma = 0.1$), the uncalibrated Raw Estimator $\hat{\rho}_{\text{raw}}$ suffers an expected diagonal attenuation of $-2(M-1)\sigma^2 = -0.06$ on $\ket{1,1}\bra{1,1}$ and leaks $+2\sigma^2 = +0.02$ of spurious population into each of the $M-1 = 3$ single-collision states $\ket{1,2}\bra{1,2}$, $\ket{1,3}\bra{1,3}$, and $\ket{1,4}\bra{1,4}$, while the remaining six population states and all 90 off-diagonal coherences remain unbiased. Summing the squared bias across all 100 entries yields the exact theoretical saturation plateau for $\ket{\psi_1}$:
\begin{equation}
\begin{split}
    \text{MSE}_{\text{plateau}}(\ket{\psi_1}) &= \frac{\sigma^4}{D^2} \Pi_1(1, M) = \frac{4M(M-1)\sigma^4}{D^2} \\
    &= \frac{48 \times 10^{-4}}{100} = 4.80 \times 10^{-5} \text{.}
\end{split}
\label{eq:psi1_plateau_exact}
\end{equation}

\begin{figure}[htbp]
    \centering
    \includegraphics[width=\linewidth]{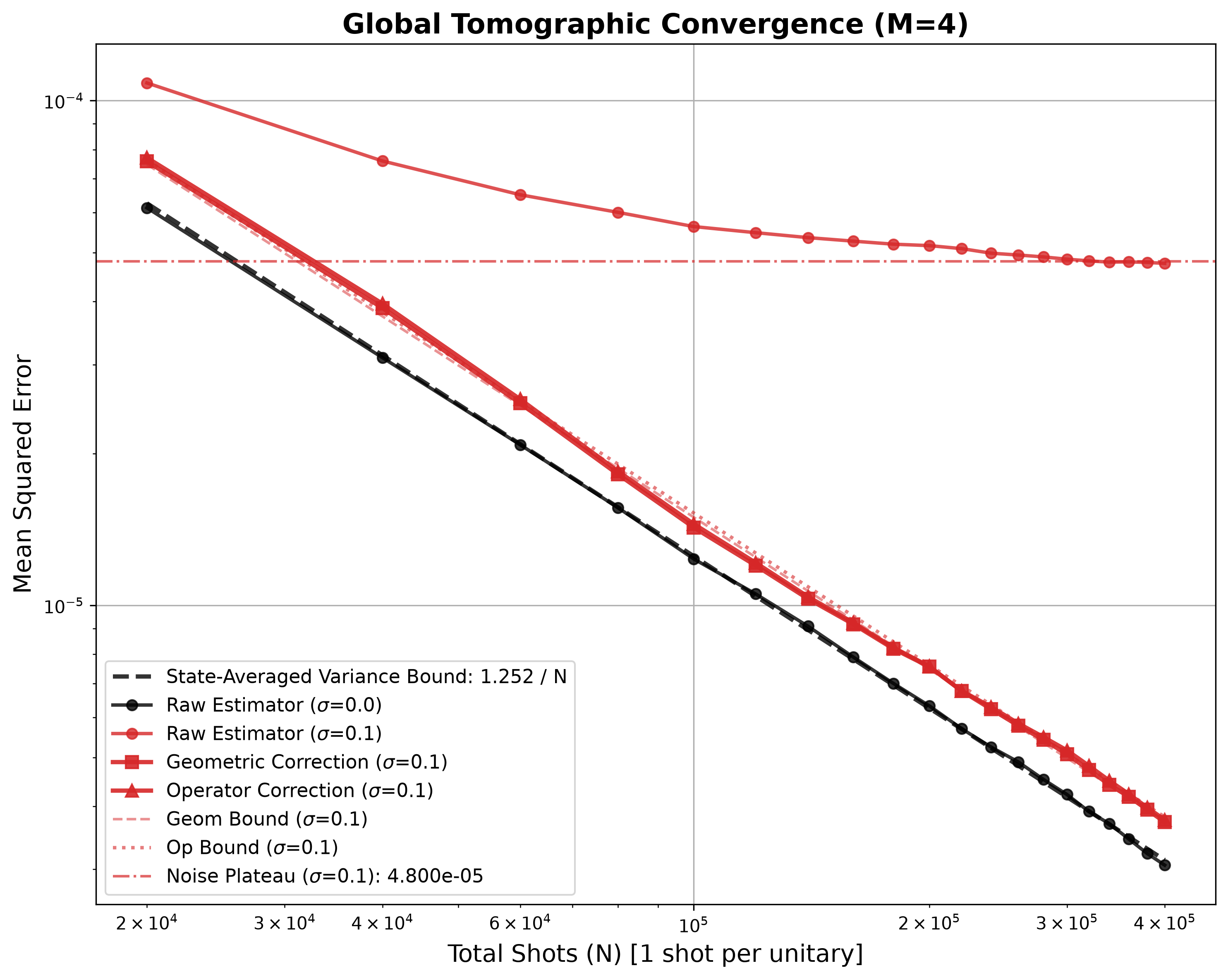}
    \caption{\textbf{Global tomographic MSE convergence for the pure collision state $\ket{\psi_1}=\ket{1,1}$ ($M=4$, $\sigma=0.1$, $N = 4 \times 10^5$ shots, 30 ensembles).} The noise-free baseline ($\sigma=0.0$, black circles) tracks the analytical variance bound $1.252/N$ [Eq.~\eqref{eq:state_averaged_variance}]. Under coherent GUE noise ($\sigma=0.1$), the uncalibrated Raw Estimator (red circles) saturates at the exact $\mathcal{O}(\sigma^4)$ theoretical plateau $4.80 \times 10^{-5}$ [Eq.~\eqref{eq:psi1_plateau_exact}]. Both the Geometric Correction (red squares) and the Operator Correction (red triangles) suppress the plateau and restore $1/N$ scaling across the sampled trajectory range.}
    \label{fig:convergence}
\end{figure}

Figure~\ref{fig:convergence} compares the simulated global MSE trajectories against these analytical predictions up to $N = 4 \times 10^5$ shots. In the shot-noise-dominated regime ($N \lesssim 2 \times 10^4$), statistical fluctuations exceed the coherent drift, and the uncalibrated Raw Estimator decays parallel to $1/N$. As $N$ increases toward $4 \times 10^5$, the Raw Estimator flattens onto the predicted $4.80 \times 10^{-5}$ plateau. The Operator Correction~\eqref{eq:operator_correction_def} eliminates this saturation floor completely, following its noise-renormalized $1/N$ variance bound over the entire range of $N$.

\subsection{Coherent Superposition States Across Multiple Superselection Sectors}
\label{subsec:superposition_benchmarks}

To test the reconstruction protocol and active noise calibration when multiple superselection blocks are simultaneously populated, we evaluate two coherent superposition states over $N = 3 \times 10^5$ shots (30 ensembles each):
\begin{align}
    \ket{\psi_2} &= \frac{1}{\sqrt{2}}\big(\ket{1,1} + i\ket{1,2}\big) \text{,} \label{eq:psi2_def} \\
    \ket{\psi_3} &= \frac{1}{\sqrt{3}}\big(\ket{1,1} + i\ket{1,2} - \ket{3,3}\big) \text{.} \label{eq:psi3_def}
\end{align}

\begin{figure}[!htbp]
    \centering
    \includegraphics[width=0.98\linewidth]{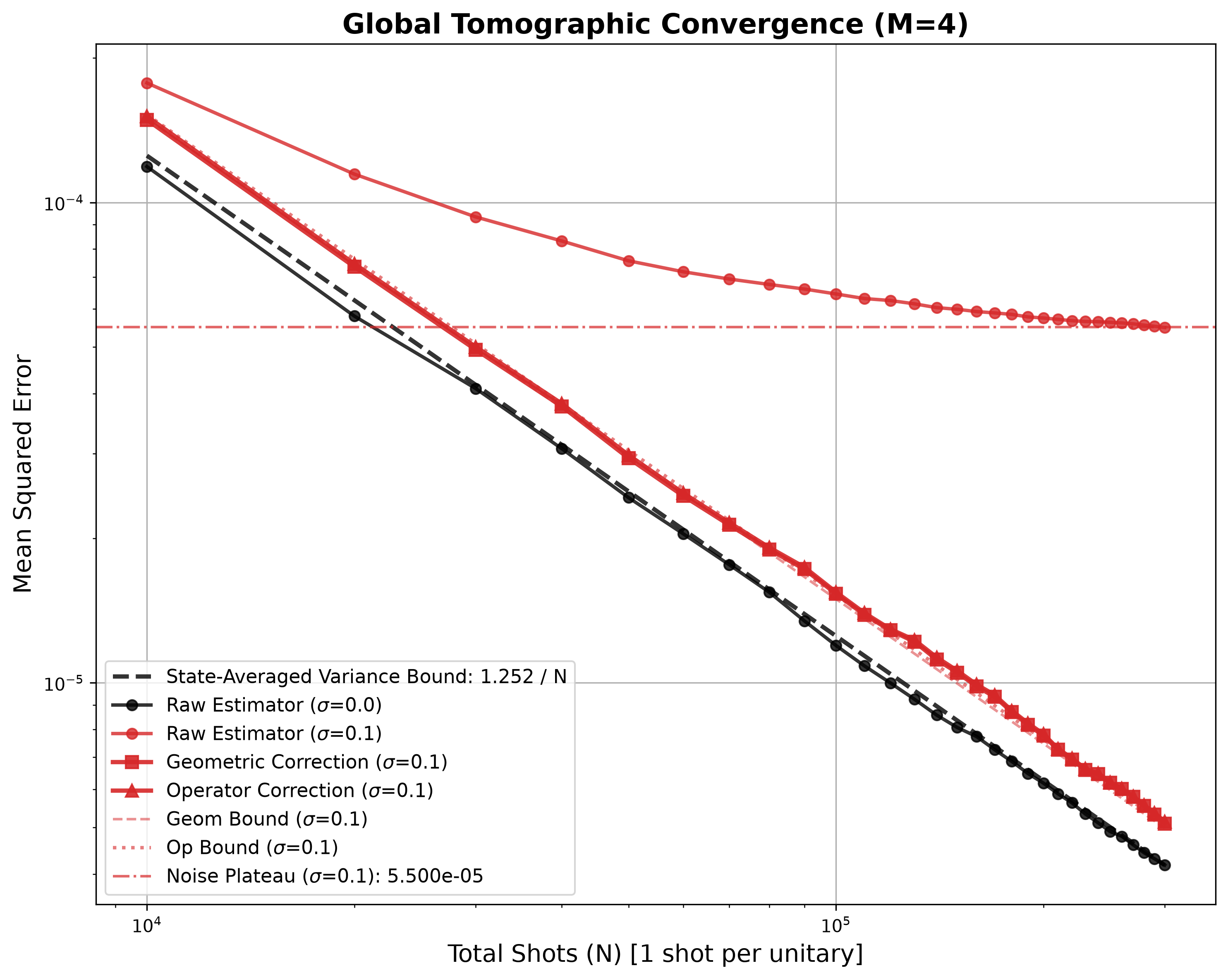}
    \vspace{0.1cm}
    \includegraphics[width=0.98\linewidth]{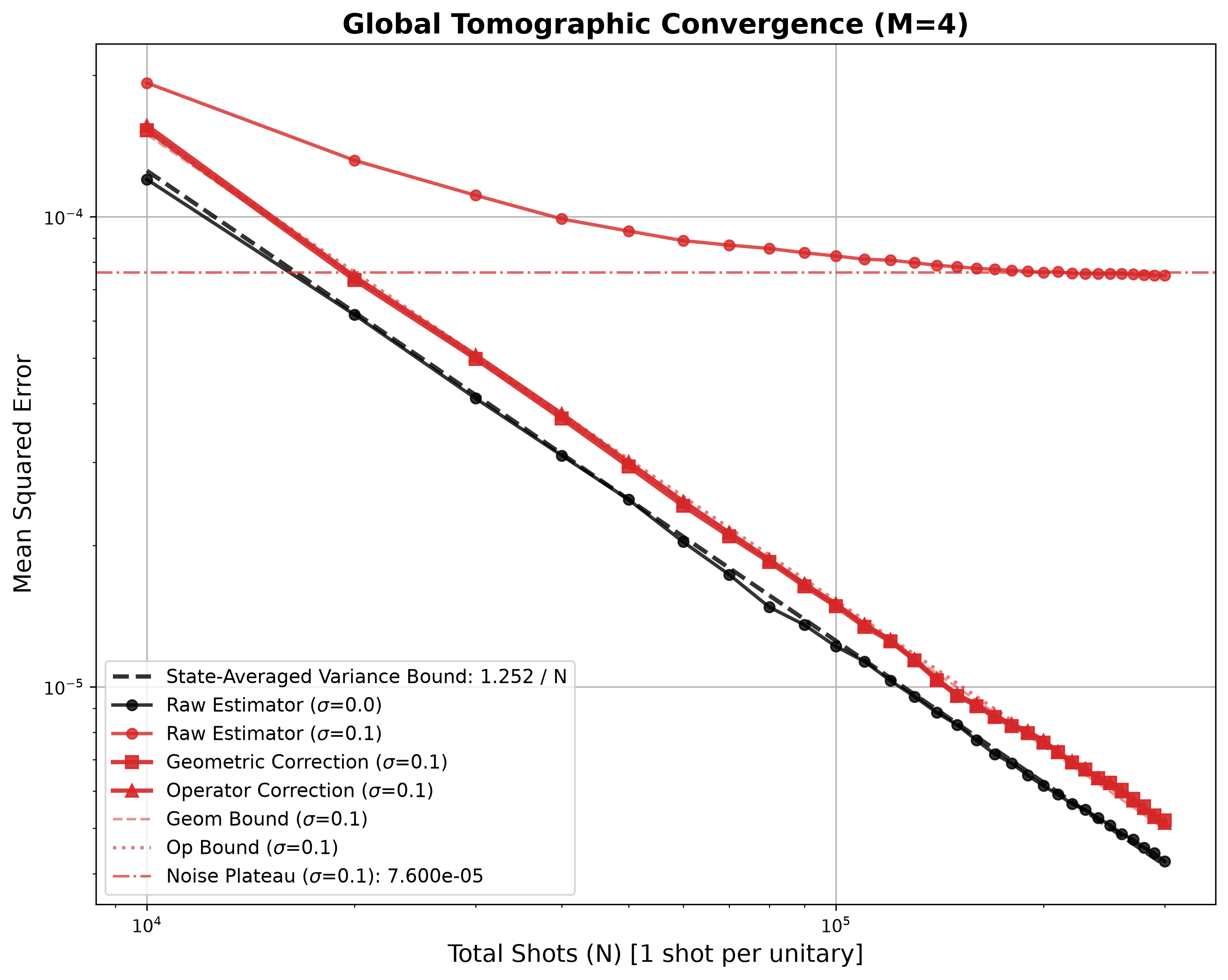}
    \caption{\textbf{Global tomographic MSE convergence for multi-sector superposition states ($M=4$, $\sigma=0.1$, $N = 3 \times 10^5$ shots, 30 ensembles).} \textbf{(Top)} Two-state superposition $\ket{\psi_2} = (\ket{1,1} + i\ket{1,2})/\sqrt{2}$, where the Raw Estimator saturates at the exact analytical plateau $5.50 \times 10^{-5}$ [Eq.~\eqref{eq:psi2_plateau_exact}]. \textbf{(Bottom)} Three-state superposition $\ket{\psi_3} = (\ket{1,1} + i\ket{1,2} - \ket{3,3})/\sqrt{3}$ spanning Population, Social, and Lonely sectors, where the Raw Estimator plateaus at $7.60 \times 10^{-5}$ [Eq.~\eqref{eq:psi3_plateau_exact}]. In both cases, the Operator Correction restores $1/N$ convergence.}
    \label{fig:convergence_superpositions}
\end{figure}

In the two-state superposition $\ket{\psi_2}$ [Eq.~\eqref{eq:psi2_def}], the density matrix $\rho_2 = \ket{\psi_2}\bra{\psi_2}$ populates the Population Block ($\vec{\Delta} = \vec{0}$) via $\frac{1}{2}\ket{1,1}\bra{1,1} + \frac{1}{2}\ket{1,2}\bra{1,2}$ and two conjugate $4 \times 4$ Social Coherence Blocks ($\Delta_1 = \pm 1, \Delta_2 = \mp 1$) via $\mp\frac{i}{2}\ket{1,1}\bra{1,2}$ and $\pm\frac{i}{2}\ket{1,2}\bra{1,1}$ (Family 2). As derived in Appendix~\ref{app:dissipator}, partial cancellation of the opposing leakage currents between $\ket{1,1}\bra{1,1}$ and $\ket{1,2}\bra{1,2}$ inside the Population Block yields a population bias sum of $19\sigma^4$, while the two conjugate Family 2 coherences contribute $2 \times \frac{1}{4}\Pi_2(\sigma, 4) = 2 \times 18\sigma^4 = 36\sigma^4$. Averaging over $D^2 = 100$ elements gives the exact plateau
\begin{equation}
    \text{MSE}_{\text{plateau}}(\ket{\psi_2}) = \frac{(19 + 36)\sigma^4}{100} = 5.50 \times 10^{-5} \text{.}
    \label{eq:psi2_plateau_exact}
\end{equation}

In the three-state superposition $\ket{\psi_3}$ [Eq.~\eqref{eq:psi3_def}], adding the $-\frac{1}{\sqrt{3}}\ket{3,3}$ component couples the bunched-coincidence pair in waveguides 1 and 2 to a two-photon NOON component in waveguide 3, activating all three dynamical architectures simultaneously: three populations in the $10 \times 10$ Population Block (contributing $\frac{140}{9}\sigma^4$), two conjugate Family 2 spectator-referenced coherences $\ket{1,1}\bra{1,2}, \ket{1,2}\bra{1,1}$ in $4 \times 4$ Social Blocks (contributing $2 \times \frac{1}{9}\Pi_2(\sigma, 4) = \frac{144}{9}\sigma^4$), and four isolated Family 5 coherences comprising the NOON transitions $\ket{1,1}\bra{3,3}, \ket{3,3}\bra{1,1} \in \mathcal{T}_1$ and the disjoint HOM transitions $\ket{1,2}\bra{3,3}, \ket{3,3}\bra{1,2} \in \mathcal{T}_2$ in $1 \times 1$ Lonely Blocks (contributing $4 \times \frac{1}{9}\Pi_5(\sigma, 4) = \frac{400}{9}\sigma^4$). Because these reference-free Family 5 coherences experience no common-mode phase cancellation, they account for nearly two-thirds of the total uncalibrated error, raising the global plateau (Appendix~\ref{app:dissipator}) to
\begin{equation}
    \text{MSE}_{\text{plateau}}(\ket{\psi_3}) = \frac{(140 + 144 + 400)\sigma^4}{9 \times 100} = 7.60 \times 10^{-5} \text{.}
    \label{eq:psi3_plateau_exact}
\end{equation}
As shown in Fig.~\ref{fig:convergence_superpositions}, the Monte Carlo simulations converge to $5.50 \times 10^{-5}$ and $7.60 \times 10^{-5}$ for the uncalibrated estimator, while the Operator Correction removes the coherent bias across all populated blocks and restores $1/N$ scaling.

% ==============================================================================
% SECTION VIII: DISCUSSION AND CONCLUSION
% ==============================================================================
\section{Discussion and Conclusion}
\label{sec:conclusion}

Solving the two-photon classical shadow channel analytically clarifies how non-interacting bosonic quantum optics differs from qubit-based randomized measurements. Because passive beam splitters and phase shifters implement single-particle transformations $U \in \mathcal{U}(M)$ without photon-photon interactions, multiphoton states scramble solely through single-photon wave interference and Hong-Ou-Mandel bunching~\cite{Hong1987}. Rather than depolarizing all operators at a uniform rate, the $D^2 \times D^2$ linear-optical shadow superoperator splits into Glauber's three optical coherence orders~\cite{Glauber1963}: conserved total photon flux ($G^{(0)}$, eigenvalue $\lambda_0 = 1$), the $(M^2-1)$-dimensional $\mathfrak{su}(M)$ subspace of single-photon first-order optical coherences ($G^{(1)}$, eigenvalue $\lambda_1 = \frac{1}{M+1}$), and the remaining $\frac{M^2(M-1)(M+3)}{4}$-dimensional space of irreducible two-photon coincidence correlations ($G^{(2)}$, eigenvalue $\lambda_2 = \frac{2}{M(M+3)}$). Because only these three physical visibility scales appear across the entire Liouville space, reconstructing any density-matrix element reduces to closed-form rational expressions in the port count $M$, eliminating numerical superoperator inversion.

The local optical phase superselection rule $\vec{\Delta}_{\alpha\beta} = \vec{\Delta}_{\mu\nu}$ explains microscopically how phase-blind photon-number-resolving detectors sense multiphoton quantum phases without an external homodyne local oscillator, and why different photonic encodings respond so differently to static interferometer imperfections. Under thermo-optic phase crosstalk and directional-coupler splitting errors~\cite{Reck1994, Clements2016, HardwareHorizon2026}, ensemble averaging over random interferometer settings converts unitary circuit distortions $\tilde{U} = U\exp(i\sigma H)$ into an intra-sector Gorini-Kossakowski-Sudarshan-Lindblad master-equation dissipator $\mathcal{D}(\rho) = 2(M+1)\rho - \sum_{p,q=1}^M L_{pq}\rho L_{pq}^\dagger$ driven by number-conserving bosonic hopping operators $L_{pq} = a_p^\dagger a_q$. When ket and bra share an occupied waveguide, that spectator photon acts as a co-propagating internal phase reference whose common-mode phase jitter cancels via the diagonal correlator $-\sum_p \hat{n}_p \rho \hat{n}_p$. Conversely, two-photon NOON states~\cite{Boto2000} ($\ket{i,i}\bra{j,j}$) and dual-rail two-qubit photonic coherences~\cite{Knill2001} ($\ket{i,j}\bra{k,l}$) share zero waveguides between ket and bra; lacking both $G^{(1)}$ single-photon visibility and an internal phase reference, they suffer the largest statistical shot noise and the highest coherent-crosstalk floor $4(M+1)^2\sigma^4$. Most notably, Hong-Ou-Mandel coalescence in $\ket{i,i}\bra{i,j}$ concentrates three of the four ket-bra photon amplitudes into a single waveguide, halving the number of parasitic hopping channels so that this phase-sensitive quantum coherence achieves both lower shot noise and a smaller crosstalk floor than a phase-insensitive coincidence population $\ket{i,j}\bra{i,j}$ for every $M \ge 2$.

These findings carry direct practical consequences for experiments on integrated silicon nitride, silica, and thin-film lithium niobate photonic processors~\cite{Flamini2018, Struchalin2021}. First, estimating single-photon $G^{(1)}$ observables or spectator-referenced coherences requires $\mathcal{O}(M)$ fewer experimental shots than certifying NOON or dual-rail entangling coherences, whose $G^{(2)}$ visibility decays as $\sim 2/M^2$. Second, characterizing a single hardware parameter, the effective circuit crosstalk amplitude $\sigma$ (measurable independently from single-photon Mach-Zehnder fringe visibility or Hong-Ou-Mandel dip contrast), determines the mode-hopping dissipator $\mathcal{D}$ completely. Inverting the noise-calibrated $d(\vec{\Delta}) \times d(\vec{\Delta})$ superselection blocks then resums multi-step intra-sector hopping to all orders, removing the $\mathcal{O}(\sigma^4)$ tomographic error floor reported in Ref.~\cite{HardwareHorizon2026} down to $\mathcal{O}(\sigma^8)$ and restoring $1/N$ convergence. Third, in uncalibrated photonic networks, encoding quantum information in bunched-coincidence superpositions (Family 2) provides greater passive immunity to static waveguide crosstalk than standard dual-rail coincidence encodings.

Finally, both the local phase superselection law~\eqref{eq:superselection_main} and the bosonic mode-hopping Lindbladian $\mathcal{D}(\rho) = n(M+n-1)\rho - \sum_{p,q=1}^M a_p^\dagger a_q \rho a_q^\dagger a_p$ [Eqs.~\eqref{eq:casimir_general_n}--\eqref{eq:dissipator_def}] hold for arbitrary photon number $n$, providing a direct physical roadmap for $n$-photon Boson Sampling~\cite{Aaronson2013} and number-conserving cold-atom interferometers~\cite{Hearth2024}. For $n$ indistinguishable bosons, passive $\mathcal{U}(M)$ interferometry partitions the observable space into $n+1$ Glauber coherence orders $G^{(0)}, G^{(1)}, \dots, G^{(n)}$ (governed in the collision-free regime $M \gg n^2$ by the $(n+1)$-class Johnson scheme $J(M,n)$ and its Eberlein eigenpolynomials), whose interferometric visibilities scale hierarchically as $\mathcal{O}(M^{-r})$ for $r$-body correlations ($0 \le r \le n$). Transitions with imbalance norm $\|\vec{\Delta}\|_1 = 2r$ retain $n - r$ shared spectator bosons as co-propagating internal phase references while the remaining $r$ bosons undergo $r$-photon interference, extending common-mode phase protection and block-level crosstalk calibration across general multiphoton quantum optics.

\begin{acknowledgments}
This research was carried out with the support of the Ministry of Culture and Innovation, funded by the National Research Development and Innovation Fund, under project number 2022-2.1.1-NL-2022-00004.
\end{acknowledgments}

% ==============================================================================
% APPENDICES
% ==============================================================================
\makeatletter
\let\saved@hangfrom@section\@hangfrom@section
\let\saved@sectioncntformat\@sectioncntformat
\makeatother

\appendix

\section{Unitary Weingarten Functions and Evaluation of the Ten Integral Groups}
\label{app:weingarten}

In this appendix, we provide the explicit degree-4 unitary Weingarten functions $\text{Wg}(\pi, M)$ appearing in Eq.~\eqref{eq:S0_weingarten_master} and demonstrate how contracting over the $D$ projective measurement outcomes $\kappa = \{m_1, m_2\}$ yields the ten Integral Groups $\mathcal{O}_1$--$\mathcal{O}_{10}$ in Table~\ref{tab:forward_integral_groups}.

For the symmetric group $S_4$ ($q=4$), the unitary Weingarten function $\text{Wg}(\pi, M)$ is a class function that depends only on the cycle partition $\lambda(\pi) \vdash 4$ of the permutation $\pi = \sigma^{-1}\tau \in S_4$. There are five conjugacy classes in $S_4$, corresponding to the integer partitions $[1,1,1,1]$ (identity, 1 element), $[2,1,1]$ (single transposition, 6 elements), $[2,2]$ (double transposition, 3 elements), $[3,1]$ (3-cycle, 8 elements), and $[4]$ (4-cycle, 6 elements). Inverting the $24 \times 24$ Gram matrix $G_{\sigma\tau} = M^{\#\text{cycles}(\sigma^{-1}\tau)}$ over the group algebra $\mathbb{C}[S_4]$ yields the five rational Weingarten functions~\cite{Collins2006}:
\begin{align}
    \text{Wg}([1,1,1,1], M) &= \frac{M^4 - 8M^2 + 6}{M^2(M^2-1)(M^2-4)(M^2-9)} \text{,} \label{eq:wg_1111} \\
    \text{Wg}([2,1,1], M) &= -\frac{1}{M(M^2-1)(M^2-9)} \text{,} \label{eq:wg_211} \\
    \text{Wg}([2,2], M) &= \frac{M^2 + 6}{M^2(M^2-1)(M^2-4)(M^2-9)} \text{,} \label{eq:wg_22} \\
    \text{Wg}([3,1], M) &= \frac{2M^2 - 3}{M^2(M^2-1)(M^2-4)(M^2-9)} \text{,} \label{eq:wg_31} \\
    \text{Wg}([4], M) &= -\frac{5}{M(M^2-1)(M^2-4)(M^2-9)} \text{.} \label{eq:wg_4}
\end{align}

In Eq.~\eqref{eq:S0_weingarten_master}, we can perform the sum over all $D$ detector outcomes $\kappa = \{m_1, m_2\}$ and all $16$ symmetrization tuples $w \in \Omega$ analytically before evaluating the column permutations $\tau \in S_4$. For any permutation $\sigma \in S_4$, let us define the detector-summed multiplicity
\begin{equation}
    K(\sigma, M) \equiv \sum_{\kappa=1}^D \frac{1}{(1+\delta_{m_1 m_2})^2} \sum_{w \in \Omega} \delta_\sigma\big(\mathbf{r}^{(w)}, {\mathbf{r}'}^{(w)}\big) \text{.}
    \label{eq:K_sigma_def}
\end{equation}
Recall from Eqs.~\eqref{eq:row_vec_U}--\eqref{eq:row_vec_Ustar} that both pairs of positions $\{1,2\}$ and $\{3,4\}$ in $\mathbf{r}^{(w)}$ and ${\mathbf{r}'}^{(w)}$ are occupied by the unordered detector pair $\{m_1, m_2\}$. Let $D_4 = (S_2 \times S_2) \rtimes \mathbb{Z}_2 \subset S_4$ denote the 8-element dihedral subgroup of permutations that preserve the pair partition $\{\{1,2\}, \{3,4\}\}$ (consisting of the 4 permutations in $S_2 \times S_2$ that map $\{1,2\}\to\{1,2\}$ and $\{3,4\}\to\{3,4\}$, and the 4 permutations that swap $\{1,2\}\leftrightarrow\{3,4\}$). For the $M$ collision outcomes $m_1 = m_2 = m$, $(1+\delta_{m_1 m_2})^2 = 4$ and $\delta_\sigma(\mathbf{r}^{(w)}, {\mathbf{r}'}^{(w)}) = 1$ for all $16$ tuples $w \in \Omega$ and all $24$ permutations $\sigma \in S_4$. For the $\binom{M}{2}$ distinct outcomes $m_1 < m_2$, $(1+\delta_{m_1 m_2})^2 = 1$, and $\delta_\sigma(\mathbf{r}^{(w)}, {\mathbf{r}'}^{(w)})$ is satisfied by $4$ tuples $w \in \Omega$ if $\sigma \in D_4$ and by $0$ tuples if $\sigma \notin D_4$. Consequently, $K(\sigma, M)$ takes only two values on $S_4$:
\begin{equation}
    K(\sigma, M) = \begin{cases}
        2M(M+1) & \text{if } \sigma \in D_4 \;\; (8 \text{ elements})\text{,} \\[6pt]
        4M & \text{if } \sigma \notin D_4 \;\; (16 \text{ elements})\text{.}
    \end{cases}
    \label{eq:K_sigma_values}
\end{equation}

Convolving Eq.~\eqref{eq:K_sigma_values} with the Weingarten functions~\eqref{eq:wg_1111}--\eqref{eq:wg_4} defines the detector-contracted Weingarten kernel $W_{\text{det}}(\tau, M) \equiv \sum_{\sigma \in S_4} \text{Wg}(\sigma^{-1}\tau, M) K(\sigma, M)$. Because $K(\sigma, M)$ is invariant under left and right multiplication by $D_4$, all unphysical poles at $M=1, 2, 3$ in Eqs.~\eqref{eq:wg_1111}--\eqref{eq:wg_4} cancel identically, and $W_{\text{det}}(\tau, M)$ reduces to a two-valued function on $S_4$:
\begin{equation}
    W_{\text{det}}(\tau, M) = \begin{cases}
        \displaystyle \frac{2}{M(M+3)} & (\tau \in D_4)\text{,} \\[8pt]
        \displaystyle \frac{M - 1}{M(M+1)(M+3)} & (\tau \notin D_4)\text{.}
    \end{cases}
    \label{eq:W_det_collapsed}
\end{equation}
Substituting Eq.~\eqref{eq:W_det_collapsed} into Eq.~\eqref{eq:S0_weingarten_master} collapses the entire degree-8 Haar integral sum to a Closed-Form Combinatorial Master Formula:
\begin{equation}
    S^{(0)}_{\mu\nu, \alpha\beta} = \frac{2(M+1) \, n_{\text{in}}(\mathbf{c}, \mathbf{c}') + (M-1) \, n_{\text{out}}(\mathbf{c}, \mathbf{c}')}{M(M+1)(M+3)\mathcal{M}_{\mu,\nu,\alpha,\beta}} \text{,}
    \label{eq:master_combinatorial_formula}
\end{equation}
where the normalization denominator is
\begin{equation}
    \mathcal{M}_{\mu,\nu,\alpha,\beta} = \sqrt{(1+\delta_{i_1 i_2})(1+\delta_{l_1 l_2})(1+\delta_{k_1 k_2})(1+\delta_{j_1 j_2})} \text{,}
    \label{eq:M_norm_factor}
\end{equation}
and $n_{\text{in}}(\mathbf{c}, \mathbf{c}') = \sum_{\tau \in D_4} \delta_\tau(\mathbf{c}, \mathbf{c}')$ and $n_{\text{out}}(\mathbf{c}, \mathbf{c}') = \sum_{\tau \notin D_4} \delta_\tau(\mathbf{c}, \mathbf{c}')$ count how many permutations inside and outside $D_4$ match $\mathbf{c} = (i_1, i_2, l_1, l_2)$ to $\mathbf{c}' = (k_1, k_2, j_1, j_2)$.

All ten Integral Groups $\mathcal{O}_1$--$\mathcal{O}_{10}$ in Table~\ref{tab:forward_integral_groups} follow immediately from Eq.~\eqref{eq:master_combinatorial_formula}:
\begin{itemize}
    \item \textbf{Group $\mathcal{O}_1$} ($\mathbf{c}=(i,i,i,i), \mathbf{c}'=(i,i,i,i)$, $\mathcal{M}=4$): $(n_{\text{in}}, n_{\text{out}}) = (8, 16) \implies \frac{16(M+1) + 16(M-1)}{4M(M+1)(M+3)} = \frac{8}{(M+1)(M+3)}$.
    \item \textbf{Group $\mathcal{O}_2$} ($\mathbf{c}=(i,i,i,j), \mathbf{c}'=(i,j,i,i)$, $\mathcal{M}=2$): $(n_{\text{in}}, n_{\text{out}}) = (2, 4) \implies \frac{4(M+1) + 4(M-1)}{2M(M+1)(M+3)} = \frac{4}{(M+1)(M+3)}$.
    \item \textbf{Group $\mathcal{O}_3$} ($\mathbf{c}=(i,i,j,j), \mathbf{c}'=(j,j,i,i)$ or $(i,i,j,j)$, $\mathcal{M}=4$): $(n_{\text{in}}, n_{\text{out}}) = (4, 0) \implies \frac{8(M+1)}{4M(M+1)(M+3)} = \frac{2}{M(M+3)}$.
    \item \textbf{Group $\mathcal{O}_4$} ($\mathbf{c}=(i,i,j,j), \mathbf{c}'=(i,j,i,j)$, $\mathcal{M}=2$): $(n_{\text{in}}, n_{\text{out}}) = (0, 4) \implies \frac{4(M-1)}{2M(M+1)(M+3)} = \frac{2(M-1)}{M(M+1)(M+3)}$.
    \item \textbf{Group $\mathcal{O}_5$} ($\mathbf{c}=(i,j,i,j), \mathbf{c}'=(i,j,i,j)$, $\mathcal{M}=1$): $(n_{\text{in}}, n_{\text{out}}) = (2, 2) \implies \frac{4(M+1) + 2(M-1)}{M(M+1)(M+3)} = \frac{2(3M+1)}{M(M+1)(M+3)}$.
    \item \textbf{Group $\mathcal{O}_6$} ($\mathbf{c}=(i,i,j,k), \mathbf{c}'=(j,k,i,i)$ or $(i,i,j,k)$, $\mathcal{M}=2$): $(n_{\text{in}}, n_{\text{out}}) = (2, 0) \implies \frac{4(M+1)}{2M(M+1)(M+3)} = \frac{2}{M(M+3)}$.
    \item \textbf{Group $\mathcal{O}_7$} ($\mathbf{c}=(i,i,j,k), \mathbf{c}'=(i,j,i,k)$, $\mathcal{M}=\sqrt{2}$): $(n_{\text{in}}, n_{\text{out}}) = (0, 2) \implies \frac{2(M-1)}{\sqrt{2}M(M+1)(M+3)} = \frac{\sqrt{2}(M-1)}{M(M+1)(M+3)}$.
    \item \textbf{Group $\mathcal{O}_8$} ($\mathbf{c}=(i,j,i,k), \mathbf{c}'=(i,k,i,j)$, $\mathcal{M}=1$): $(n_{\text{in}}, n_{\text{out}}) = (1, 1) \implies \frac{2(M+1) + (M-1)}{M(M+1)(M+3)} = \frac{3M+1}{M(M+1)(M+3)}$.
    \item \textbf{Group $\mathcal{O}_9$} ($\mathbf{c}=(i,j,k,l), \mathbf{c}'=(k,l,i,j)$ or $(i,j,k,l)$, $\mathcal{M}=1$): $(n_{\text{in}}, n_{\text{out}}) = (1, 0) \implies \frac{2(M+1)}{M(M+1)(M+3)} = \frac{2}{M(M+3)}$.
    \item \textbf{Group $\mathcal{O}_{10}$} ($\mathbf{c}=(i,j,j,l), \mathbf{c}'=(i,j,j,l)$ with $c_2 = c_3 = c'_2 = c'_3 = j$, $\mathcal{M}=1$): $(n_{\text{in}}, n_{\text{out}}) = (0, 1) \implies \frac{M-1}{M(M+1)(M+3)}$.
\end{itemize}

\section{Step-by-Step Derivation of the Analytical Block Inverses}
\label{app:inversion}

Here we derive the closed-form inverse formulas for the $D \times D$ Population Block [Eq.~\eqref{eq:S0_pop_inverse_closed}] and the $M \times M$ Social Coherence Blocks [Eq.~\eqref{eq:S0_soc_inverse_closed}].

\subsection{Inversion of the \texorpdfstring{$D \times D$}{D x D} Population Block}

Let $\mathbf{N} \in \mathbb{R}^{D \times M}$ be the mode-occupation incidence matrix with entries $N_{\alpha m} = n_m(\alpha) = \delta_{m,i} + \delta_{m,j}$ for $\alpha = \{i, j\}$. Summing across a row gives the total photon number $\sum_{m=1}^M N_{\alpha m} = 2$, so $\mathbf{N}\mathbf{1}_M = 2\mathbf{1}_D$. Summing down column $m$ gives $\sum_{\alpha=1}^D N_{\alpha m} = 2 + (M-1)\cdot 1 = M+1$, so $\mathbf{N}^\top \mathbf{1}_D = (M+1)\mathbf{1}_M$. Moreover, the $M \times M$ matrix $\mathbf{N}^\top \mathbf{N}$ has diagonal entries $\sum_\alpha N_{\alpha m}^2 = 4 + (M-1)\cdot 1 = M+3$ and off-diagonal entries $\sum_\alpha N_{\alpha m} N_{\alpha m'} = 1$ ($m \neq m'$), so
\begin{equation}
    \mathbf{N}^\top \mathbf{N} = (M+2)I_M + \mathbf{J}_M \text{.}
    \label{eq:app_NTN}
\end{equation}
From these relations, we compute the products of $\mathbf{J}_D = \mathbf{1}_D \mathbf{1}_D^\top$ and $\mathbf{N}\mathbf{N}^\top$:
\begin{align}
    \mathbf{J}_D^2 &= D\,\mathbf{J}_D = \frac{M(M+1)}{2}\,\mathbf{J}_D \text{,} \\
    (\mathbf{N}\mathbf{N}^\top)\mathbf{J}_D &= \mathbf{J}_D(\mathbf{N}\mathbf{N}^\top) = 2(M+1)\,\mathbf{J}_D \text{,} \\
    (\mathbf{N}\mathbf{N}^\top)^2 &= (M+2)\mathbf{N}\mathbf{N}^\top + 4\mathbf{J}_D \text{.}
\end{align}
Thus $\{I_D, \mathbf{J}_D, \mathbf{N}\mathbf{N}^\top\}$ spans a 3-dimensional commutative algebra with orthogonal idempotents $E_0, E_1, E_2$ given in Eqs.~\eqref{eq:idempotent_E0}--\eqref{eq:idempotent_E2}. Rewriting $\mathbf{S}^{(0)}_{\vec{0}}$ in Eq.~\eqref{eq:S0_pop_gram} in the idempotent basis,
\begin{equation}
    \mathbf{S}^{(0)}_{\vec{0}} = E_0 + \frac{1}{M+1}E_1 + \frac{2}{M(M+3)}E_2 \text{,}
\end{equation}
and substituting $E_2 = I_D - E_0 - E_1$ into $(\mathbf{S}^{(0)}_{\vec{0}})^{-1} = E_0 + (M+1)E_1 + \frac{M(M+3)}{2}E_2$ gives
\begin{equation}
\begin{split}
    \left(\mathbf{S}^{(0)}_{\vec{0}}\right)^{-1} &= \frac{M(M+3)}{2} I_D \\
    &\quad + \left(1 - \frac{M(M+3)}{2}\right)\frac{2}{M(M+1)}\mathbf{J}_D \\
    &\quad + \left(M+1 - \frac{M(M+3)}{2}\right) \\
    &\qquad \times \frac{1}{M+2}\left(\mathbf{N}\mathbf{N}^\top - \frac{4}{M}\mathbf{J}_D\right) \\
    &= \frac{M(M+3)}{2} I_D - \frac{M-1}{2}\mathbf{N}\mathbf{N}^\top + \frac{M-3}{M+1}\mathbf{J}_D \text{,}
\end{split}
\end{equation}
which establishes Eq.~\eqref{eq:S0_pop_inverse_closed} and all entries of Part I of Table~\ref{tab:exact_inverse_formulas_final}.

\subsection{Inversion of the \texorpdfstring{$M \times M$}{M x M} Social Coherence Blocks}

For the Social Coherence Block $\mathbf{S}^{(0)}_{\text{soc}} = a I_M + b \mathbf{v}\mathbf{v}^\top$ in Eq.~\eqref{eq:S0_soc_rank1} with $a = \frac{2}{M(M+3)}$, $b = \frac{M-1}{M(M+1)(M+3)}$, and $\mathbf{v}^\top\mathbf{v} = M+2$, the Sherman-Morrison formula gives
\begin{equation}
    \big(a I_M + b \mathbf{v}\mathbf{v}^\top\big)^{-1} = \frac{1}{a} I_M - \frac{b}{a\big(a + b(\mathbf{v}^\top\mathbf{v})\big)} \mathbf{v}\mathbf{v}^\top \text{.}
\end{equation}
Since $a + b(M+2) = \frac{2(M+1) + (M-1)(M+2)}{M(M+1)(M+3)} = \frac{1}{M+1}$, the rank-one coefficient simplifies to $\frac{b}{a(a+b(M+2))} = \frac{M-1}{2}$, proving Eq.~\eqref{eq:S0_soc_inverse_closed} and all entries of Part II of Table~\ref{tab:exact_inverse_formulas_final}.

\section{Action of the Casimir Dissipator and Analytical Bias Plateaus}
\label{app:dissipator}

In this appendix, we evaluate the action of the Casimir scattering dissipator $\mathcal{D}(\rho) = 2(M+1)\rho - C(\rho)$ [Eq.~\eqref{eq:dissipator_def}], where $C(\rho) = \sum_{p,q=1}^M a_p^\dagger a_q \rho a_q^\dagger a_p$, on each of the five Quantum Memory Families, deriving the exact bias plateau bounds $\Pi_i(\sigma, M)$ in Table~\ref{tab:quantum_memory_families} and the multi-state benchmarks in Section~\ref{sec:numerical_verification}.

\begin{enumerate}
    \item \textbf{Family 1 ($\ket{i,i}\bra{i,i}$, Topology $\mathcal{T}_6$):} Only $q=i$ annihilates $\ket{i,i}$. For $p=i$, $a_i^\dagger a_i \ket{i,i} = 2\ket{i,i}$, contributing $4\ket{i,i}\bra{i,i}$ to $C$. For each of the $M-1$ modes $p \neq i$, $a_p^\dagger a_i \ket{i,i} = \sqrt{2}\ket{i,p}$, contributing $2\ket{i,p}\bra{i,p}$ to $C$. Thus
    \begin{equation}
        \mathcal{D}\big(\ket{i,i}\bra{i,i}\big) = 2(M-1)\ket{i,i}\bra{i,i} - 2 \sum_{p \neq i} \ket{i,p}\bra{i,p} \text{.}
        \label{eq:app_D_fam1}
    \end{equation}
    Summing the squares of the coefficients gives $\Pi_1(\sigma, M) = \big([2(M-1)]^2 + (M-1)(-2)^2\big)\sigma^4 = 4M(M-1)\sigma^4$.
    \item \textbf{Family 2 ($\ket{i,i}\bra{i,j}$ with $i \neq j$, Topology $\mathcal{T}_4$):} Because $a_q$ must annihilate both $\ket{i,i}$ and $\ket{i,j}$, only $q=i$ contributes to $C(\ket{i,i}\bra{i,j})$. For $p=i$, $(2\ket{i,i})(\bra{i,j})$ gives $2\ket{i,i}\bra{i,j}$. For $p=j$, $(\sqrt{2}\ket{i,j})(\sqrt{2}\bra{j,j})$ gives $2\ket{i,j}\bra{j,j}$. For the $M-2$ modes $p \notin \{i, j\}$, $(\sqrt{2}\ket{i,p})(\bra{p,j})$ gives $\sqrt{2}\ket{i,p}\bra{p,j}$. Thus
    \begin{equation}
    \begin{split}
        \mathcal{D}\big(\ket{i,i}\bra{i,j}\big) &= 2M\ket{i,i}\bra{i,j} - 2\ket{i,j}\bra{j,j} \\
        &\quad - \sqrt{2} \sum_{p \notin \{i,j\}} \ket{i,p}\bra{p,j} \text{.}
    \end{split}
    \label{eq:app_D_fam2}
    \end{equation}
    Summing the squared coefficients gives $\Pi_2(\sigma, M) = \big((2M)^2 + (-2)^2 + (M-2)(-\sqrt{2})^2\big)\sigma^4 = 2M(2M+1)\sigma^4$.
    \item \textbf{Family 3 ($\ket{i,j}\bra{i,j}$ with $i \neq j$, Topology $\mathcal{T}_7$):} Both $q=i$ and $q=j$ contribute to $C(\ket{i,j}\bra{i,j})$. For $q=i$: $p=i$ gives $\ket{i,j}\bra{i,j}$; $p=j$ gives $2\ket{j,j}\bra{j,j}$; and $p \notin \{i,j\}$ gives $\ket{p,j}\bra{p,j}$. Symmetrically for $q=j$: $p=j$ gives $\ket{i,j}\bra{i,j}$; $p=i$ gives $2\ket{i,i}\bra{i,i}$; and $p \notin \{i,j\}$ gives $\ket{i,p}\bra{i,p}$. Thus
    \begin{equation}
    \begin{split}
        \mathcal{D}\big(\ket{i,j}\bra{i,j}\big) &= 2M\ket{i,j}\bra{i,j} \\
        &\quad - 2\big(\ket{i,i}\bra{i,i} + \ket{j,j}\bra{j,j}\big) \\
        &\quad - \sum_{p \notin \{i,j\}} \big(\ket{i,p}\bra{i,p} + \ket{p,j}\bra{p,j}\big) \text{.}
    \end{split}
    \label{eq:app_D_fam3}
    \end{equation}
    Summing the squared coefficients gives $\Pi_3(\sigma, M) = \big((2M)^2 + 2(-2)^2 + 2(M-2)(-1)^2\big)\sigma^4 = 2(2M^2 + M + 2)\sigma^4$.
    \item \textbf{Family 4 ($\ket{i,k}\bra{j,k}$ with $i, j, k$ distinct, Topology $\mathcal{T}_5$):} Only the shared spectator mode $q=k$ contributes to $C(\ket{i,k}\bra{j,k})$. For $p=k$, we obtain $\ket{i,k}\bra{j,k}$; for $p=i$ and $p=j$, bosonic enhancement produces $\sqrt{2}\ket{i,i}\bra{i,j}$ and $\sqrt{2}\ket{i,j}\bra{j,j}$; for the $M-3$ modes $p \notin \{i,j,k\}$, we obtain $\ket{i,p}\bra{j,p}$. Thus
    \begin{equation}
    \begin{split}
        \mathcal{D}\big(\ket{i,k}\bra{j,k}\big) &= (2M+1)\ket{i,k}\bra{j,k} \\
        &\quad - \sqrt{2}\big(\ket{i,i}\bra{i,j} + \ket{i,j}\bra{j,j}\big) \\
        &\quad - \sum_{p \notin \{i,j,k\}} \ket{i,p}\bra{j,p} \text{.}
    \end{split}
    \label{eq:app_D_fam4}
    \end{equation}
    Summing the squared coefficients gives $\Pi_4(\sigma, M) = \big((2M+1)^2 + 2(-\sqrt{2})^2 + (M-3)(-1)^2\big)\sigma^4 = (4M^2 + 5M + 2)\sigma^4$.
    \item \textbf{Family 5 (Disjoint bra and ket $\alpha \cap \beta = \emptyset$, Topologies $\mathcal{T}_{1,2,3}$):} Since no mode $q$ is occupied in both $\ket{\alpha}$ and $\ket{\beta}$, $C(\ket{\alpha}\bra{\beta}) = 0$, so
    \begin{equation}
        \mathcal{D}\big(\ket{\alpha}\bra{\beta}\big) = 2(M+1)\ket{\alpha}\bra{\beta} \text{,}
        \label{eq:app_D_fam5}
    \end{equation}
    giving $\Pi_5(\sigma, M) = 4(M+1)^2\sigma^4$.
\end{enumerate}

Finally, combining Eqs.~\eqref{eq:app_D_fam1}--\eqref{eq:app_D_fam5} at $M=4$ immediately yields the exact multi-state plateaus in Eqs.~\eqref{eq:psi2_plateau_exact} and~\eqref{eq:psi3_plateau_exact}. For $\rho_2 = \frac{1}{2}(\ket{1,1}\bra{1,1} + \ket{1,2}\bra{1,2} - i\ket{1,1}\bra{1,2} + i\ket{1,2}\bra{1,1})$, applying Eqs.~\eqref{eq:app_D_fam1} and~\eqref{eq:app_D_fam3} to the population part $\rho_{2,\text{pop}} = \frac{1}{2}(\ket{1,1}\bra{1,1} + \ket{1,2}\bra{1,2})$ gives
\begin{equation}
\begin{split}
    \mathcal{D}(\rho_{2,\text{pop}}) &= 2\ket{1,1}\bra{1,1} + 3\ket{1,2}\bra{1,2} - \ket{2,2}\bra{2,2} \\
    &\quad - \frac{3}{2}\big(\ket{1,3}\bra{1,3} + \ket{1,4}\bra{1,4}\big) \\
    &\quad - \frac{1}{2}\big(\ket{2,3}\bra{2,3} + \ket{2,4}\bra{2,4}\big) \text{,}
\end{split}
\end{equation}
whose squared norm is $2^2 + 3^2 + (-1)^2 + 2(-\frac{3}{2})^2 + 2(-\frac{1}{2})^2 = 4 + 9 + 1 + 4.5 + 0.5 = 19$. Adding the two orthogonal Family 2 sectors ($2 \times \frac{1}{4}\Pi_2(1,4) = 36$) gives $19 + 36 = 55$. Similarly, for $\rho_3$, the population part $\frac{1}{3}(\ket{1,1}\bra{1,1} + \ket{1,2}\bra{1,2} + \ket{3,3}\bra{3,3})$ has squared norm $\frac{1}{9}(4^2 + (-2)^2 + 6^2 + 6^2 + (-5)^2 + (-3)^2 + (-3)^2 + (-1)^2 + (-2)^2) = \frac{140}{9}$, which together with the two Family 2 sectors ($\frac{144}{9}$) and four Family 5 sectors ($\frac{400}{9}$) sums to $\frac{684}{9} = 76$.


% ==============================================================================
% SUPPLEMENTAL MATERIAL
% ==============================================================================
\clearpage
\onecolumngrid

\makeatletter
\let\@hangfrom@section\saved@hangfrom@section
\let\@sectioncntformat\saved@sectioncntformat
\@removefromreset{equation}{section}
\def\theequation@prefix{}
\makeatother

\setcounter{equation}{0}
\setcounter{figure}{0}
\setcounter{table}{0}
\setcounter{section}{0}
\renewcommand{\theequation}{S\arabic{equation}}
\renewcommand{\thefigure}{S\arabic{figure}}
\renewcommand{\thetable}{S\arabic{table}}
\renewcommand{\thesection}{S-\Roman{section}}
\renewcommand{\theHsection}{S\arabic{section}}
\renewcommand{\theHequation}{S\arabic{equation}}
\renewcommand{\theHfigure}{S\arabic{figure}}
\renewcommand{\theHtable}{S\arabic{table}}

\begin{center}
\textbf{\large Supplemental Material for: Multiphoton interference and internal phase references in classical shadow tomography of noisy linear-optical networks}\\[0.8em]
Attila Baumann,$^{1}$ Zsolt Kis,$^{2}$ J\'anos Koltai,$^{1}$ and G\'abor Vattay$^{1,*}$\\[0.3em]
\textit{$^{1}$Institute for Physics and Astronomy, E\"otv\"os Lor\'and University, P\'azm\'any P\'eter s\'et\'any 1/A, Budapest 1117, Hungary}\\
\textit{$^{2}$Institute for Solid State Physics and Optics, HUN-REN Wigner Research Centre for Physics, Konkoly-Thege Mikl\'os \'ut 29-33, Budapest 1121, Hungary}
\end{center}
\vspace{0.5em}
\hrule
\vspace{1.2em}

% ==============================================================================
% SECTION S-I: EXPLICIT 16-INTEGRAL EXPANSION AND PARTITION TABLES
% ==============================================================================
\section{Explicit 16-Term Haar Integral Expansion and the 22 Eight-Index Partitions}
\label{sec:supp_partitions}

In Sec.~\ref{sec:shadow_protocol} of the main text, the expected classical shadow signal $\overline{I}_{\mu\nu}$ for a two-photon state ($n=2$) in an $M$-mode linear optical network is expressed as a contraction over four matrix elements of the symmetrized two-photon unitary $V(U) = \Gamma(U)$,
\begin{equation}
    \overline{I}_{\mu\nu} = \sum_{\kappa=1}^D \sum_{\alpha=1}^D \sum_{\beta=1}^D \rho_{\alpha\beta} \, \mathbb{E}_{U \sim \text{Haar}(M)} \left[ V_{\kappa\alpha}(U) V_{\kappa\nu}(U) V^*_{\kappa\mu}(U) V^*_{\kappa\beta}(U) \right] \text{,}
    \label{eq:supp_accumulated_signal}
\end{equation}
where $D = \binom{M+1}{2} = M(M+1)/2$ is the dimension of the two-photon symmetric Fock subspace $\mathcal{H}_{\text{sym}}^{(2)}$. Assigning unordered two-element mode multisets to the five Fock indices, namely $\kappa = \{m_1, m_2\}$, $\alpha = \{i_1, i_2\}$, $\beta = \{j_1, j_2\}$, $\mu = \{k_1, k_2\}$, and $\nu = \{l_1, l_2\}$, and factoring out the bosonic normalization denominator
\begin{equation}
    \mathcal{N}_{\kappa, \mu, \nu, \alpha, \beta} = (1+\delta_{m_1 m_2})^2 \sqrt{(1+\delta_{i_1 i_2})(1+\delta_{l_1 l_2})(1+\delta_{k_1 k_2})(1+\delta_{j_1 j_2})} \text{,}
    \label{eq:supp_normalization}
\end{equation}
the product of the four two-photon permanents expands into 16 degree-8 Haar integral monomials over $\mathcal{U}(M)$:
\begin{equation}
\begin{split}
    \overline{I}_{\mu\nu} &= \sum_{\kappa=1}^D \sum_{\alpha=1}^D \sum_{\beta=1}^D \frac{\rho_{\alpha\beta}}{\mathcal{N}_{\kappa, \mu, \nu, \alpha, \beta}} \sum_{\pi_1, \pi_2, \pi_3, \pi_4 \in S_2} \\
    &\quad \times \int_{\mathcal{U}(M)} U_{m_{\pi_1(1)} i_1} U_{m_{\pi_1(2)} i_2} U_{m_{\pi_2(1)} l_1} U_{m_{\pi_2(2)} l_2} \\
    &\quad \times U^*_{m_{\pi_3(1)} k_1} U^*_{m_{\pi_3(2)} k_2} U^*_{m_{\pi_4(1)} j_1} U^*_{m_{\pi_4(2)} j_2} \, d\mu(U) \text{.}
\end{split}
\label{eq:supp_16_integrals}
\end{equation}

Because the Haar measure $d\mu(U)$ is invariant under permutations of the $M$ spatial mode labels, the value of $S^{(0)}_{\mu\nu,\alpha\beta}$ depends only on the coincidence pattern of the eight external mode indices $S_8 = \alpha \uplus \beta \uplus \mu \uplus \nu$. Table~\ref{tab:supp_22_partitions} lists all 22 integer partitions of 8 that classify the multiset $S_8$. Applying the $U(1)^{\otimes M}$ gauge condition $\alpha \uplus \nu = \mu \uplus \beta$ (equivalently $\vec{\Delta}_{\alpha\beta} = \vec{\Delta}_{\mu\nu}$) eliminates all partitions containing odd multiplicities or mismatched mode assignments, leaving only the five all-even partitions $[8]$, $[6,2]$, $[4,4]$, $[4,2,2]$, and $[2,2,2,2]$. Within these five partitions, 26 non-zero directed transition geometries survive, listed together with their ideal transition amplitudes $S^{(0)}_{\mu\nu,\alpha\beta}$ and second-order GUE noise corrections $S^{(2)}_{\mu\nu,\alpha\beta}$ in Table~\ref{tab:supp_26_complete}.

\begin{table}[htbp]
\centering
\caption{The 22 integer partitions of the 8-index multiset $S_8 = \alpha \uplus \beta \uplus \mu \uplus \nu$. Under the $U(1)^{\otimes M}$ superselection rule $\alpha \uplus \nu = \mu \uplus \beta$, only the five all-even partitions (Nos. 1, 3, 8, 10, and 18) can support non-zero superoperator elements.}
\label{tab:supp_22_partitions}
\begin{ruledtabular}
\begin{tabular}{ccc|ccc}
No. & Partition Class & Multiset Structure ($S_8$) & No. & Partition Class & Multiset Structure ($S_8$) \\
\hline
1  & $[8]$               & $\{i,i,i,i,i,i,i,i\}$ & 12 & $[4,1,1,1,1]$     & $\{i,i,i,i,j,k,l,m\}$ \\
2  & $[7,1]$             & $\{i,i,i,i,i,i,i,j\}$ & 13 & $[3,3,2]$         & $\{i,i,i,j,j,j,k,k\}$ \\
3  & $[6,2]$             & $\{i,i,i,i,i,i,j,j\}$ & 14 & $[3,3,1,1]$       & $\{i,i,i,j,j,j,k,l\}$ \\
4  & $[6,1,1]$           & $\{i,i,i,i,i,i,j,k\}$ & 15 & $[3,2,2,1]$       & $\{i,i,i,j,j,k,k,l\}$ \\
5  & $[5,3]$             & $\{i,i,i,i,i,j,j,j\}$ & 16 & $[3,2,1,1,1]$     & $\{i,i,i,j,j,k,l,m\}$ \\
6  & $[5,2,1]$           & $\{i,i,i,i,i,j,j,k\}$ & 17 & $[3,1,1,1,1,1]$   & $\{i,i,i,j,k,l,m,p\}$ \\
7  & $[5,1,1,1]$         & $\{i,i,i,i,i,j,k,l\}$ & 18 & $[2,2,2,2]$       & $\{i,i,j,j,k,k,l,l\}$ \\
8  & $[4,4]$             & $\{i,i,i,i,j,j,j,j\}$ & 19 & $[2,2,2,1,1]$     & $\{i,i,j,j,k,k,l,m\}$ \\
9  & $[4,3,1]$           & $\{i,i,i,i,j,j,j,k\}$ & 20 & $[2,2,1,1,1,1]$   & $\{i,i,j,j,k,l,m,p\}$ \\
10 & $[4,2,2]$           & $\{i,i,i,i,j,j,k,k\}$ & 21 & $[2,1,1,1,1,1,1]$ & $\{i,i,j,k,l,m,p,q\}$ \\
11 & $[4,2,1,1]$         & $\{i,i,i,i,j,j,k,l\}$ & 22 & $[1,1,1,1,1,1,1,1]$ & $\{i,j,k,l,m,p,q,r\}$ \\
\end{tabular}
\end{ruledtabular}
\end{table}

\clearpage

\begin{table}[p]
\centering
\caption{Complete catalog of the 26 non-zero transition equivalence classes ($\vec{\Delta}_{\alpha\beta} = \vec{\Delta}_{\mu\nu}$), grouped by 8-index partition class and Weingarten Integral Group $\mathcal{O}_1$--$\mathcal{O}_{10}$. Both the ideal forward tensor $S^{(0)}_{\mu\nu,\alpha\beta}$ and the second-order GUE noise correction tensor $S^{(2)}_{\mu\nu,\alpha\beta}$ are given as closed-form rational functions of the mode count $M$.}
\label{tab:supp_26_complete}
\small
\setlength{\tabcolsep}{4pt}
\renewcommand{\arraystretch}{1.2}
\begin{ruledtabular}
\begin{tabular}{ccccc}
Partition Class & Integral Group & Transition Geometry: $\alpha \to \beta \mid \mu \to \nu$ & Ideal Tensor $S_{\mu\nu, \alpha\beta}^{(0)}$ & Quadratic Correction $S_{\mu\nu, \alpha\beta}^{(2)}$ \\
\hline
$[8]$ & $\mathcal{O}_{1}$ & $(i,i)\to(i,i) \mid (i,i)\to(i,i)$ & $\frac{8}{(M+1)(M+3)}$ & $-\frac{8(M-1)}{(M+1)(M+3)}$ \\
\hline
$[6, 2]$ & $\mathcal{O}_{2}$ & $(i,i)\to(i,i) \mid (i,j)\to(i,j)$ & $\frac{4}{(M+1)(M+3)}$ & $-\frac{2(M-5)}{(M+1)(M+3)}$ \\
 & & $(i,j)\to(i,j) \mid (i,i)\to(i,i)$ & $\frac{4}{(M+1)(M+3)}$ & $-\frac{2(M-5)}{(M+1)(M+3)}$ \\
 & & $(i,i)\to(i,j) \mid (i,i)\to(i,j)$ & $\frac{4}{(M+1)(M+3)}$ & $-\frac{2(3M+1)}{(M+1)(M+3)}$ \\
 & & $(i,j)\to(i,i) \mid (i,j)\to(i,i)$ & $\frac{4}{(M+1)(M+3)}$ & $-\frac{2(3M+1)}{(M+1)(M+3)}$ \\
\hline
$[4, 4]$ & $\mathcal{O}_{3}$ & $(i,i)\to(i,i) \mid (j,j)\to(j,j)$ & $\frac{2}{M(M+3)}$ & $\frac{4(M-1)}{M(M+1)(M+3)}$ \\
 & & $(i,i)\to(j,j) \mid (i,i)\to(j,j)$ & $\frac{2}{M(M+3)}$ & $-\frac{4(M+1)}{M(M+3)}$ \\
\cline{2-5}
 & $\mathcal{O}_{4}$ & $(i,i)\to(i,j) \mid (i,j)\to(j,j)$ & $\frac{2(M-1)}{M(M+1)(M+3)}$ & $-\frac{2(M^2 - 3M - 2)}{M(M+1)(M+3)}$ \\
 & & $(i,j)\to(i,i) \mid (j,j)\to(i,j)$ & $\frac{2(M-1)}{M(M+1)(M+3)}$ & $-\frac{2(M^2 - 3M - 2)}{M(M+1)(M+3)}$ \\
 & & $(i,j)\to(j,j) \mid (i,i)\to(i,j)$ & $\frac{2(M-1)}{M(M+1)(M+3)}$ & $-\frac{2(M^2 - 3M - 2)}{M(M+1)(M+3)}$ \\
\cline{2-5}
 & $\mathcal{O}_{5}$ & $(i,j)\to(i,j) \mid (i,j)\to(i,j)$ & $\frac{2(3M+1)}{M(M+1)(M+3)}$ & $-\frac{2(3M^2 - M + 2)}{M(M+1)(M+3)}$ \\
\hline
$[4, 2, 2]$ & $\mathcal{O}_{6}$ & $(i,i)\to(i,i) \mid (j,k)\to(j,k)$ & $\frac{2}{M(M+3)}$ & $\frac{4(M-1)}{M(M+1)(M+3)}$ \\
 & & $(i,j)\to(i,j) \mid (k,k)\to(k,k)$ & $\frac{2}{M(M+3)}$ & $\frac{4(M-1)}{M(M+1)(M+3)}$ \\
 & & $(i,i)\to(j,k) \mid (i,i)\to(j,k)$ & $\frac{2}{M(M+3)}$ & $-\frac{4(M+1)}{M(M+3)}$ \\
 & & $(i,j)\to(k,k) \mid (i,j)\to(k,k)$ & $\frac{2}{M(M+3)}$ & $-\frac{4(M+1)}{M(M+3)}$ \\
\cline{2-5}
 & $\mathcal{O}_{7}$ & $(i,i)\to(i,j) \mid (i,k)\to(j,k)$ & $\frac{\sqrt{2}(M-1)}{M(M+1)(M+3)}$ & $-\frac{\sqrt{2}(M^2 - 3M - 2)}{M(M+1)(M+3)}$ \\
 & & $(i,j)\to(i,i) \mid (j,k)\to(i,k)$ & $\frac{\sqrt{2}(M-1)}{M(M+1)(M+3)}$ & $-\frac{\sqrt{2}(M^2 - 3M - 2)}{M(M+1)(M+3)}$ \\
 & & $(i,j)\to(i,k) \mid (j,k)\to(k,k)$ & $\frac{\sqrt{2}(M-1)}{M(M+1)(M+3)}$ & $-\frac{\sqrt{2}(M^2 - 3M - 2)}{M(M+1)(M+3)}$ \\
 & & $(i,j)\to(j,k) \mid (i,i)\to(i,k)$ & $\frac{\sqrt{2}(M-1)}{M(M+1)(M+3)}$ & $-\frac{\sqrt{2}(M^2 - 3M - 2)}{M(M+1)(M+3)}$ \\
 & & $(i,j)\to(j,k) \mid (i,k)\to(k,k)$ & $\frac{\sqrt{2}(M-1)}{M(M+1)(M+3)}$ & $-\frac{\sqrt{2}(M^2 - 3M - 2)}{M(M+1)(M+3)}$ \\
\cline{2-5}
 & $\mathcal{O}_{8}$ & $(i,j)\to(i,j) \mid (i,k)\to(i,k)$ & $\frac{3M+1}{M(M+1)(M+3)}$ & $-\frac{M^2-7M+2}{M(M+1)(M+3)}$ \\
 & & $(i,j)\to(i,k) \mid (i,j)\to(i,k)$ & $\frac{3M+1}{M(M+1)(M+3)}$ & $-\frac{5M^2+5M+2}{M(M+1)(M+3)}$ \\
\hline
$[2, 2, 2, 2]$ & $\mathcal{O}_{9}$ & $(i,j)\to(i,j) \mid (k,l)\to(k,l)$ & $\frac{2}{M(M+3)}$ & $\frac{4(M-1)}{M(M+1)(M+3)}$ \\
 & & $(i,j)\to(k,l) \mid (i,j)\to(k,l)$ & $\frac{2}{M(M+3)}$ & $-\frac{4(M+1)}{M(M+3)}$ \\
\cline{2-5}
 & $\mathcal{O}_{10}$ & $(i,j)\to(i,k) \mid (j,l)\to(k,l)$ & $\frac{M-1}{M(M+1)(M+3)}$ & $-\frac{M^2-3M-2}{M(M+1)(M+3)}$ \\
 & & $(i,j)\to(j,k) \mid (i,l)\to(k,l)$ & $\frac{M-1}{M(M+1)(M+3)}$ & $-\frac{M^2-3M-2}{M(M+1)(M+3)}$ \\
\end{tabular}
\end{ruledtabular}
\end{table}

\clearpage

% ==============================================================================
% SECTION S-II: FULL-PAGE MATRIX ATLASES (M=4)
% ==============================================================================
\section{Full-Page Matrix Atlases of the Forward and Inverse Superoperator \texorpdfstring{($M=4$)}{(M=4)}}
\label{sec:supp_matrix_atlases}

\begin{figure}[htbp]
    \centering
    \includegraphics[width=0.92\linewidth]{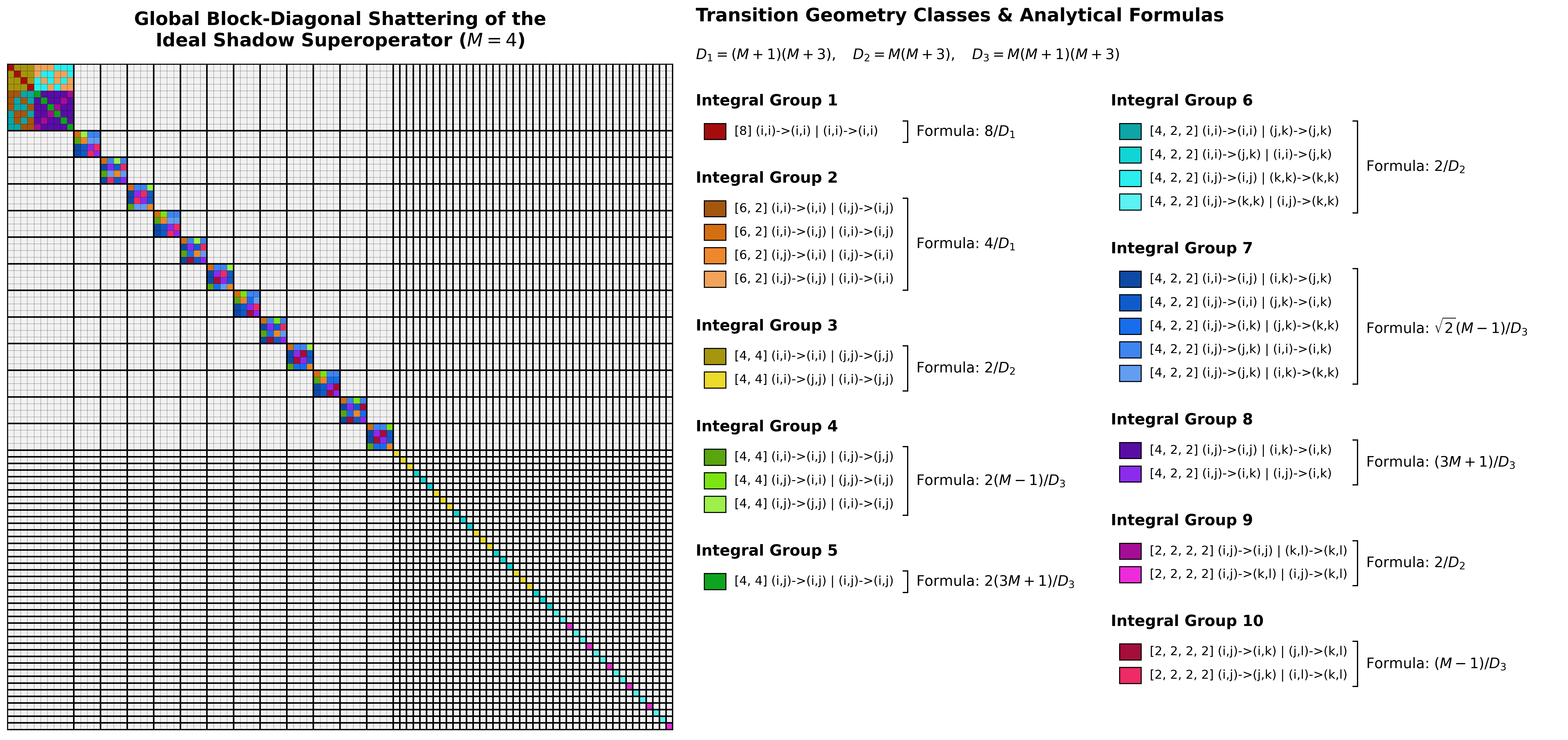}
    \caption{\textbf{Global forward transition superoperator $\mathbf{S}^{(0)}$ for $M=4$ ($100 \times 100$).} Entries vanish outside blocks sharing a common photon imbalance vector $\vec{\Delta}$. Non-zero entries are color-coded by their equivalence class in Table~\ref{tab:supp_26_complete}.}
    \label{fig:supp_global_s_matrix}
\end{figure}

\begin{figure}[htbp]
    \centering
    \includegraphics[width=0.68\textwidth]{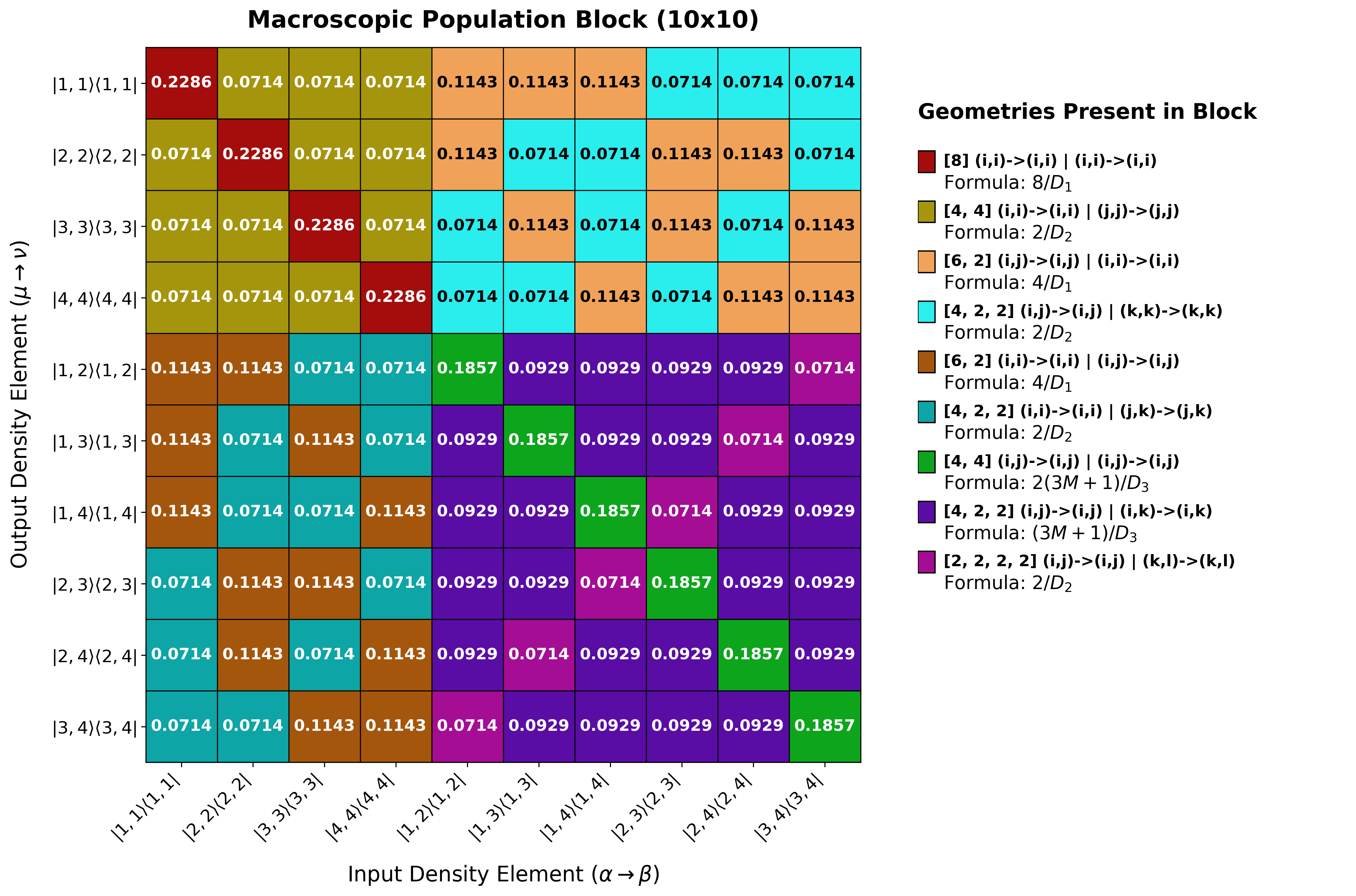}
    \vspace{0.3cm}
    \includegraphics[width=0.68\textwidth]{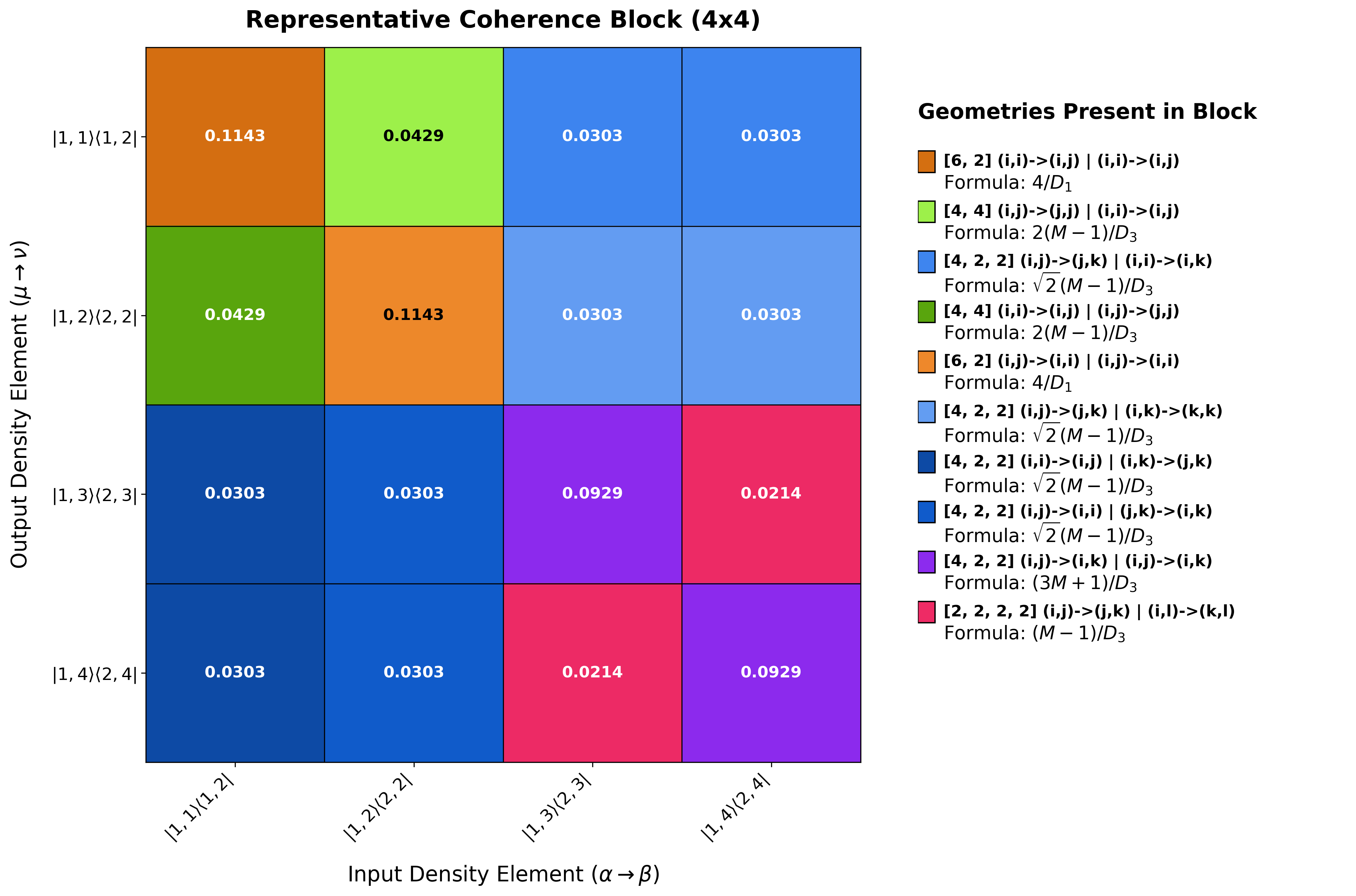}
    \caption{\textbf{Isolated forward sub-blocks for $M=4$.} \textbf{(Top)} The $10 \times 10$ Population Block $\mathbf{S}^{(0)}_{\vec{0}}$ ($d = D = 10$), ordered with the four collision states $\ket{i,i}$ followed by the six distinct states $\ket{i,j}$. \textbf{(Bottom)} A representative $4 \times 4$ Social Coherence Block $\mathbf{S}^{(0)}_{\vec{\Delta}}$ ($d = M = 4$) with $\Delta_1 = +1$, $\Delta_2 = -1$, spanned by the two collision-assisted transitions ($\ket{1,1}\bra{1,2}$, $\ket{1,2}\bra{2,2}$) and the two spectator-mode transitions ($\ket{1,3}\bra{2,3}$, $\ket{1,4}\bra{2,4}$).}
    \label{fig:supp_isolated_blocks}
\end{figure}

\begin{figure}[htbp]
    \centering
    \includegraphics[width=0.92\linewidth]{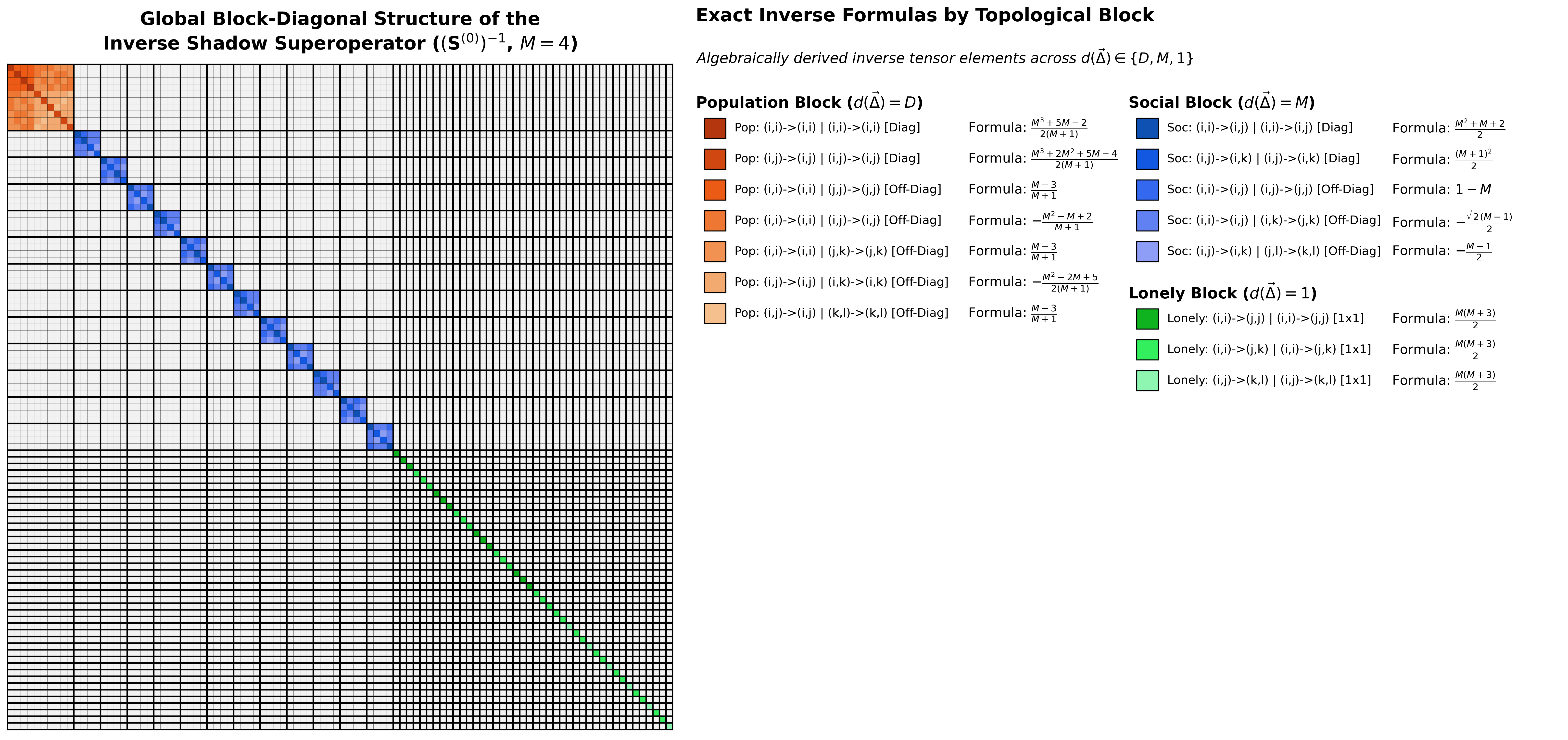}
    \caption{\textbf{Global inverse transition superoperator $(\mathbf{S}^{(0)})^{-1}$ for $M=4$ ($100 \times 100$).} Closure of the commutative Bose-Mesner algebra under inversion preserves the exact superselection topology $d(\vec{\Delta}) \in \{D, M, 1\}$.}
    \label{fig:supp_global_inverse_s_matrix}
\end{figure}

\begin{figure}[htbp]
    \centering
    \includegraphics[width=0.68\textwidth]{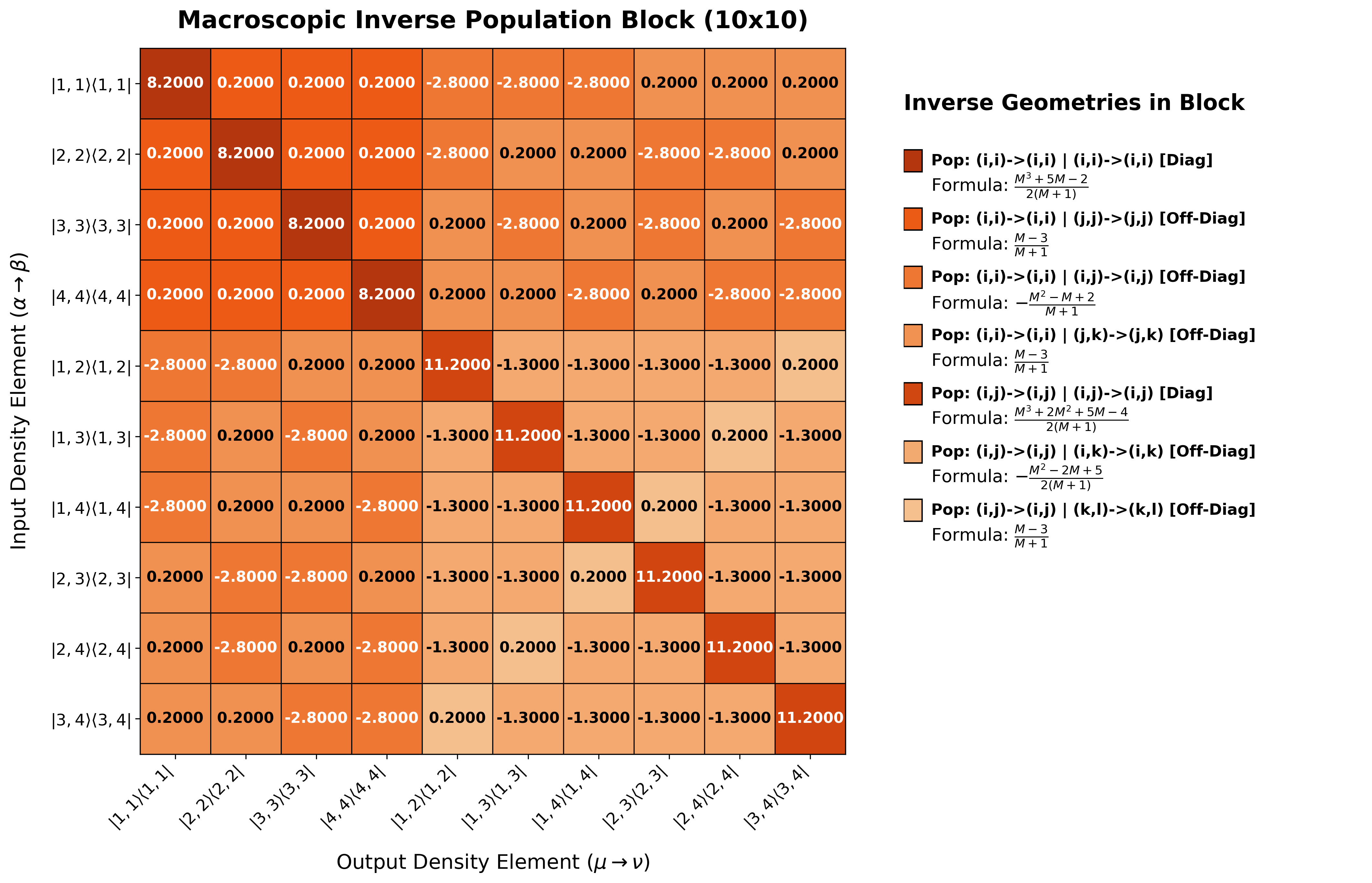}
    \vspace{0.3cm}
    \includegraphics[width=0.68\textwidth]{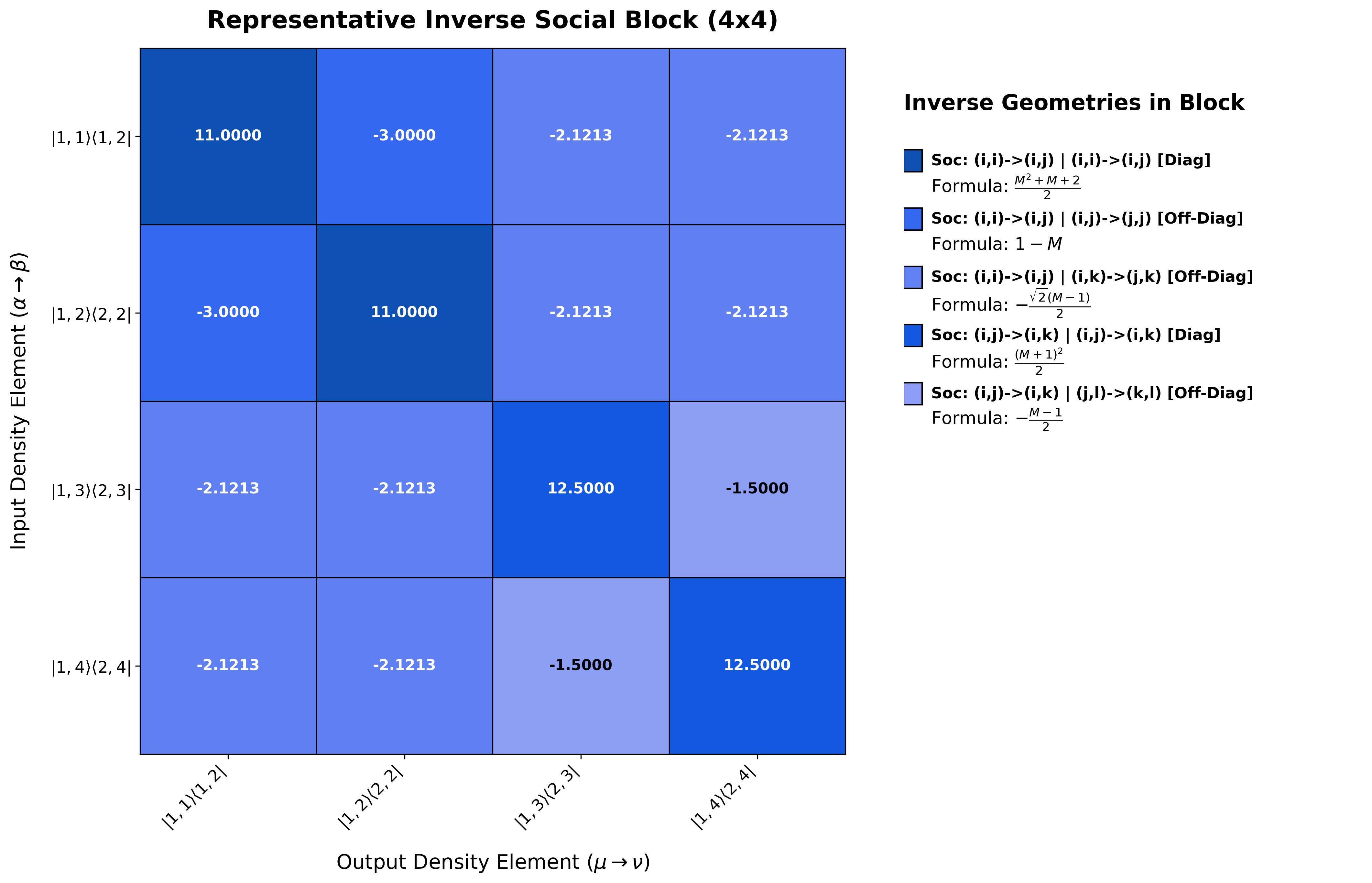}
    \caption{\textbf{Isolated inverse sub-blocks for $M=4$.} \textbf{(Top)} The $10 \times 10$ Inverse Population Block $(\mathbf{S}^{(0)}_{\vec{0}})^{-1}$. \textbf{(Bottom)} A representative $4 \times 4$ Inverse Social Coherence Block $(\mathbf{S}^{(0)}_{\vec{\Delta}})^{-1}$. Exact rational values match Table~\ref{tab:exact_inverse_formulas_final} of the main text evaluated at $M=4$.}
    \label{fig:supp_isolated_inverse_blocks}
\end{figure}

To visualize the block-diagonal structure induced by the $U(1)^{\otimes M}$ superselection rule, we flatten the rank-4 superoperator $\mathbf{S}^{(0)}$ into a $D^2 \times D^2$ matrix by mapping input transitions $\ket{\alpha}\bra{\beta}$ to column index $C = (\alpha - 1)D + \beta$ and output measurements $\ket{\mu}\bra{\nu}$ to row index $R = (\mu - 1)D + \nu$. For $M=4$ modes ($D = 10$), the $100 \times 100$ matrix decomposes into one $10 \times 10$ Population Block ($\vec{\Delta} = \vec{0}$), twelve $4 \times 4$ Social Coherence Blocks ($d(\vec{\Delta}) = 4$), and 42 isolated $1 \times 1$ Lonely Coherence Blocks ($d(\vec{\Delta}) = 1$), satisfying $10 + 12 \times 4 + 42 \times 1 = 100$. Figures~\ref{fig:supp_global_s_matrix}--\ref{fig:supp_isolated_inverse_blocks} display the complete forward and inverse matrix atlases and their isolated sub-blocks.

\clearpage

% ==============================================================================
% SECTION S-III: COMPLETE DENSITY-MATRIX RECONSTRUCTION HEATMAPS
% ==============================================================================
\section{Complete Density-Matrix Reconstruction Heatmaps \texorpdfstring{($M=4$)}{(M=4)}}
\label{sec:supp_heatmaps}

\begin{figure}[htbp]
    \centering
    \includegraphics[width=0.72\textwidth]{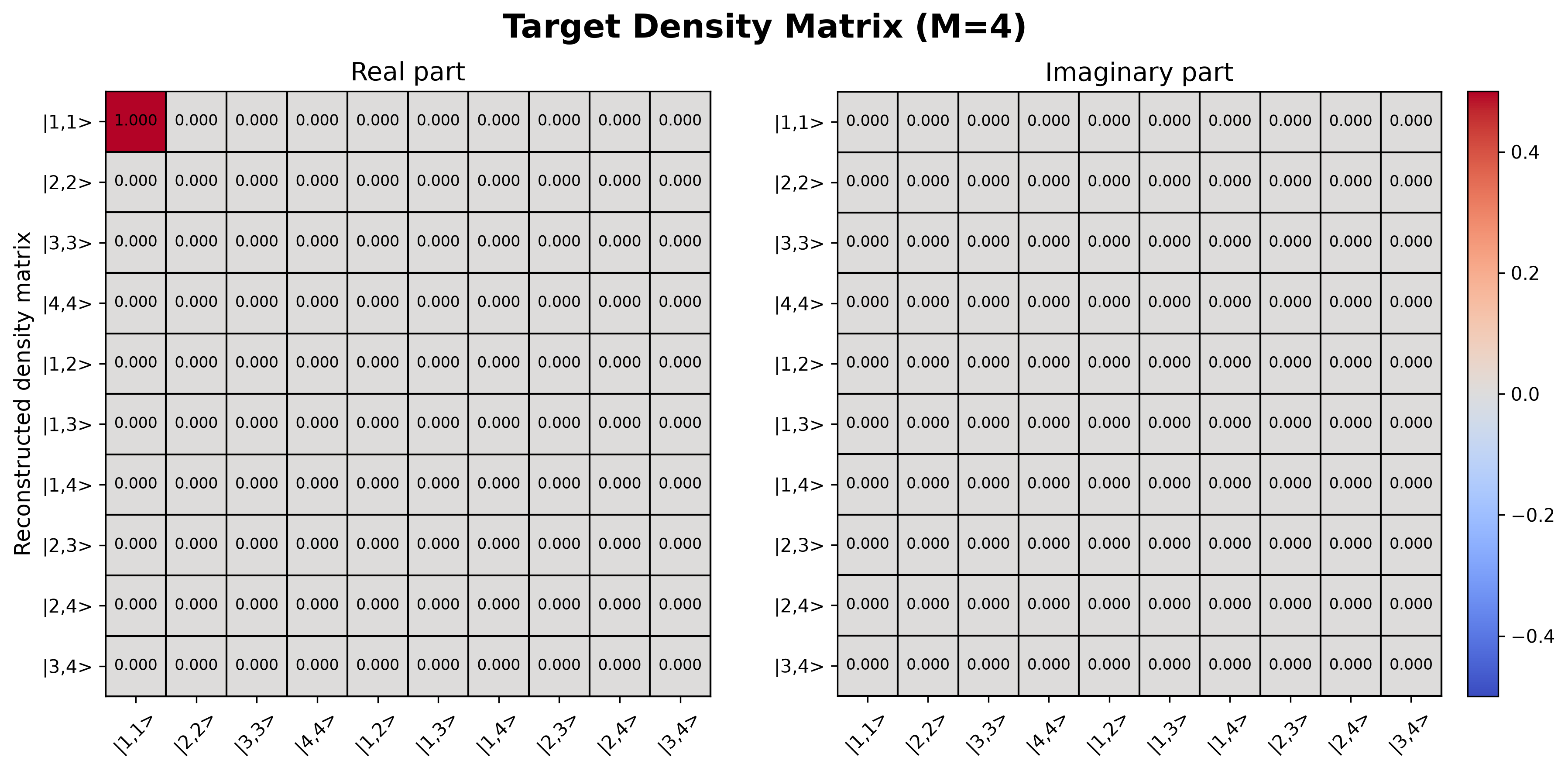}
    \caption{\textbf{Analytical Target State ($M=4$).} Real and imaginary parts of $\rho_{\text{target}} = \ket{1,1}\bra{1,1}$ in the 10-dimensional two-photon Fock basis.}
    \label{fig:supp_heatmap_target}
\end{figure}

\begin{figure}[htbp]
    \centering
    \includegraphics[width=0.72\textwidth]{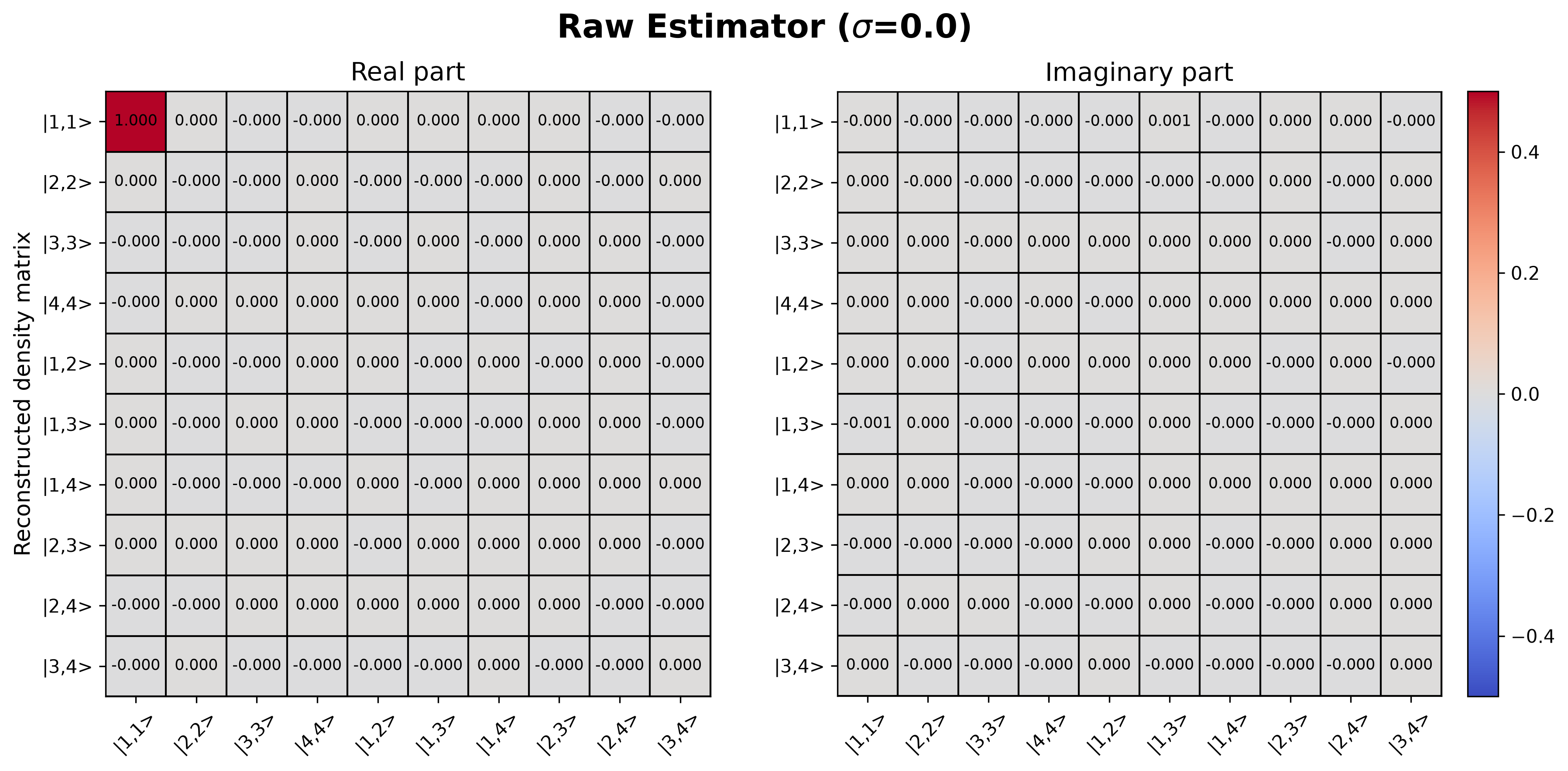}
    \caption{\textbf{Noise-Free Baseline ($\sigma = 0.0$, $M=4$).} Ensemble average of 30 independent runs ($N = 4 \times 10^5$ trajectories each) using the ideal inverse $(\mathbf{S}^{(0)})^{-1}$.}
    \label{fig:supp_heatmap_raw_00}
\end{figure}

\begin{figure}[htbp]
    \centering
    \includegraphics[width=0.72\textwidth]{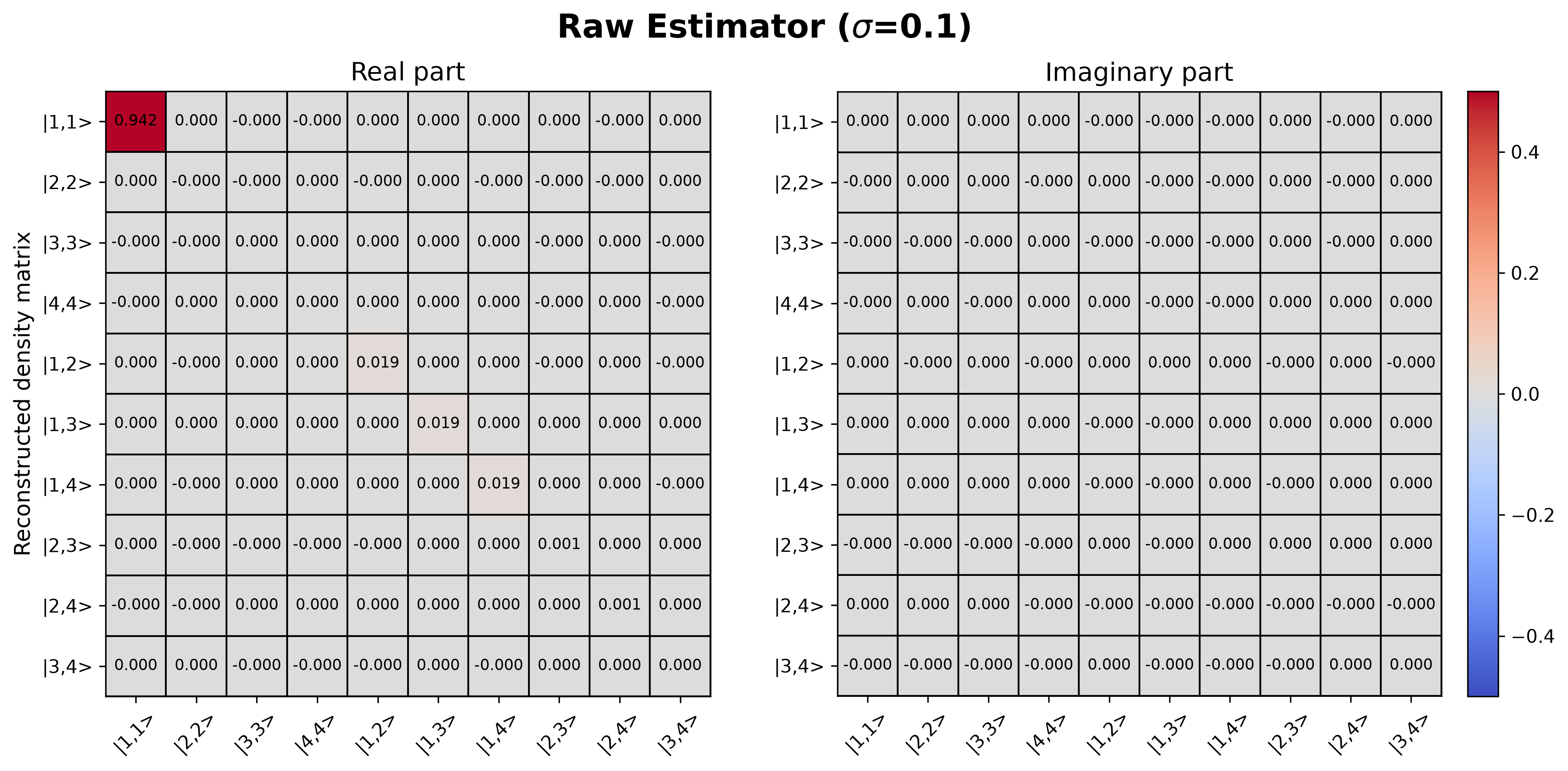}
    \caption{\textbf{Uncalibrated Raw Estimator under Coherent Noise ($\sigma = 0.1$, $M=4$).} Probability leaks from $\ket{1,1}\bra{1,1}$ into the three states $\ket{1,2}\bra{1,2}$, $\ket{1,3}\bra{1,3}$, and $\ket{1,4}\bra{1,4}$ within the $\vec{\Delta} = \vec{0}$ population sector, matching $-\sigma^2 \mathcal{D}(\ket{1,1}\bra{1,1})$.}
    \label{fig:supp_heatmap_raw_01}
\end{figure}

\begin{figure}[htbp]
    \centering
    \includegraphics[width=0.72\textwidth]{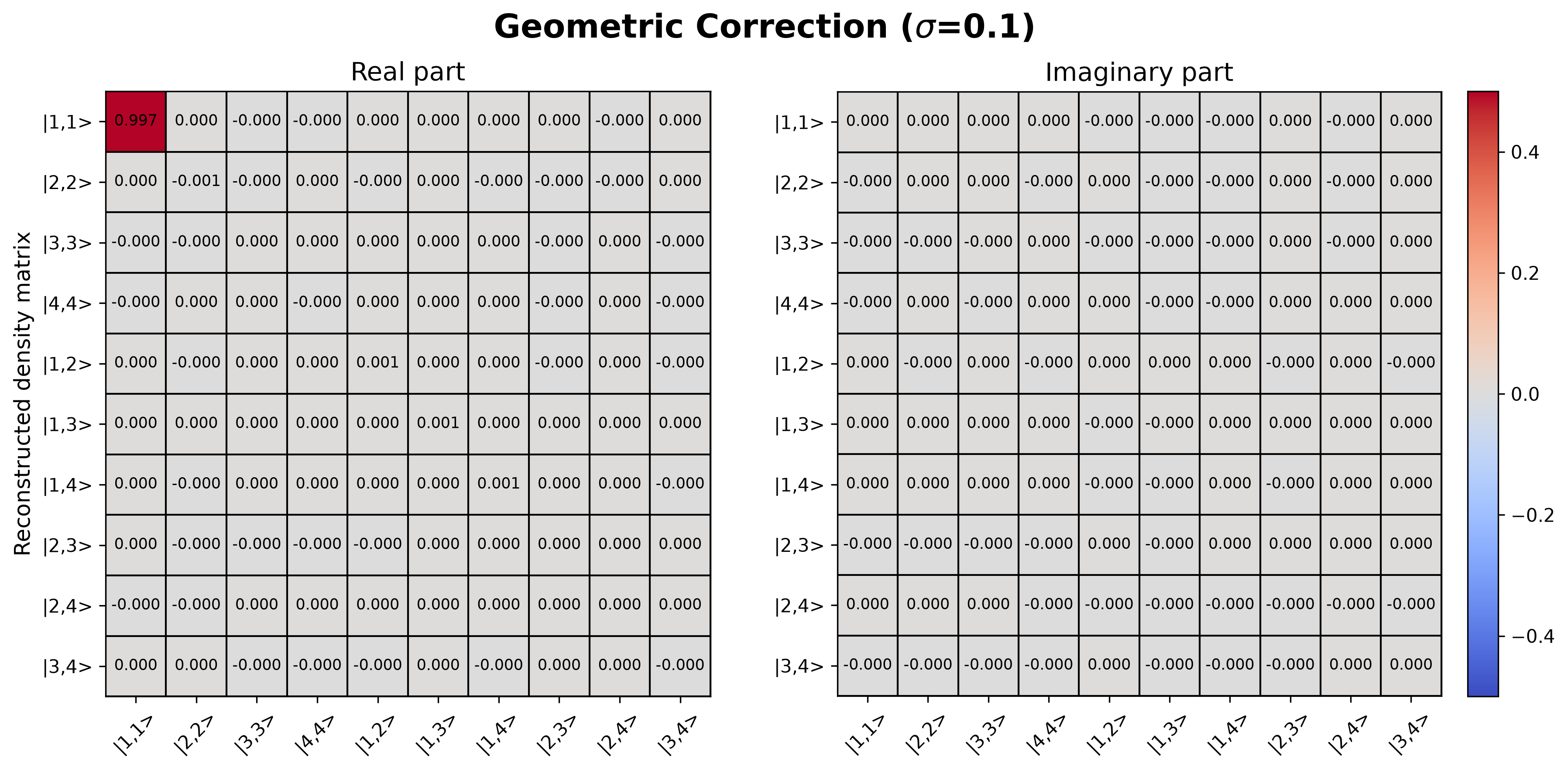}
    \caption{\textbf{Linear Geometric Correction ($\sigma = 0.1$, $M=4$).} First-order Neumann calibration $\hat{\rho}^{(\text{geom})} = (\mathcal{I} + \sigma^2 \mathcal{D})\hat{\rho}_{\text{raw}}$ suppresses leading $\mathcal{O}(\sigma^2)$ intra-block diffusion.}
    \label{fig:supp_heatmap_geom}
\end{figure}

\begin{figure}[htbp]
    \centering
    \includegraphics[width=0.72\textwidth]{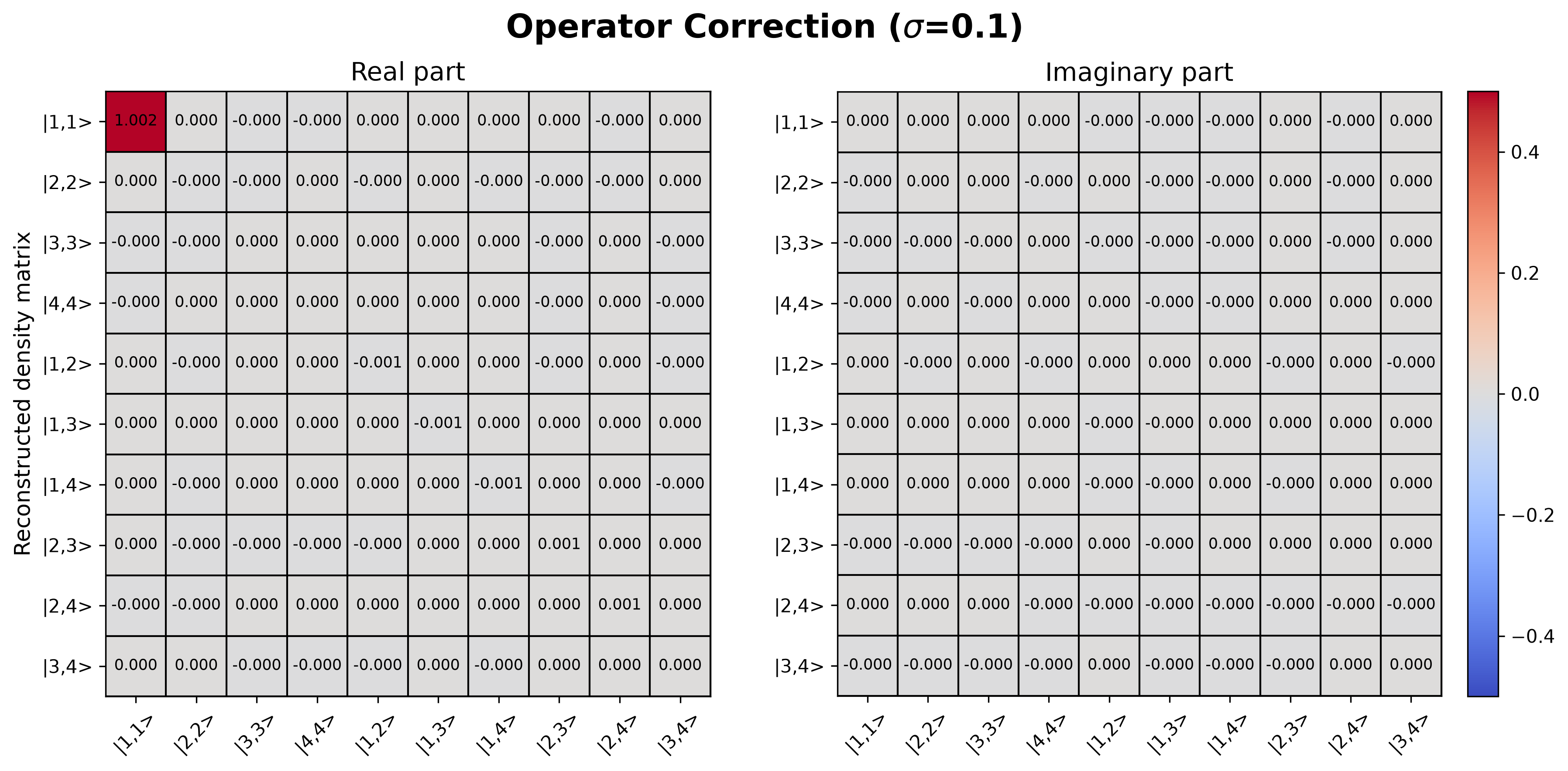}
    \caption{\textbf{Exact Operator Correction ($\sigma = 0.1$, $M=4$).} Full block inversion $\hat{\rho}^{(\text{op})} = (\mathbf{S}^{(0)}_{\vec{\Delta}} + \sigma^2 \mathbf{S}^{(2)}_{\vec{\Delta}})^{-1}\overline{I}$ eliminates intra-block leakage to all geometric orders.}
    \label{fig:supp_heatmap_op}
\end{figure}

In Sec.~\ref{sec:numerical_verification} of the main text, we benchmark the classical shadow reconstruction protocol and active noise-mitigation estimators on an $M=4$ interferometer ($D=10$) initialized in the pure collision state $\ket{\psi_1} = \ket{1,1}$, averaging over 30 independent Monte Carlo ensembles of $N = 4 \times 10^5$ trajectories each. Figures~\ref{fig:supp_heatmap_target}--\ref{fig:supp_heatmap_op} show the full $10 \times 10$ real and imaginary parts of the reconstructed density matrix across all five benchmark conditions:
\begin{enumerate}
    \item \textbf{Analytical Target State (Fig.~\ref{fig:supp_heatmap_target}):} $\rho_{\text{target}} = \ket{1,1}\bra{1,1}$, with unit amplitude in the $(1,1)$ corner and zero elsewhere.
    \item \textbf{Noise-Free Baseline at $\sigma = 0.0$ (Fig.~\ref{fig:supp_heatmap_raw_00}):} Uncalibrated estimator $\hat{\rho}_{\text{raw}} = (\mathbf{S}^{(0)})^{-1}\overline{I}$ over $N = 4 \times 10^5$ shots, confirming unbiased convergence to $\rho_{\text{target}}$ within $\mathcal{O}(N^{-1/2})$ statistical fluctuations.
    \item \textbf{Uncalibrated Raw Estimator at $\sigma = 0.1$ (Fig.~\ref{fig:supp_heatmap_raw_01}):} Under GUE coherent perturbations ($\sigma = 0.1$), the diagonal amplitude $\hat{\rho}_{\text{raw}, (1,1),(1,1)}$ drops from $1.000$ to $1 - 2(M-1)\sigma^2 \approx 0.94$, while positive spurious populations of magnitude $2\sigma^2 \approx 0.02$ appear in the three single-collision states $\ket{1,2}\bra{1,2}$, $\ket{1,3}\bra{1,3}$, and $\ket{1,4}\bra{1,4}$ that share mode $1$. Off-diagonal coherences ($\vec{\Delta} \neq \vec{0}$) remain unaffected in expectation, confirming that GUE noise scattering is strictly confined within the $\vec{\Delta} = \vec{0}$ superselection sector.
    \item \textbf{Linear Geometric Correction at $\sigma = 0.1$ (Fig.~\ref{fig:supp_heatmap_geom}):} Applying the first-order Neumann correction $\hat{\rho}^{(\text{geom})} = (\mathcal{I} + \sigma^2 \mathcal{D})\hat{\rho}_{\text{raw}}$ removes the $\mathcal{O}(\sigma^2)$ leakage into $\ket{1,2}$, $\ket{1,3}$, and $\ket{1,4}$, leaving only small $\mathcal{O}(\sigma^4)$ residuals.
    \item \textbf{Exact Operator Correction at $\sigma = 0.1$ (Fig.~\ref{fig:supp_heatmap_op}):} Inverting the calibrated block $\hat{\rho}^{(\text{op})} = (\mathbf{S}^{(0)}_{\vec{\Delta}} + \sigma^2 \mathbf{S}^{(2)}_{\vec{\Delta}})^{-1}\overline{I}$ sums intra-block scattering to all geometric orders, restoring the reconstructed density matrix to the noise-free baseline.
\end{enumerate}

\end{document}